\documentclass[aps,prd,nofootinbib,showkeys,floatfix,twocolumn,bibnotes,preprintnumbers]{revtex4-1}
\usepackage{graphicx,amsmath,amssymb,hyperref,natbib,slashed}
\usepackage[dvipsnames]{xcolor}
\usepackage{xspace}
\usepackage{bbm} 

\usepackage[utf8]{inputenc}
\usepackage[normalem]{ulem}

\newcommand{\be}{\begin{equation}}
\newcommand{\ee}{\end{equation}}
\newcommand{\bea}{\begin{eqnarray}}
\newcommand{\eea}{\end{eqnarray}}

\begin{document}
\title{Converting non-relativistic dark matter to radiation}

\preprint{DESY-18-032, TTK-18-10}

\newcommand{\AddrOslo}{%
Department of Physics, University of Oslo, Box 1048, N-0371 Oslo, Norway}
\newcommand{\AddrDESY}{%
Deutsches Elektronen-Synchrotron DESY,   Notkestra\ss e 85, D-22607 Hamburg, Germany}
\newcommand{\AddrAachen}{%
Institute for Theoretical Particle Physics and Cosmology (TTK), RWTH Aachen University, D-52056 Aachen, Germany}

 \author{Torsten Bringmann}
 \email{torsten.bringmann@fys.uio.no}
 \affiliation{\AddrOslo}
 
  \author{Felix Kahlhoefer}
 \email{kahlhoefer@physik.rwth-aachen.de}
 \affiliation{\AddrAachen}
 
 \author{Kai Schmidt-Hoberg}
 \email{kai.schmidt-hoberg@desy.de}
 \affiliation{\AddrDESY}
 
  \author{Parampreet Walia}
 \email{p.s.walia@fys.uio.no}
 \affiliation{\AddrOslo~}

\begin{abstract}
Dark matter in the cosmological concordance model is parameterised by a single number,
describing the covariantly conserved energy density of a non-relativistic fluid. 
Here we test this assumption in a model-independent and conservative way by considering the 
possibility that, at any point during the cosmological evolution, dark matter may be 
converted into a non-interacting form of radiation. This scenario encompasses, but is more general than,
the cases where dark matter decays or annihilates into these states. We show that observations of the cosmic 
microwave background allow to strongly constrain this scenario for {any conversion time} after big bang nucleosynthesis.
We discuss in detail, both from a Bayesian and frequentist point of view, in which sense 
adding large-scale structure observations may even provide a certain preference for
a conversion of dark matter to radiation at late times. Finally we apply our general results to a 
specific particle physics 
realisation of such a  scenario, featuring late kinetic decoupling and Sommerfeld-enhanced dark matter
annihilation. We identify a small part of parameter space that both
mitigates the tension between cosmic microwave and large-scale structure data and allows for
velocity-dependent dark matter self-interactions strong enough to address the small-scale problems of structure 
formation.

\end{abstract}


\maketitle

\section{Introduction}

There is overwhelming evidence for the existence of dark matter (DM) in our Universe from various astrophysical and 
cosmological observations. While many of its particle physics properties are completely unknown, the amount of DM 
at the time of recombination has been precisely determined through observations of the Cosmic Microwave 
Background (CMB)~\cite{Ade:2015xua}.  The corresponding DM relic abundance is typically assumed to have been 
set early on, at  temperatures comparable to the DM mass in the most commonly considered scenario of 
thermally produced DM particles~\cite{Lee:1977ua,Gondolo:1990dk}, such that the comoving DM density is 
constant throughout the subsequent cosmological evolution. 

In this work we analyse how cosmological observations constrain deviations from the simple picture of a comovingly
constant DM density. An interesting example for a possible underlying mechanism
is if all or a part of the DM is 
unstable.
If the decay products are Standard Model (SM) states such as electrons or photons, a scenario of this type will be 
strongly constrained by a variety of cosmological and astrophysical probes (see 
e.g.~\cite{Cirelli:2012ut,Ibarra:2013cra,Poulin:2016anj}).
It is however an interesting possibility that the decay products are new massless or very light states in the dark 
sector, such that effectively a fraction of DM is converted into relativistic `dark' radiation 
(DR)~{\cite{Zentner:2001zr,Ichiki:2004vi, Lattanzi:2007ux,Peter:2010au,Wang:2010ma,Bjaelde:2012wi,Allahverdi:2014bva,Audren:2014bca,Poulin:2016nat}}. 
Such a conversion has received some interest lately as it has been argued to alleviate a possible tension between 
measurements of the CMB and large scale structure (LSS) 
observables~\cite{Enqvist:2015ara,Berezhiani:2015yta,Blackadder:2015uta,Pourtsidou:2016ico,Poulin:2016nat,Chudaykin:2016yfk,Hamaguchi:2017ihw}. 

\begin{figure*}[t!]
\includegraphics[trim={0 12 0 35},clip,width=1.01\columnwidth]{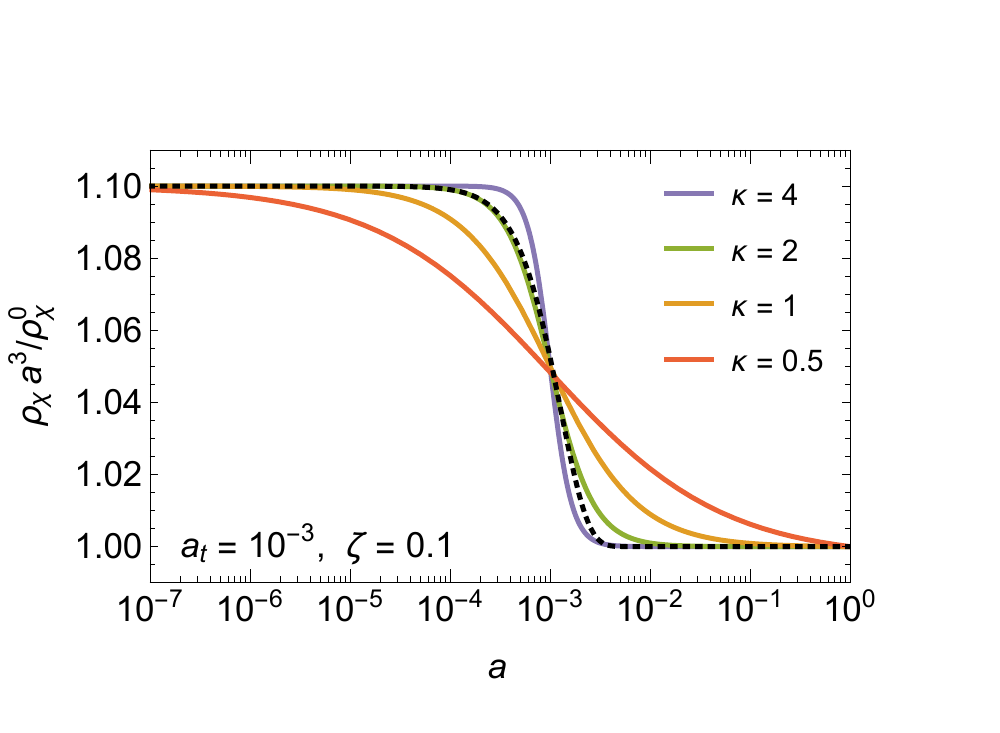}
\includegraphics[trim={0 13 0 35},clip,width=1.01\columnwidth]{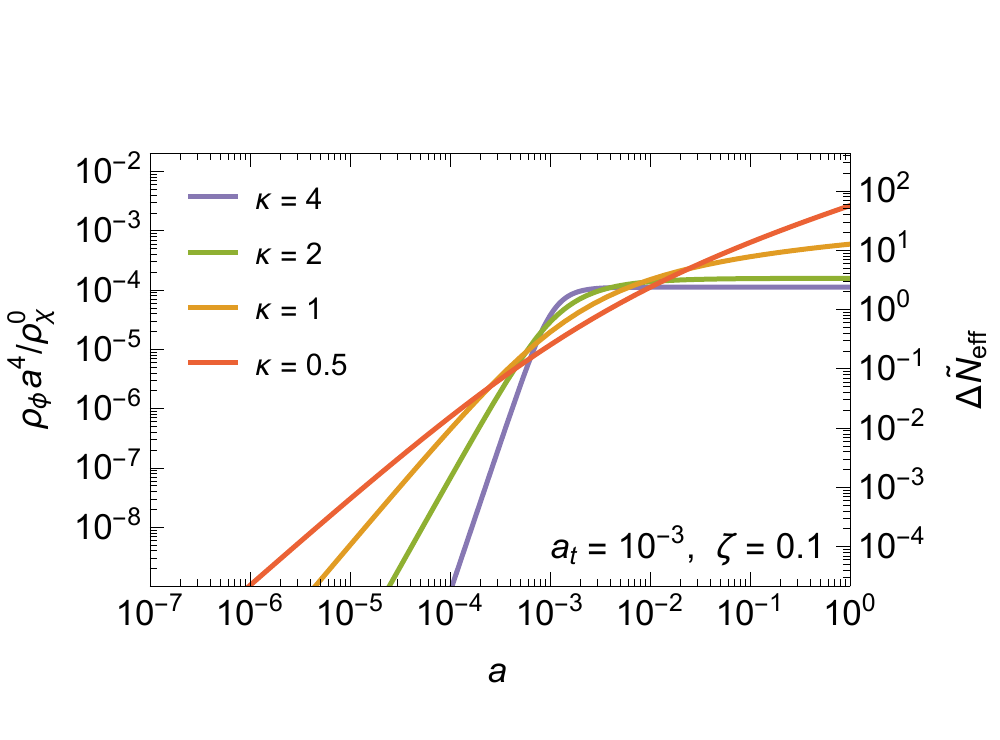}
\caption{{\it Left panel}. Evolution of comoving DM density for the step-like transition described by 
Eq.~(\ref{rhodmeq}), for a transition redshift of $a_t=10^{-3}$,  a 
conversion factor of $1+\zeta=1.1$ and,
as indicated, four values of the parameter $\kappa$ characterising the steepness of the transition.
For comparison, we also show the case of {\it decaying DM} (dotted line), assuming that a
fraction $\zeta/(1+\zeta)$ of the initial DM abundance decays with a rate $\Gamma= 0.15 H_{\rm eq}$. 
{\it Right panel}. Resulting evolution of the comoving DR density as given in Eq.~(\ref{rhodreq}). 
This assumes that there is no additional (e.g.~constant) source of DR and, for the translation to
$\Delta \tilde N_{\rm eff}$ as defined in Eq.~(\ref{eg:neff}), we have here chosen $\rho_\chi^0$ to 
agree with the value of $\Omega_\chi^0 h^2=0.1198$ measured by Planck.}
\label{rhodmfig}
\end{figure*}

A second example in which the comoving dark matter density can change is if the DM annihilation rate becomes relevant 
at late times, which may happen if the annihilations experience a sufficiently strong Sommerfeld 
enhancement~\cite{Dent:2009bv,Zavala:2009mi,Feng:2010zp,vandenAarssen:2012ag,
ArmendarizPicon:2012mu,Binder:2017lkj}. 
Yet another case where DM may be converted into DR is given by 
merging primordial black holes emitting gravitational waves~\cite{Nakamura:1997sm,Raidal:2017mfl}, 
a scenario currently receiving a lot of interest due to the observations by advanced LIGO~\cite{Abbott:2016blz}. 
We note that also ordinary astrophysical processes can convert matter into radiation, but only at rates
below the sensitivity of (near) future observation \cite{Torres:2016fmj}.

In this work we employ data from the CMB as well as LSS observables to constrain the possibility of DM 
being converted into DR in a model-independent way. Clearly the {\it amount} of DM which is allowed to be converted 
into DR will depend on the {\it time} of this conversion, given that the relative contributions of matter and radiation to 
the overall energy density change as the Universe evolves. Also the {\it rate} of this conversion is expected to have 
an impact on the constraints.
We will concentrate on conversion times well after the end of primordial 
nucleosynthesis, as sufficiently early transitions can always be mapped onto a cosmology with a 
constant additional radiation component, $\Delta N_\text{eff} > 0$.\footnote{BBN constraints of a possible 
DM-DR conversion have recently been studied in Ref.~\cite{Hufnagel:2017dgo}.} 

This article is structured as follows: In the next section we will discuss how we implement the DM-DR transition. In 
Sec.~\ref{sec:cmb} we will discuss the effects on the CMB as well as the resulting constraints, while 
Sec.~\ref{sec:lss} is devoted to the discussion of low redshift observables. In Sec.~\ref{sec:sommer} we will 
map our general constraints to the case of Sommerfeld enhanced annihilation, before we conclude in 
Sec.~\ref{sec:discussion}.

\section{Converting dark matter to dark radiation}
\label{sec:DMtoDR}

As motivated in the introduction, our aim is to quantify in rather general terms {\it i)} {\it how much} DM 
can be converted to DR, as well as how this depends on the {\it ii) time} and {\it iii) rate} of this conversion. 
Phenomenologically we are thus interested in a step-like transition in the comoving DM density as shown 
in the left panel of Fig.~\ref{rhodmfig} where, at least for the moment, we choose to remain completely agnostic
about the underlying mechanism that causes such a transition. {Nevertheless, we emphasise that the parametrisation is sufficiently general to capture a range of interesting scenarios, such as the case of a decaying DM sub-component (indicated by a black dotted line in Fig.~\ref{rhodmfig}) and Sommerfeld-enhanced DM annihilations. The latter case will be the subject of Sec.~\ref{sec:sommer}, where we will discuss in detail how to map the underlying particle physics parameters onto the effective parametrisation discussed in this section.}

\subsection{Evolution of background densities}
\label{background}

In the following, we will adopt a simple parametric form for the DM density $\rho_\chi (a)$ as shown
in Fig.~\ref{rhodmfig}, namely
\begin{align}
\rho_\chi (a)&= \frac{\rho^0_\chi}{a^3}\left[1+ \zeta\frac{1-a^\kappa}{1 +(a/a_t)^\kappa}\right]\,.
\label{rhodmeq}
\end{align}
Here $a$ denotes the scale factor of the Friedman-Robertson-Walker (FRW) metric, $\rho^0_\chi \equiv \rho_\chi(1)$
the DM density today, and the three parameters
$(\zeta,a_t,\kappa)$ directly relate to the points {\it i) -- iii)} raised above. Specifically,
the comoving DM density decreases in total by a factor of $1+\zeta$, the transition is centred at 
$a=a_t$, and the parameter $\kappa$ 
determines how fast the transition occurs.

{This parametrisation enables us in particular to understand which properties of DM-DR conversion are constrained observationally. For example, we will see below that for a conversion after recombination constraints are largely independent on when and how quickly the transition occurs, but {mostly} depend only on the total amount of DM converted to DR.  A similar observation was previously made for the case of a sub-dominant component of DM decaying into DR~\cite{Poulin:2016nat}, and our findings generalise this result. Conversely, for a very early transition, we find constraints to depend only on the total amount of DR produced, which can be described by the effective number of neutrino {specie}s $N_\text{eff}$.}
{For transitions around matter-equality, on the other hand, the constraints can no longer be understood in terms
of these simple limiting behaviours, and depend in a more complicated way on when and how quickly the conversion takes place.}

{As already stressed, the phenomenological parametrisation suggested above allows to capture a significant
range of cosmologically interesting scenarios.} 
{For example, we find that the case of a decaying DM sub-component can be accurately described by setting $\kappa = 2$ and choosing $a_t$ such that the Hubble expansion rate at the transition is comparable to the decay rate. Sommerfeld-enhanced annihilations, on the other hand, can be accurately matched by setting $\kappa = 1$ (see Sec.~\ref{sec:sommer}).}

By assumption, we demand that this transition occurs because DM is being converted to radiation.
The rates of change of the {\it comoving} DM and DR densities must thus be of equal size, 
and opposite in sign:
\be
\label{eq:Econs}
\frac{1}{a^3}\frac{\mathrm{d}}{\mathrm{d}t}\left(a^3\rho_\chi\right) 
= - \frac{1}{a^4}\frac{\mathrm{d}}{\mathrm{d}t}\left(a^4\rho_\phi\right)\,.
\ee
Alternatively, we can write this statement in terms of coupled Boltzmann equations for the
two fluid components:
\bea
\frac{\mathrm{d}\rho_\chi}{\mathrm{d}t} + 3 H \rho_\chi &\equiv& -\mathcal{Q} \label{eq:Boltz_DM}\\
\frac{\mathrm{d}\rho_\phi}{\mathrm{d}t} + 4 H \rho_\phi &=& \mathcal{Q}\,,\label{eq:Boltz_DR}
\eea
where $H=\dot a/a$ is the Hubble rate and $\mathcal{Q}>0$ describes the (momentum-integrated) collision term.
In this formulation, being agnostic about the underlying mechanism of the DM to DR transition simply
means, as indicated, that we start from Eq.~(\ref{rhodmeq}) and view Eq.~(\ref{eq:Boltz_DM}) as 
a {\it definition} for  $\mathcal{Q}$ -- rather than determining $\rho_\chi$ from 
a given collision term.

\begin{figure*}[t]
\includegraphics[width=0.95\columnwidth]{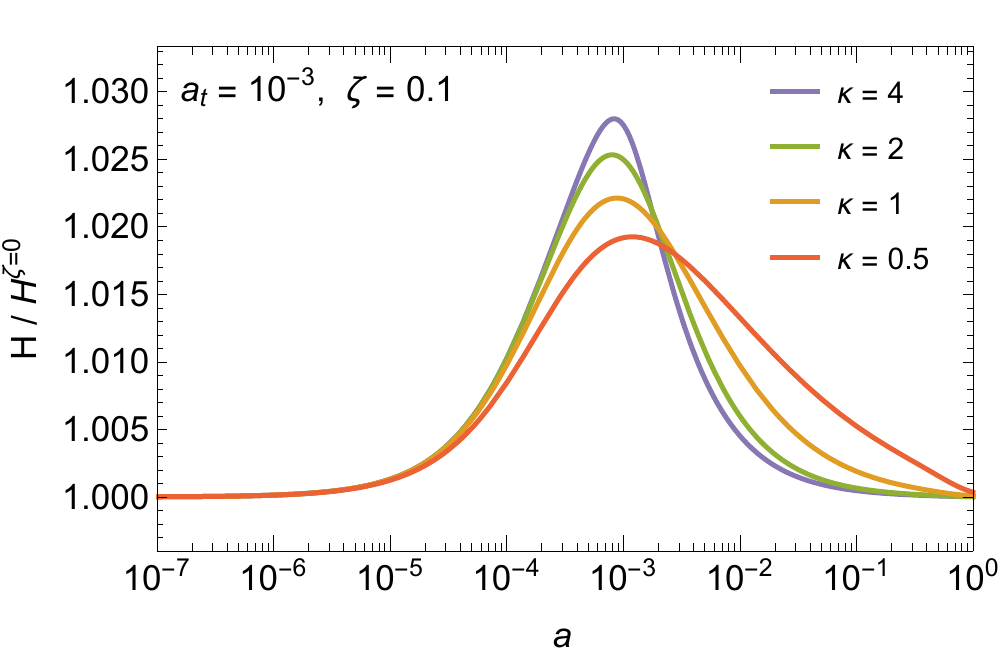}
~~~~~~~\includegraphics[width=0.95\columnwidth]{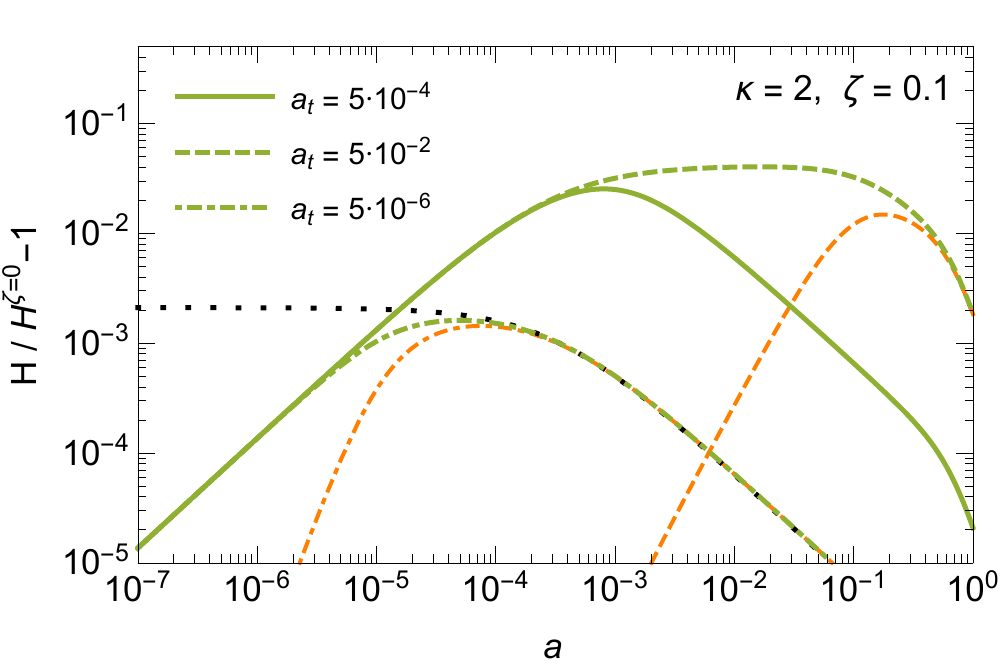}
\caption{ 
{\it Left panel.} Evolution of Hubble rate for the same scenarios as shown in Fig.~\ref{rhodmfig},
compared to the $\Lambda$CDM Hubble rate $H^{\zeta=0}$ (which in our scenarios 
is obtained for $\zeta=0$),
{\it Right panel.} Impact of changing $a_t$ on the Hubble rate, for $\kappa=2$. Orange (thinner) lines
indicate the impact of the produced DR alone. For the $a_t=5\cdot 10^{-6}$ case we show, for comparison,
also how the Hubble rate is affected by a {\it constant} DR contribution, characterised by a constant
$\Delta N_{\rm eff}$ (black dotted line).}
\label{fig:deltaH}
\end{figure*}

We can now obtain the DR energy density by integrating Eq.~(\ref{eq:Econs}), with the boundary 
condition $\rho_\phi(a\to0)=0$. This leads to
\begin{align}
\label{rhodreq}
\rho_\phi(a)&=\zeta \frac{\rho_\chi^0}{a^3}   \frac{\left(1+a_t^\kappa\right)}{\left(a^\kappa+a_t^\kappa\right)}\\
&\times 
   \left(\left(a^\kappa+a_t^\kappa\right) \,
   _2F_1\left[1,\frac{1}{\kappa};1+\frac{1}{\kappa};-\left(\frac{a}{a_t}\right)^\kappa\right]-a_t^\kappa\right)\,,\nonumber
\end{align}
{where $_2F_1$ denotes the ordinary Hypergeometric functions.}
Let us stress that the above solution for 
the DR energy density $\rho_\phi(a)$ does not explicitly depend on the form of $H$, which is one of the
advantages of our parametrisation for $\rho_\chi(a)$. This implies that also the transition from radiation to 
matter domination is fully and consistently covered in this approach (at least at the level
of the evolution of background densities). In the right panel of Fig.~\ref{rhodmfig}, we show how the 
DR density evolves, according to Eq.~(\ref{rhodreq}), for the  $\rho_\chi(a)$ scenarios plotted in the left
panel. To facilitate comparison with the literature, we also indicate the amount of DR in terms
of an effective number of additional neutrino species, by defining
\begin{equation}
\label{eg:neff}
\Delta \tilde{N}_{\text{eff}}(a) \equiv   \frac{\rho_\phi (a)}{\rho_{1\nu}(a)}=\frac87 \left(\frac{11}{4}\right)^{4/3}  
\frac{\rho_\phi (a)}{\rho_\gamma(a)} \;,
\end{equation}
where the last equality is only valid for sufficiently late times (after $e^\pm$ annihilation).
For $\rho_\phi\propto a^{-4}$, this reduces to the standard definition of the effective number of 
additional neutrino species, $\Delta \tilde{N}_{\text{eff}}\to\Delta {N}_{\text{eff}}$, 
typically used to describe a (comovingly) constant contribution of DR. In the scenarios that we describe
here, the comoving DR density is {\it not} constant (but saturates for $a\gg a_t$ if $\kappa>1$).

We note that the large range of transition histories that we consider here essentially also includes the case of 
decaying DM, which much of the literature has focused on so far. To illustrate this, we include in the same 
figure the case of a 2-component DM model, where one component is stable and the other decays (dotted 
lines). To make the comparison more straight-forward for the purpose of this figure, we have adjusted the 
decaying component to make up a fraction $\zeta/(1+\zeta)$ of the initial DM density and tuned the decay 
rate $\Gamma$ such that the total DM density intersects with the other curves at $a=a_t$.

Let us conclude the discussion of how the DM and DR densities evolve in our transition scenarios  
by showing in Fig.~\ref{fig:deltaH} the induced effect on the expansion rate of the Universe. For the
purpose of this figure, we compute the Hubble rate $H^2=8\pi G\rho/3$ by fixing the density parameters 
for the various components to the mean  $\Lambda$CDM values resulting from the Planck TTTEEE+lowP 
analysis~\cite{Ade:2015xua}, taking $\Omega_\chi^0 h^2=0.1198$ to correspond to the DM density {\it today}, 
and compare it to the Hubble rate in the $\Lambda$CDM case that is obtained for $\zeta\to0$.
During radiation  domination, as seen in the left panel, the Hubble rate starts to be visibly affected as 
soon as the {\it additional} comoving DM density compared to its value today, $\zeta \rho_\chi^0$,
contributes sufficiently to the total energy density;
for the small values of $\zeta$ shown here, this happens not much earlier than the transition at 
$a=a_t$. The largest deviation of the Hubble rate occurs at $a\sim a_t$ during matter domination,
or somewhat earlier during radiation domination (right panel). As indicated by the thin orange lines, 
furthermore, the DR density always starts to change the Hubble rate only at later times; as expected,
its relative impact (compared to that of DM), is largest if the transition takes place during radiation 
domination (and then, for $\kappa=2$ and $\kappa=4$, mimics the impact of a constant $\Delta N_{\rm eff}$ 
after equality, cf.~the black dotted line).

\subsection{Perturbations}
\label{sec:pert}
In order to study the impact of our modified cosmological scenario on CMB and LSS observables, we must
not only account for the modified evolution of the background densities, but also include the effect of
perturbations. In {\it synchronous gauge}~\cite{Ma:1995ey}, the perturbed line element of the FRW metric  
is given by 
\begin{align}
\mathrm{d}s^2 &= g_{\mu\nu} \mathrm{d}x^\mu \mathrm{d}x^\nu
=a^2 \left[-\mathrm{d}\tau^2 + (\delta_{ij}+h_{ij})\mathrm{d}x^i \mathrm{d}x^j\right]\,, 
\label{syngaugemetric}
\end{align}
where $\tau$ is the conformal time and $h_{ij}$ are the metric perturbations (we will denote its trace
as $h\equiv h_{ii}$). 

The above form of the line element leaves a residual gauge freedom, which we remove by working in
\textit{comoving synchronous gauge} (as also used, e.g., in {\sf CAMB}~\cite{Lewis:1999bs,Howlett:2012mh}).
In this gauge, the DM fluid remains at rest and its four-velocity is thus given by 
$u^\chi_{\,\mu}=a\,(1,\mathbf{0})$ just as in the unperturbed case.
The full DM and DR energy momentum tensors are then of the form
\begin{align}
T^{\chi}_{\mu \nu} &= \rho_{\chi} u^\chi_\mu u^\chi_\nu\,, \label{tmunuchi} \\
T^{\phi}_{\mu \nu} &= \frac43 \rho_{\phi} u^\phi_\mu u^\phi_\nu + \frac{\rho_\phi}{3} g_{\mu \nu} 
+\Pi_{\mu\nu}^\phi\,,
\end{align}
where $u^{\phi}_{\,\mu}=a\,(1,\mathbf{v^\phi})$ denotes the DR four-velocity, and $\rho_\chi$ and $\rho_\phi$ 
now refer to the full (perturbed) energy densities. $\Pi_{\mu\nu}^\phi$ describes the anisotropic stress of the DR 
component, i.e.~perturbations away from the perfect fluid form (as, e.g., caused by free-streaming).

As before, we demand that {any decrease in DM is fully compensated by an increase in DR}. Covariant conservation
of energy thus implies $\nabla^{\nu} \left(T^{\chi}_{\mu\nu} + T^{\phi}_{\mu\nu}\right)=0$,
which we can formally split and rewrite as
\begin{eqnarray}
\nabla^{\nu} T^{\chi}_{\mu\nu} =-\nabla^{\nu}T^{\phi}_{\mu\nu}\equiv -\mathcal{Q} u^\chi_\mu\,,
\label{energymomeq}
\end{eqnarray}
where $\nabla^\mu$ denotes the covariant derivative with respect to the full (perturbed) metric $g_{\mu\nu}$
given in Eq.~(\ref{syngaugemetric}). 
To leading order, as expected, this simply reproduces Eqs.~(\ref{eq:Econs}--\ref{eq:Boltz_DR}).
Demanding the DM density to evolve as in Eq.~(\ref{rhodmeq}) thus 
provides the same definition of $\mathcal{Q}\propto\zeta$ {\it  at leading order}.  

At next order in the perturbed quantities,  the DM part of Eq.~(\ref{energymomeq}) becomes 
\begin{align}
\delta'_{\chi} + \frac{1}{2} h' 
= \frac{a}{\rho_\chi}  \left(\mathcal{Q} \delta_\chi - \delta\mathcal{Q} \right)\,. \label{ddm}
\end{align}
Here, the prime $'$ denotes a derivative with respect to conformal time and 
$\delta_{\chi}= \delta \rho_\chi/\rho_\chi$  is the usual dimensionless  perturbation 
in the DM density.
 The perturbation $\delta  \mathcal{Q}$ to $\mathcal{Q}$ would, in analogy to the leading order result,
 be defined by an extension of our ansatz in Eq.~(\ref{rhodmeq}) to include perturbations.
 The minimal option for such an extension, in some sense, is that the perturbations only affect the 
 volume expansion (and hence not the comoving DM density). 
 In other words,  one would have to replace only the leading factor in Eq.~(\ref{rhodmeq}),\footnote{
 A simple heuristic way of seeing this is to consider the determinant of the spatial part of the metric,
 $\det g_{ij}=a^6\exp {\rm Tr} \ln (\delta_{ij}+h_{ij})$. 
 Expanding to first order, the `perturbed' scale factor is thus  given by $(\det g_{ij})^{1/6}=a(1+h/6)$.
 }
 \be
 \rho_\chi = \frac{\rho^0_\chi}{(a+ah/6)^3}\left[1+ \zeta\frac{1-a^\kappa}{1 +(a/a_t)^\kappa}\right]\,.
 \label{eq:comoving_pert}
 \ee
 Such an ansatz for the DM density implies $\delta'_{\chi} =-\frac{1}{2} h'$, 
 as can easily be verified, and is hence equivalent to setting
 \be
 \delta \mathcal{Q}\equiv\mathcal Q \delta_\chi\,.
 \label{eq:dQansatz}
 \ee
 While we will adopt this choice in the following, for simplicity, we stress that it is model-dependent
 and a full discussion is beyond the scope of this work. 
 We will, however, get explicitly 
 back to this issue in Section \ref{sec:sommer} when we try to motivate $\mathcal{Q}$ from 
 the collision term in the Boltzmann equation for a specific scenario (rather than by directly
 postulating the evolution of the DM density).  
 In general, it is worth noting that any deviation 
 from Eq.~(\ref{eq:dQansatz})  must be proportional to $\mathcal{Q}$ which, 
as we will see, is strongly constrained already from the evolution of the background densities
(unless $a_t$ is very small -- in which case the scale of the horizon, and hence of any perturbation 
that can be affected, is much smaller than what can be probed by the CMB). For the case of 
decaying DM, furthermore, Eq.~(\ref{eq:dQansatz})  is exactly satisfied~\cite{Poulin:2016nat}.

To first order in the perturbed quantities related to DR, on the other hand, Eq.~(\ref{energymomeq}) takes
the form
\begin{align}
\delta'_{\phi} + \frac{2}{3} h' + \frac{4}{3}\theta_\phi  &= - \frac{a}{ \rho_\phi}\left(  \mathcal{Q} \delta_\phi 
-   \delta\mathcal{Q}\right)\,, \label{ddr}\\
\theta'_\phi +\frac{1}{4}\nabla^2\delta_\phi  +\frac{1}{2\rho_\phi} \nabla^4\Pi^\phi 
&=-\frac{ a}{\rho_\phi}  \mathcal{Q} \theta_\phi\,. \label{vdr}
\end{align}
Here, $\nabla^2$ is the Laplacian operator, $\delta_\phi\equiv\delta\rho_\phi/\rho_\phi$ is 
defined in analogy to the DM case, and $\theta_\phi\equiv \partial_i v_\phi^i$ is the scalar part
of the DR velocity.  In the second equation, we have as usual only considered the scalar
part of $\nabla^{\nu}T^{\phi}_{i\nu}$, by taking its divergence, because the vector part of
the perturbations only have decaying modes. This is the reason why only the scalar part 
of the anisotropic stress enters, defined as
$\Pi_{ij}^{\phi,\text{scalar}}\equiv\left(\partial_i\partial_j-\frac13\delta_{ij}\nabla^2\right)\Pi^\phi$.
We implement this part as for an additional neutrino species, where $\Pi^\phi$ arises due to
the effect of free-streaming~\cite{Weinberg:2003ur}.

Let us point out that for $\mathcal{Q}=0$ 
Eqs.~(\ref{ddr},\ref{vdr}) simply describe the standard way of including non-interacting 
relativistic degrees of freedom, e.g.~in the form of (sterile) neutrinos, and for the choice 
of $\delta\mathcal{Q}$ made in Eq.~(\ref{eq:dQansatz}) we recover exactly the case of 
decaying dark matter (assuming an appropriate choice of $\mathcal{Q}$, cf.~Fig.~\ref{rhodmfig}).
We re-iterate that we expect a small effect from including perturbations because 
$\mathcal{Q}$ (and hence $\delta Q$) is already strongly constrained from the evolution
of the background densities.

\section{Generic effects on the cosmic microwave background}
\label{sec:cmb}
\subsection{Changes in the temperature anisotropy spectrum}
\label{sec:CMBspectrum}
\begin{figure*}[t]
\centering
\includegraphics[width=\columnwidth]{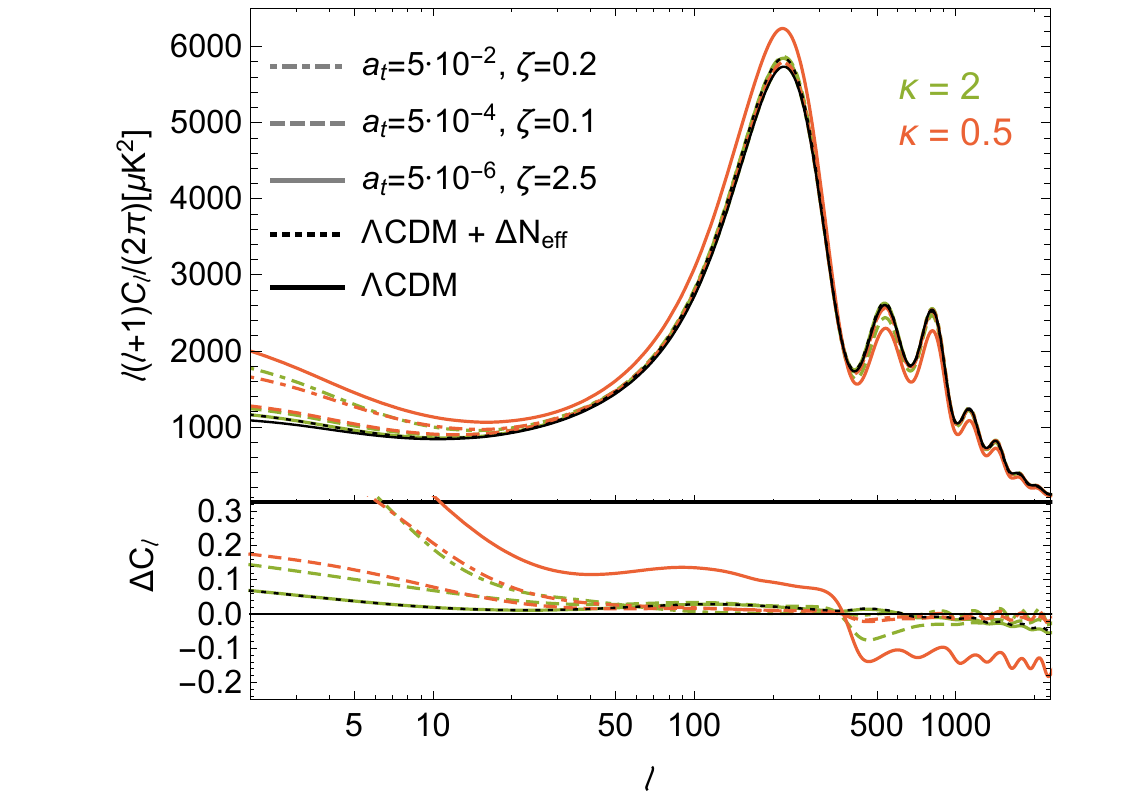}
\includegraphics[width=\columnwidth]{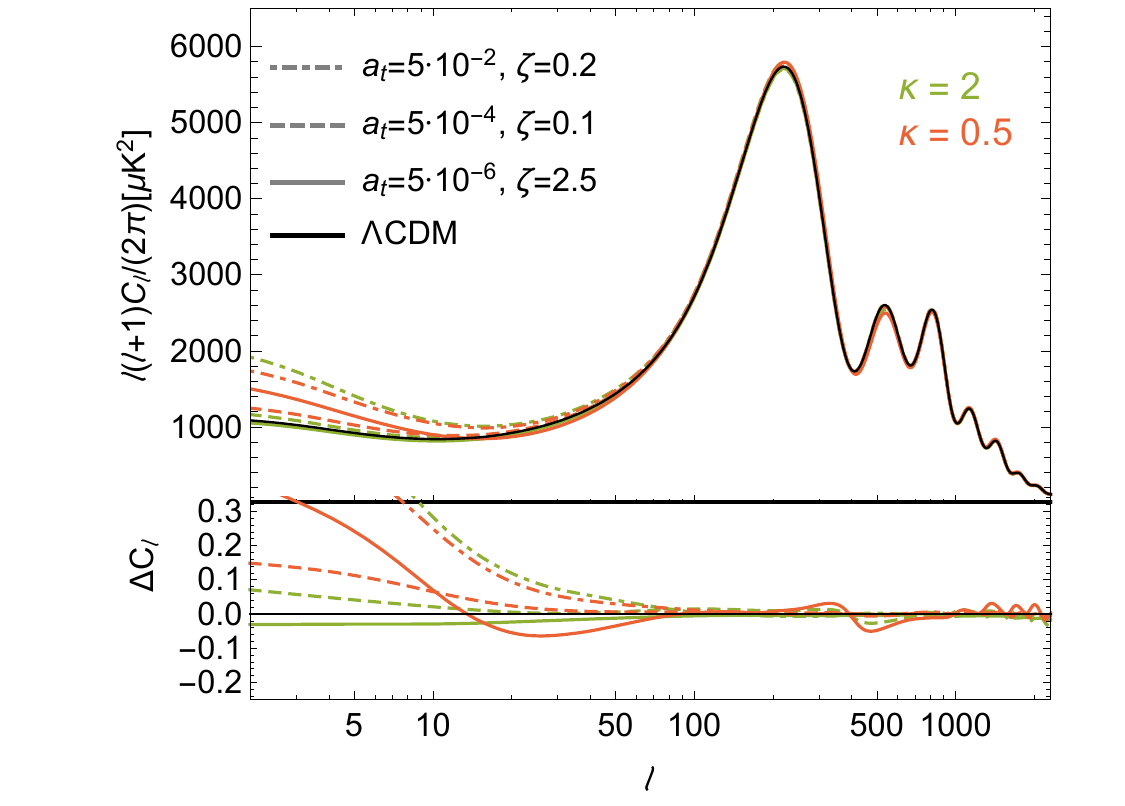}
\vspace{-2mm}
\caption{Lensed TT spectra for transition rates of $\kappa=2$ (green) and $\kappa=1/2$ (orange) for three different 
transition times $a_t=5 \cdot 10^{-6}$, $5 \cdot 10^{-4}$, $5 \cdot 10^{-2}$ for fixed $\Lambda$CDM parameters (left) 
and for the respective best-fit points (right).
For comparison we show the $\Lambda$CDM spectrum (solid black line) as well as 
$\Lambda$CDM $+ \Delta N_\text{eff}$ (dashed black line) for comparison with the early transition case.
In the bottom panels we show the fractional difference between the different scenarios and the $\Lambda$CDM 
case. See text for the remaining parameter values of the models used to obtain these spectra.}
\label{fig:TT_spectra}
\end{figure*}

The spectrum of the CMB is sensitive to the amount of matter and radiation from time-scales starting at around 
recombination until late times (e.g.\ through lensing effects). In addition, even earlier epochs may be constrained if 
they leave an imprint at later times such as an extra DR component. Let us start the discussion of CMB constraints 
by an evaluation of the possible imprints of the scenario described in Sec.~\ref{sec:DMtoDR} on the CMB 
spectrum. 

The $\Lambda$CDM model is described by only six parameters, which may be chosen as
{\it i) the amount of baryons} $\Omega_b h^2$ and {\it ii) dark matter} $\Omega_\chi h^2$, the {\it iii) approximate 
angular size of the sound horizon} $\theta_{\text{MC}}$,\footnote{The parameter, $\theta_{\text{MC}}$ is used in {\sf 
CosmoMC}~\cite{Lewis:2002ah,Lewis:2013hha} and is an approximate measure of the angular size of the sound 
horizon at the surface of last scattering. See \url{http://cosmologist.info/cosmomc/} or Ref.~\cite{Ade:2013zuv} for details.} 
the {\it iv) re-ionisation optical depth} $\tau$, the {\it v) amplitude of scalar perturbations} $\ln (10^{10}A_s)$ and the 
{\it vi) scalar spectral index} $n_s$. 
Given that the $\Lambda$CDM cosmology provides an excellent fit to the CMB data, any deviations should be very 
tightly constrained.

To calculate CMB as well as LSS observables, we use a modified version of the publicly available Boltzmann code 
{\sf CAMB}\footnote{\url{http://camb.info}}~\cite{Lewis:1999bs,Howlett:2012mh}.
In particular we have implemented the non-standard time evolution of energy densities of DM and DR according to 
Eqs.~(\ref{rhodmeq}) and (\ref{rhodreq}) to investigate and constrain the imprints of our scenario on the CMB.
As described in Section \ref{sec:pert}, furthermore, we treat DR as an extra neutrino species.

As discussed in the last section, the qualitative features of the DM to DR conversion depend on the time $a_t$ as 
well as the rate $\kappa$ of the conversion. To capture the relevant effects for the different regimes, we consider 
three different transition times  $a_t=5\cdot 10^{-6}, 5\cdot 10^{-4}$ and $5\cdot 10^{-2}$ 
as well as two different conversion rates 
$\kappa=2$ and ${1}/{2}$. The transition times are chosen such that we cover radiation domination as well as matter 
domination before and after recombination, while the choices of $\kappa$ describe, respectively, a fast and a slow 
conversion scenario.

To illustrate the effect on the CMB spectrum we fix five of the six $\Lambda$CDM parameters to their Planck 2015 
TTTEEE +low-P~\cite{Ade:2015xua} mean values, i.e., $\Omega_b h^2=0.02225$,  
$100\theta_{\text{MC}}=1.04077$ , $\tau=0.0790$,  $\ln (10^{10}A_s)=3.094$ and $n_s=0.9645$. The DM density is 
naturally evolving within our scenario and we fix $\Omega_\chi h^2$ such that for any $\kappa$, $\zeta$ and $a_t$ 
we have $\Omega_\chi h^2 = \Omega_\chi h^2 \vert_{\Lambda \text{CDM}}$ at 
$z_{\text{rec}}\equiv 1100$, i.e.~we require the same amount of DM as inferred for the $\Lambda$CDM model 
around recombination. This choice essentially ensures that the first peak of the CMB spectrum resembles that of 
the $\Lambda$CDM model and therefore agrees well with observations. We show the TT spectra of our scenario as 
well as the fractional difference from the usual $\Lambda$CDM paradigm, with parameters fixed in the way just
described, in the left panel of Fig.~\ref{fig:TT_spectra}. In the right panel of Fig.~\ref{fig:TT_spectra}, for comparison,
we show the spectra for the same values of our model parameters ($\kappa$, $a_t$, $\zeta$), but with the 
$\Lambda$CDM parameters fixed to the respective best-fit values in these scenarios.

Let us begin our discussion with a couple of simple observations:
For a rather quick transition $(\kappa=2$) which happens
rather early $(a_t=5\cdot 10^{-6})$, the transition will be complete before the onset 
of matter domination and thus the only significant change compared to the $\Lambda$CDM case is due to a 
remaining extra component of DR from 
the conversion. Given that the conversion takes place during radiation domination where the DM energy density is 
sub-leading, rather large values of $\zeta$ are consistent with data (for the chosen value of $\zeta=2.5$ we obtain 
$\Delta \tilde{N}_{\text{eff}}(1) \simeq 0.42$). Once the conversion is complete the comoving energy density of DR 
will remain constant. We thus expect this model to have a  spectrum which is very similar to the $\Lambda$CDM 
case with a constant additional $\Delta N_{\text{eff}} = 0.42$.  We illustrate this case with a dashed black line in the 
plot. As expected the spectrum is almost identical, and only very small differences are visible for high values of $\ell$, 
which are most sensitive to early times. We have confirmed that for even earlier transition times the two cases are 
indistinguishable. For a very slow transition $(\kappa=0.5)$ on the other hand, a significant part of the matter density 
will be converted to radiation
much later, implying that a larger fraction of the initial matter density will end up in radiation such that the effect on the 
CMB will be significantly larger, which can also clearly be seen in Fig.~\ref{fig:TT_spectra}. We therefore expect this
case to be much more strongly constrained.
For very late transitions, $(a_t=5\cdot 10^{-2})$, the cosmic history is the same as for the 
$\Lambda$CDM case until recombination. We accordingly observe that the spectrum resembles the $\Lambda$CDM 
case for high multipoles as expected. 

A more detailed understanding of the different effects on the power spectra requires knowledge about the evolution of 
the different energy densities $\Omega_i$. Given that we fix the value of 
$\Omega_\chi h^2 = \Omega_\chi h^2 \vert_{\Lambda \text{CDM}}$ at $z=z_\text{rec}$ 
(for the left panel in Fig.~\ref{fig:TT_spectra}) while having 
at the same time a somewhat increased value of $h$ due to the extra radiation component, $\Omega_\chi$ will be 
correspondingly smaller. Requiring the Universe to remain flat,  $\sum \Omega_i=1$, the energy density within some 
other components needs to be increased to compensate the decrease in $\Omega_\chi$. The way in which the 
different components change depends on which parameters we keep fixed in the analysis. For instance fixing 
$\theta_{\text{MC}}$ as we have done in the left panel of Fig.~\ref{fig:TT_spectra} will lead to an enhancement in 
$\Omega_\Lambda$, because the enhancement of the Hubble rate prior to recombination decreases the size of the 
sound horizon at the surface of last scattering $r_s$, which implies a simultaneous decrease of the angular distance 
to the last scattering surface $D_A$ in order to keep $\theta_{\text{MC}}$ fixed. The required decrease in $D_A$ in 
turn is achieved by increasing the vacuum energy $\Omega_\Lambda$. Overall this will lead to an enhanced 
\textit{Late time Integrated Sachs Wolfe} (LISW) effect, that is (relatively speaking) more power on very large scales 
(small values of $\ell$). 

As these types of effects strongly depend on what we keep fixed, we will refrain
from describing the changes of the temperature anisotropies compared to the $\Lambda$CDM case in more detail.
To construct the bounds on the model parameters in the next section, all $\Lambda$CDM parameters will be varied, 
allowing for a partial compensation of the effects of the matter to radiation transition. This partial compensation can 
already be anticipated by comparing the left and right panels of Fig.~\ref{fig:TT_spectra}.


\subsection{CMB constraints}

In this section, we will constrain our model with CMB observations. The concrete 
data set that we use for this purpose, with likelihoods as implemented in the publicly available 
Markov Chain Monte-Carlo (MCMC) code {\sf CosmoMC}
~\cite{Lewis:2002ah,Lewis:2013hha}, we will denote as follows:
\begin{itemize}
\item \textbf{CMB:} Planck TTTEEE + lowTEB~\cite{Aghanim:2015xee} 
\end{itemize}
At this stage, in particular, we do not add information from the Planck lensing power spectrum 
reconstruction~\cite{Ade:2015zua} because this effectively adds a measurement implicitly related to the
matter power spectrum (which we will discuss in more detail in the next section).

In order to explore the parameter space of our model, we modify {\sf CosmoMC} to 
communicate our additional model parameters to the modified {\sf CAMB} version described above. 
We run chains using the fast/slow sampling method~\cite{Neal,Lewis:2013hha}, as recommended for a large 
parameter space. We assume the chains to be converged if the Gelman-Rubin 
criterion ($R$)~\cite{Gelman:1992zz} satisfies $R-1< 0.01$. Along with a large number of Planck 
nuisance parameters, we scan over the six $\Lambda$CDM parameters with 
flat priors as follows:
\begin{align}
\Omega_b h^2 \in (0.01,0.1),~ &   \Omega_\chi^0  h^2 \in (0.01,0.5) \nonumber \\
100\theta_{\rm MC} \in (0.8,2),~&\tau \in (0.01,0.2) \nonumber \\
\ln (10^{10}A_s) \in (2,4), ~& ~n_s \in (0.8,1.2) \,.
\end{align}

Let us first have a look at very early transitions. In this case, as discussed above, CMB constraints on
our model should be equivalent to those for a model with constant $\Delta N_{\text{eff}}$ 
(at least for large values of $\kappa$, since for $\kappa\leq1$ the comoving DR energy density does not 
saturate, cf.~Fig.~\ref{rhodmfig}). To check this expectation, we fix $a_t=10^{-7}$ and scan over the six 
$\Lambda$CDM parameters and $\zeta \geq 0$ (with a flat prior). 
For comparison with the constant $\Delta N_{\text{eff}}$ case, we 
use the default {\sf CosmoMC/CAMB} implementation with $N_{\rm eff}$ as a free parameter in addition to the 
$\Lambda$CDM parameters. For this scan we have set the (flat) prior for $N_{\rm eff}$ to be greater than 3.046,
in order to be comparable to the prior choice for our model parameter $\zeta$. 

\begin{figure}[t!]
\centering
\includegraphics[width=0.95\columnwidth]{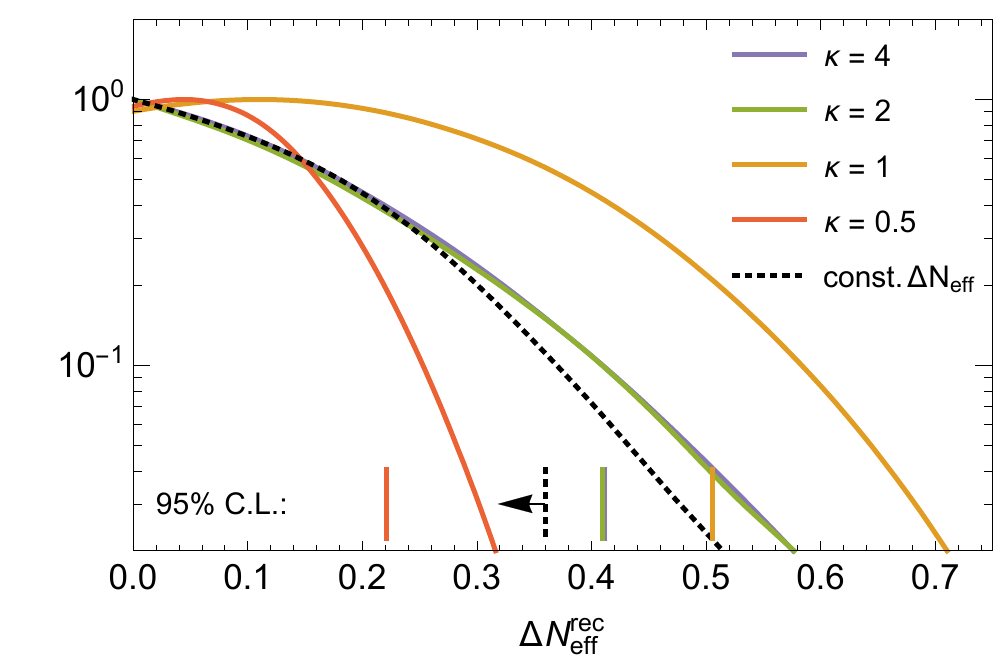}
\caption{Marginalised 1D posterior pdfs for $\Delta \tilde{N}_{\text{eff}}(a_{\text{rec}})$, normalised such that the 
maximum value is 1,
 using the \textbf{CMB} dataset only. The solid lines are for 
$\kappa=0.5,1,2,4$ with fixed $a_t=10^{-7}$.  Note that for $\kappa=2,4$, but not for smaller values of 
$\kappa$, we have $\Delta \tilde{N}_{\text{eff}}(a_{\text{rec}})=\Delta \tilde{N}_{\text{eff}}^{\rm today}$, 
cf.~Fig.~\ref{rhodmfig}.
For comparison, we also include the standard case of a constant $\Delta N_{\text{eff}} \geq 0$ 
(dashed black line). 
The vertical lines indicate the corresponding 95\% C.L.\ limits. For a constant $\Delta N_{\text{eff}}$, our limit 
is in good agreement with the Planck limit  of $0.35$~\cite{Ade:2015xua} 
(obtained with a flat prior on $\Delta N_{\rm eff}$ that, unlike in our case, also allows $\Delta N_{\rm eff}<0$).}
\label{fig:delneff}
\end{figure}

\begin{figure*}[t!]
\centering
\includegraphics[width=\columnwidth]{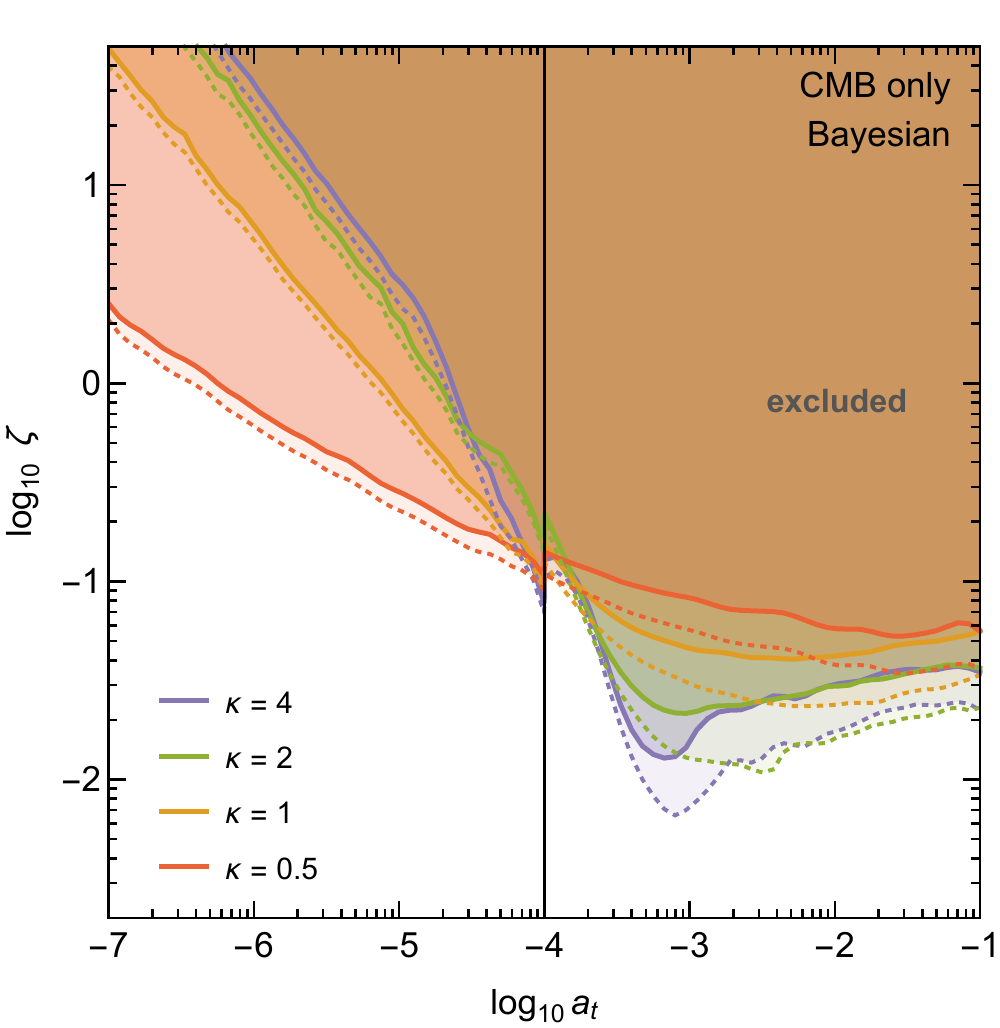}\quad
\includegraphics[width=\columnwidth]{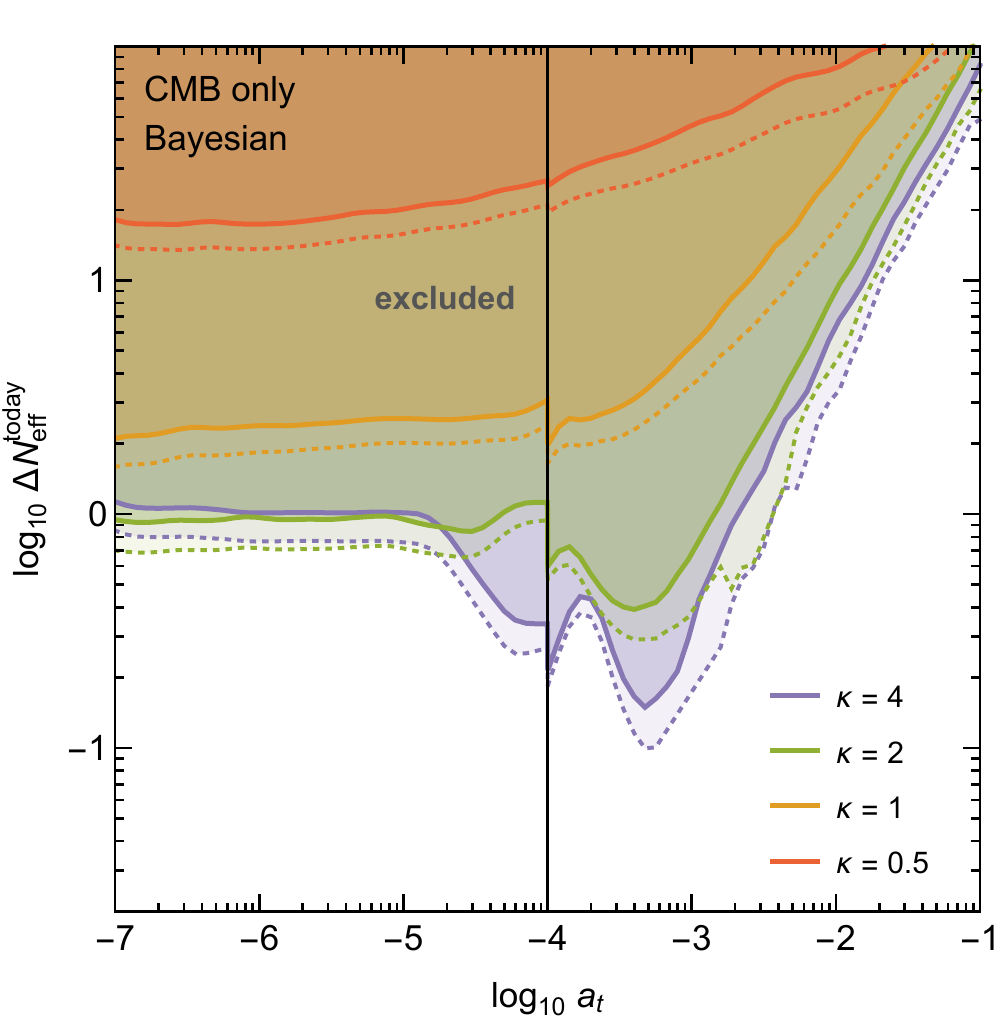}
\caption{95\%\,C.L.~(dotted lines) and 99\%\,C.L.~(solid lines) Bayesian limits from \textbf{CMB} only; the coloured 
region above each line is excluded.
{\it Left panel}. Constraints on the amount of converted 
DM, cf.~Eq.~(\ref{rhodmeq}). {\it Right panel}. Constraints on the amount of DR {\it today}, expressed in terms
of $\Delta \tilde N_{\rm eff}$ as given in Eq.~(\ref{eg:neff}).
For both cases, we adopted a flat prior on $\Delta N_{\rm eff}^{\rm today}$ for $a_t<10^{-4}$, and a flat
prior on $\zeta$ for $a_t>10^{-4}$.
}
\label{fig:CMB_limits}
\end{figure*}

In Fig.~\ref{fig:delneff}, we show the marginalised 1D posterior probability density functions (pdfs)  
for $\Delta \tilde{N}_{\rm eff}(a_{\text{rec}})$  that result from the \textbf{CMB}
likelihood, for $\kappa=0.5,1,2,4$.
For $\kappa=2,4$, the posteriors are indeed similar to the case of a constant $\Delta N_{\text{eff}}$ 
(shown as a black dashed line). 
The discrepancy at larger values of $\Delta \tilde{N}_{\rm eff}(a_{\text{rec}})$ can be traced
back to how the Helium abundance $Y_{\rm He}$ enters in the CMB code. Concretely, $Y_{\rm He}$ is
a derived parameter that depends not only on the baryon density but also on the DR density 
{\it at the time of big bang nucleosynthesis} (BBN), because a non-zero value of the latter affects 
the Hubble expansion rate during  that time~\cite{Peebles:1966zz,Hou:2011ec}. In our case, unlike for a constant 
$\Delta N_{\text{eff}}$, there is no DR present during BBN because we always assume that the 
DM to DR transition occurs only much later. We checked explicitly that we get exact agreement 
between our $\kappa=2,4$ limits and constant $\Delta N_{\text{eff}}$, up to 99\,\% C.L., 
if we use a numerical value of $Y_{\rm He}$ as calculated from 
$\Delta \tilde{N}_{\rm eff}({\text{BBN}})=\Delta \tilde{N}_{\rm eff}({\text{today}})$.
Lastly, let us mention that these limits also agree to a good approximation with the Planck limits on a 
constant $N_{\rm eff}$~\cite{Ade:2015xua} -- though such a comparison should be taken with 
a grain of salt given that those limits are based on a slightly different prior choice 
(allowing for $\Delta N_{\rm eff}<0$) than what we have adopted here. 

We now turn to the CMB constraints when scanning freely over our model parameters.
For this, we choose a flat prior on $\log a_t$, constraining the scan to 
$-7 \leq \log_{10} a_t \leq -1$ in order to focus on the case where BBN constraints are negligible
(lower bound) and to ensure that we can neglect the effect of structure formation and still treat the perturbations 
at the linear level (upper bound). We note that the upper bound here is somewhat optimistic in this respect, 
so results presented for $a_t\gtrsim 10^{-2}$ should be interpreted with care
(what actually matters is of course not the value of $a_t$, but whether the transition is largely
completed while perturbations still are at the linear level, cf.~Fig.~\ref{rhodmfig}). 
For $\zeta$ we choose a more complicated prior to optimise the sampling efficiency
of the Metropolis-Hastings algorithm implemented in {\sf CosmoMC}. Concretely, in anticipation of
our results, we choose a prior for $\zeta$ that corresponds to a flat prior on $\Delta N_{\rm eff}^{\rm today}$
for $a_t<10^{-4}$ and a prior that is flat in $\zeta$ itself for $a_t>10^{-4}$.
Since for fixed $a_t$ and fixed cosmological parameters $\Delta N_{\rm eff}^{\rm today}$ is directly 
proportional to $\zeta$, the two regions are expected to smoothly connect to each other at 
$a_t = 10^{-4}$.\footnote{%
The normalisation of the posterior pdfs are independent in the two regions, so one needs to apply an 
appropriate rescaling before the two regions can be connected. To minimise the impact of numerical 
inaccuracies, we require that the maxima of the respective posterior pdfs agree at the transition.
}

We show our results in Fig.~\ref{fig:CMB_limits}, as a function of $a_t$,
both expressed in terms of limits on $\log_{10}\zeta$ (left panel) and in terms of limits on 
$\log_{10}\Delta \tilde N_{\rm eff}$ {\it today} (right panel). For the sake of our later discussion, let
us stress that these are {\it Bayesian limits} constructed in the standard way, i.e.~curves of 
constant 2D (marginalised) posterior probabilities chosen such that the integral over the 
enclosed area (which includes the point of maximum pdf) results in 0.95 and 0.99, respectively.
For very small values of $a_t$, as discussed above, we expect that the CMB cannot distinguish
between our model and the case of a constant $\Delta N_{\rm eff}$. This implies that the bound
on $\zeta$, as a function of $a_t$, must simply be inversely proportional to the total amount of DR 
that is created prior to recombination. 
For a fast transition ($\kappa=2$ and $\kappa=4$) the latter is roughly proportional to the ratio of the 
amount of converted DM to the total amount of radiation, which in turn is proportional to $\zeta \, a_t$. 
This explains the approximate $\zeta \propto a_t^{-1}$ slope visible in the figure.

Closer inspection reveals that
 the simple requirement of a fixed total amount of DR just before recombination indeed gives a 
 qualitatively very good description of the limits for $a_t\lesssim 10^{-3}$.
 We note that the limits in this range can also be reproduced, within reasonable accuracy,
 just by using the fact that the CMB peak positions are tightly constrained observationally.\footnote{%
Technically we checked that we can roughly reproduce these limits by allowing the angular size of the
sound horizon close to recombination, $\theta^*$, to vary within observational bounds~\cite{Ade:2013zuv},
in analogy to what was done in Ref.~\cite{Binder:2017lkj}.
 } 
 For large values of
 $a_t$, on the other hand, the constraints are less and less affected by the additional radiation
 component and rather driven by the reduced CDM component -- which explains why the maximally allowed
 value of $\zeta$ becomes almost independent of $a_t$ at very late times.  Physically, it is a combination
 of various mechanisms that sets the constraints in this case, with the ISW effect becoming more 
 and more relevant with increasing $a_t$. While we refrain from attempting a detailed discussion here,
 we therefore expect that simple prescriptions for estimating these constraints are likely to fail.
 For example, demanding the peak positions not to change (which gave a very good estimate of the
 full results for  $a_t\lesssim10^{-3}$) would result in constraints that are too strong 
 and feature a qualitatively wrong dependence on $a_t$. 

 \begin{figure*}[t]
\centering
\includegraphics[width=\columnwidth]{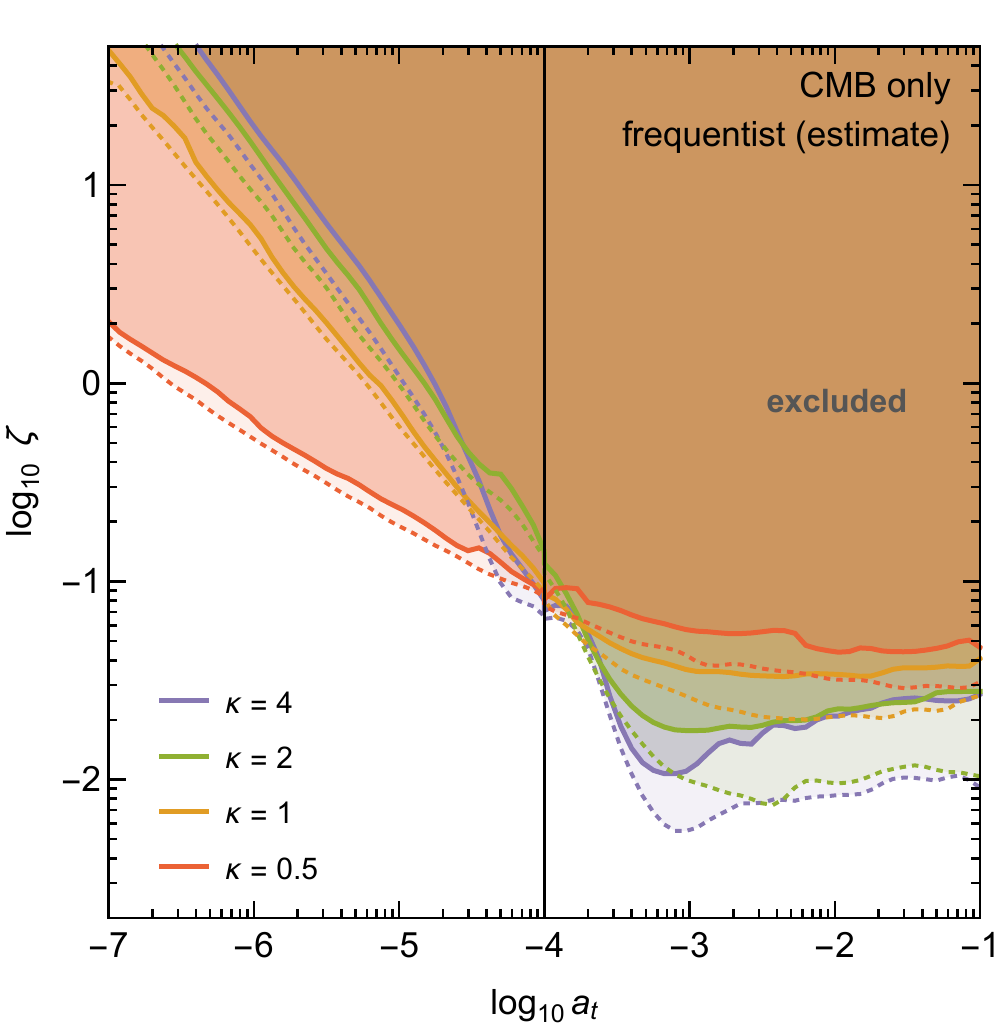}\quad
\includegraphics[width=\columnwidth]{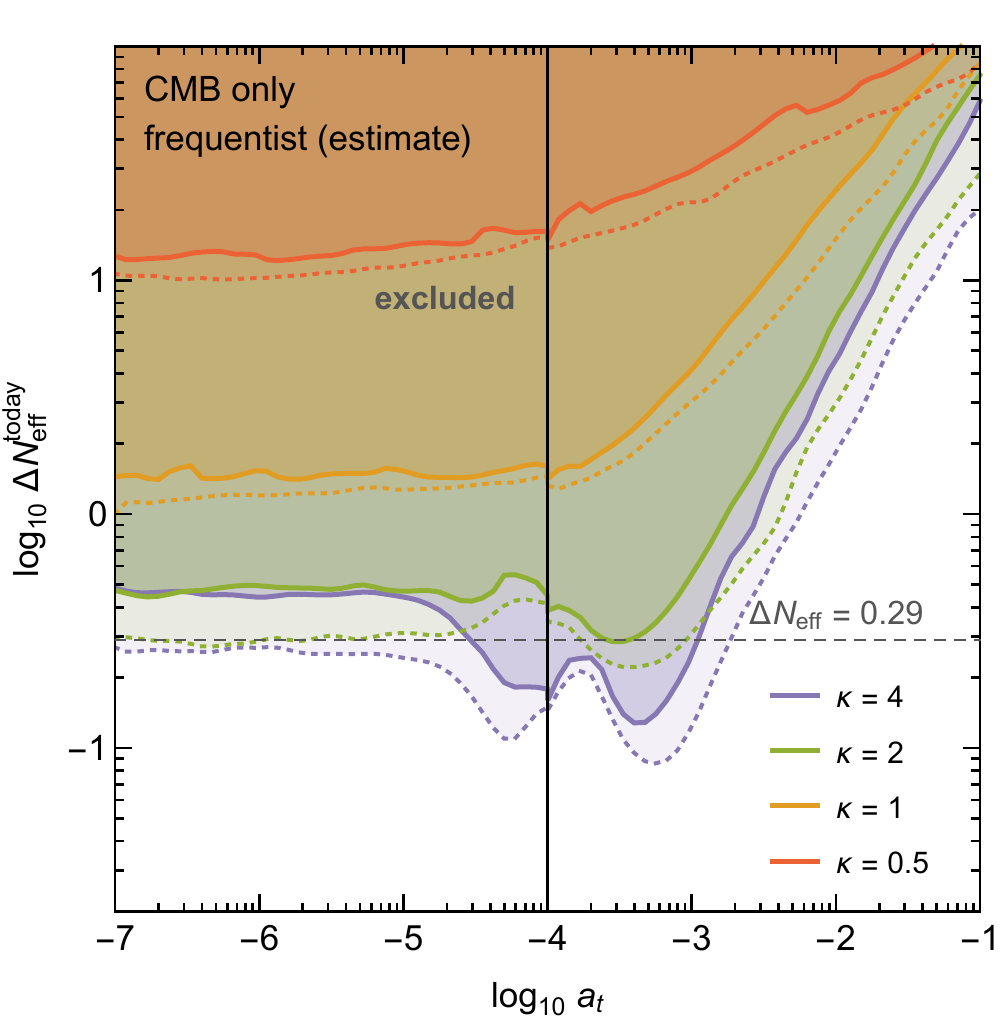}
\caption{95\%\,C.L.~(dotted lines) and 99\%\,C.L.~(solid lines) approximate frequentist constraints from \textbf{CMB} only; 
the coloured 
region above each line is excluded. {\it Left panel}. Constraints on the amount of converted 
DM. {\it Right panel}. Constraints on the amount of DR {\it today}, expressed in terms
of $\Delta \tilde N_{\rm eff}$. For comparison we indicate the frequentist 95\% C.L. bound on $\Delta \tilde N_{\rm eff}$ 
obtained from a scan with flat prior on $\Delta N_\text{eff}$ and $a_t = 10^{-7}$ 
(derived from the 1D posterior shown in Fig.~\ref{fig:delneff}).}
\label{fig:CMB_limits_frequentist}
\end{figure*}

The discussion in the preceding paragraph has focussed on a qualitative understanding 
of the constraints on $\zeta$ shown in the left panel of Fig.~\ref{fig:CMB_limits}. With the additional
input from Fig.~\ref{rhodmfig}, it is straightforward to achieve a similar understanding concerning the
qualitative shape of the constraints on $\Delta N_{\rm eff}$ as presented in the right panel of 
Fig.~\ref{fig:CMB_limits}. In particular, the fact that these constraints are flat for small values 
of $a_t$ should not come as a surprise given that in this limits our model is expected to be indistinguishable
from the case of a constant $\Delta N_{\rm eff}$. {\it Quantitatively}, however, the situation is less clear at first sight.
In particular we infer from the right panel of Fig.~\ref{fig:CMB_limits} that for $\kappa = 2,4$ and small $a_t$ 
values of $\Delta N_\text{eff}^\text{today} \gtrsim 0.7-0.8$ are excluded at 95\%\,C.L. 
The reason for the difference between this value and the bound 
$\Delta N_\text{eff}^\text{today} \lesssim 0.4$ inferred from Fig.~\ref{fig:delneff} is that here we consider the 
posterior pdf as a function of $\log_{10} \Delta N_\text{eff}^\text{today}$ rather than $\Delta N_\text{eff}^\text{today}$, which 
disfavours small values of $\log_{10} \Delta N_\text{eff}^\text{today}$ 
and hence introduces an overall bias towards larger values.

The prior dependence of the bounds shown in Fig.~\ref{fig:CMB_limits} makes it difficult to interpret them in a 
model-independent way. After all, $a_t$ and $\zeta$ are only effective parameters introduced to describe the 
evolution of the DM density, and the appropriate priors may depend sensitively on how this effect is realised in a 
more fundamental theory. A way to avoid this ambiguity is to consider frequentist rather than Bayesian exclusion 
limits. This is possible in a rather straight-forward manner thanks to the following two observations: 
First, since we consider flat priors on $\log_{10} a_t$ and $\zeta$ (or equivalently $\Delta N_\text{eff}^\text{today}$ 
for small $a_t$), the marginalised posterior as a function of these two parameters is directly proportional to the 
marginalised likelihood. Second, since all parameters apart from $a_t$ and $\zeta$ (or 
$\Delta N_\text{eff}^\text{today}$) are very well constrained by the CMB, the marginalised likelihood is 
expected to be similar to the profile likelihood (where for each value of $a_t$ and $\zeta$, or 
$\Delta N_\text{eff}^\text{today}$, all other parameters have been fixed to their best-fit 
value)~\cite{Hamann:2011hu}. We can therefore use the posterior probability to construct approximate 
profile likelihood ratios.

To construct frequentist upper bounds on the amount of DM that can be converted into DR, we determine the values of 
$a_t$ and $\zeta$ that give the best fit to the data, i.e.\ that maximise the posterior probability. For the data sets that we 
study in this section there is at most a very mild preference for non-zero $\zeta$, so that we typically find 
$\zeta_\text{best} \approx 0$. We then consider the test statistic
\begin{equation}
t = -2 \, \Delta \log \mathcal{L} \approx -2 \log \left[\frac{p(\zeta, a_t)}{p(\zeta_\text{best}, a_{t,\text{best}})}\right] \; , 
\label{eq:ts}
\end{equation}
where $p$ denotes the posterior probability. We expect that for random fluctuations in the data, $t$ will 
approximately follow a $\chi^2$ distribution with two degrees of freedom. We thus can exclude parameter 
points with $t > 5.99$ ($t > 9.21$) at 95\% (99\%) C.L.

We show the resulting estimate of frequentist exclusion limits on $\zeta$ in the left panel of 
Fig.~\ref{fig:CMB_limits_frequentist}. By construction, the frequentist exclusion limits follow lines of constant posterior 
probability and therefore have the same shape as the Bayesian exclusion limits shown in Fig.~\ref{fig:CMB_limits}. In other 
words, the difference between the frequentist and the Bayesian exclusion limits is the confidence level associated to a 
specific posterior probability, i.e.\ frequentist exclusion limits correspond to Bayesian exclusion limits at a \emph{different} 
confidence level. More specifically, we find the frequentist exclusion limits to be somewhat stronger. 

The advantage of using frequentist exclusion limits is illustrated in the right panel of Fig.~\ref{fig:CMB_limits_frequentist}, 
which shows the bounds on $\Delta N_{\rm eff}$ calculated from the frequentist exclusion limits on $\zeta$ for $\kappa = 2$ 
and $\kappa = 4$. The only cosmological parameter required to perform this translation is $\Omega_\chi h^2$. Ideally, 
$\Delta N_{\rm eff}$ should be calculated using the respective best-fit value of $\Omega^0_\chi h^2$ for each value of 
$a_t$ and $\zeta$. However, given the precision of CMB constraints on this combination of DM density and expansion rate 
during recombination, it is sufficient 
to simply  require $\Omega_\chi h^2 = \Omega_\chi h^2 \vert_{\Lambda \text{CDM}}$ at $z_{\text{rec}}\equiv 1100$.

In contrast to the bounds on $\Delta N_{\rm eff}$ shown in the right panel of Fig.~\ref{fig:CMB_limits}, the bounds derived 
from the frequentist exclusion limits on $\zeta$ do not depend on the choice of priors for $\zeta$ and $a_t$.\footnote{%
We observe some residual prior dependence due to the way in which the parameter space is sampled, which leads to a 
less efficient exploration of the tails for the case of logarithmic priors.
} 
As a result, the bounds on $\Delta N_{\rm eff}$ obtained for small $a_t$ are much closer to the frequentist bounds 
derived from the 1D posterior shown in Fig.~\ref{fig:delneff} (based on $a_t = 10^{-7}$ and 
a flat prior on $\Delta N_\text{eff}$), which gives  $\Delta N_\text{eff} < 0.29$ for both $\kappa = 2$ 
and $\kappa = 4$ (indicated by the black dashed line). {We will therefore from now on focus on frequentist exclusion limits. The corresponding Bayesian exclusion limits can be found in Appendix~\ref{app:Bayesian}.}

\section{Generic imprints on low-redshift observables}
\label{sec:lss}

Let us now turn to the implications of converting DM to DR for low-redshift observables. We will focus here 
on the two most important late-time effects, namely a modified expansion 
rate and a change of the linear matter power spectrum $P(k)$. The former is something we briefly discussed
already in Section \ref{background}, cf.~Fig.~\ref{fig:deltaH}. Such a late-time enhancement of the Hubble 
rate may in principle help to reconcile a known discrepancy between low- and high-redshift 
observables~\cite{Ade:2013zuv, Raveri:2015maa, Bernal:2016gxb,Riess:2016jrr}.
In terms of possible physics realisations, such an option has so far mostly been 
discussed in terms of a constant DR (or sub-dominant hot DM) contribution~\cite{Hou:2011ec,Mehta:2012hh,Wyman:2013lza,Hamann:2013iba,Battye:2013xqa,Costanzi:2014tna} or decaying
DM scenarios~\cite{Enqvist:2015ara,Berezhiani:2015yta,Blackadder:2015uta,Poulin:2016nat,Chudaykin:2016yfk,Hamaguchi:2017ihw}. 
By making the connection to our more general conversion scenario from
DM to DR, we will revisit this question in a broader context.

\begin{figure*}[t!]
\centering
\includegraphics[width=\columnwidth]{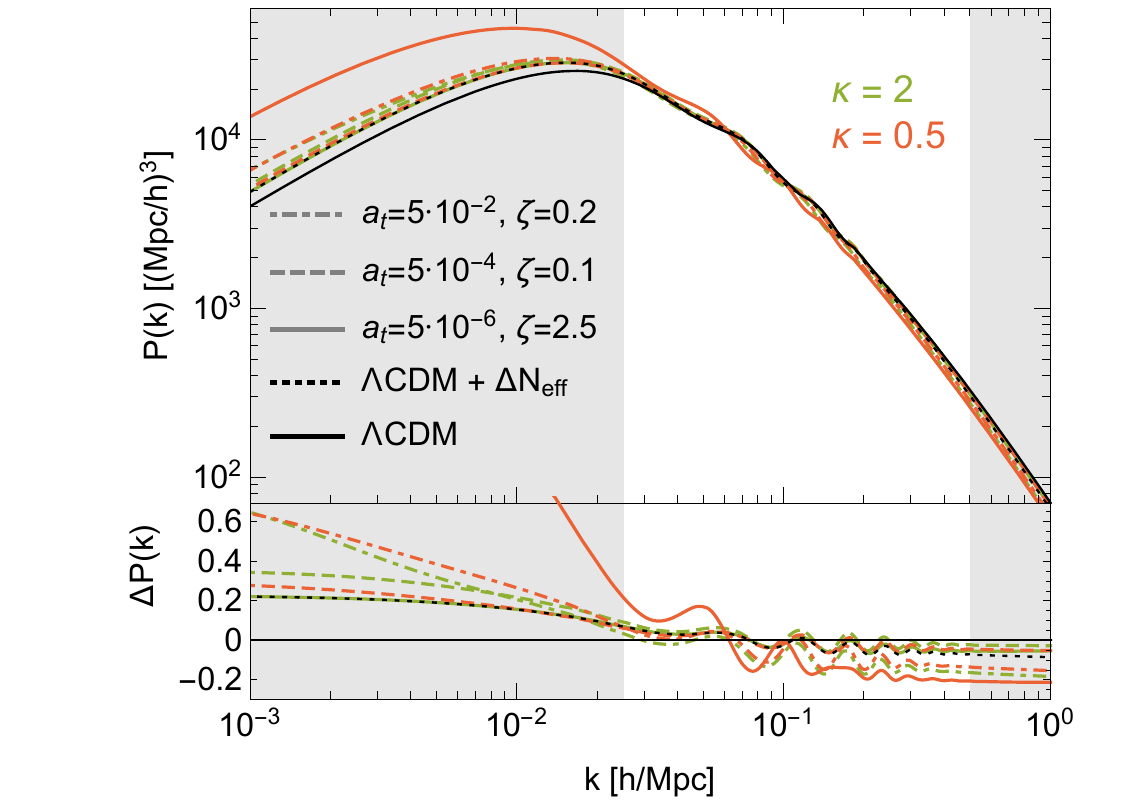}
\includegraphics[width=\columnwidth]{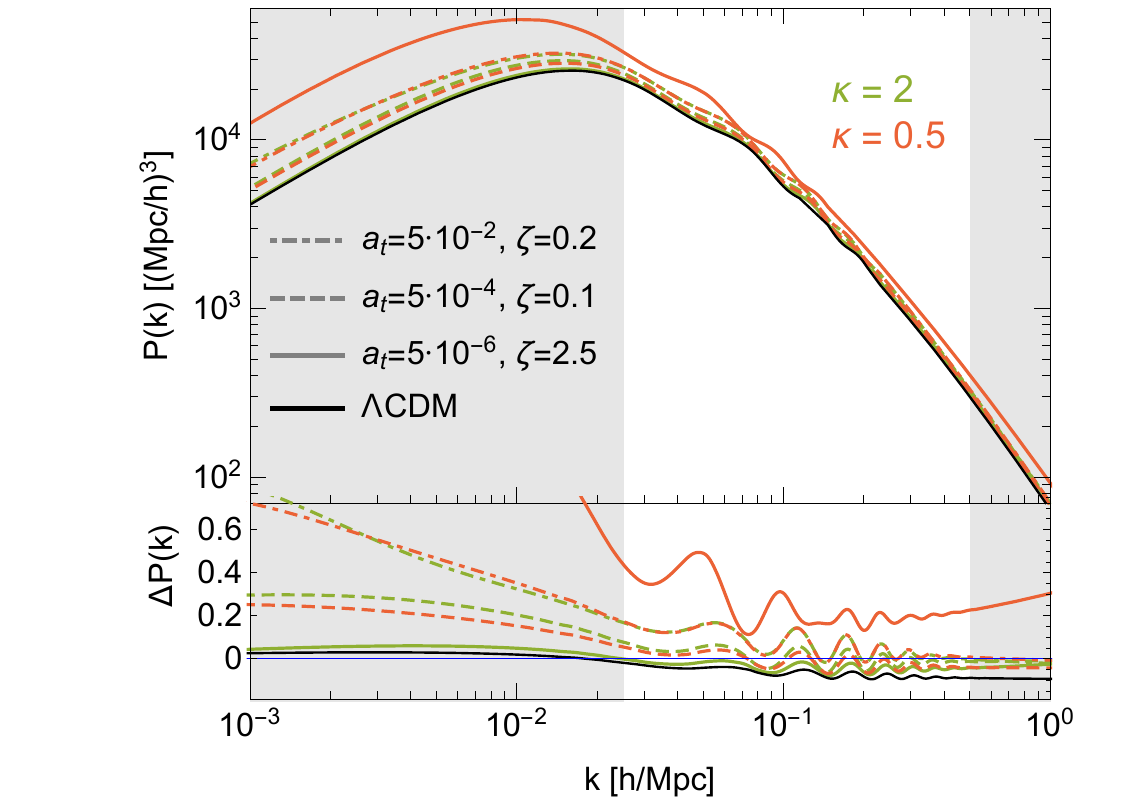}
\caption{Linear matter power spectrum for the same set of benchmark models that we considered in 
Fig.~\ref{fig:TT_spectra}. The range of wavenumbers that is {\it not} shaded gives the dominant contribution to
$\sigma_8$. {\it Left panel.}  $\Lambda$CDM parameters fixed to best-fit values from {\bf CMB} only (as in left panel 
of Fig.~\ref{fig:TT_spectra}). {\it Right panel.} $\Lambda$CDM parameters fixed to best-fit values from 
{\bf CMB+Lensing+HST+PC}.
Here the difference plot is still normalised to the $\Lambda$CDM power spectrum shown in the {\it left} panel. 
\label{fig:Pk}
}
\end{figure*}

Before doing so, however, let us briefly discuss the expected imprint on $P(k)$. To this end, we show in 
the left panel of Fig.~\ref{fig:Pk} how the {\it linear} matter power spectrum changes, with respect to 
the $\Lambda$CDM case, for the same set of benchmark models (and  $\Lambda$CDM parameters) 
that we considered in the left panel of Fig.~\ref{fig:TT_spectra}. 
Note that the full {\it non-linear} power spectrum would be needed to make a meaningful comparison to data for 
large values of the wave-number $k$. For the present study we will therefore mostly limit ourselves to discussing the 
parameter combination $\sigma_8 \Omega_m^{0.3}$, for which direct measurements exist~\cite{Ade:2013lmv} and to
 which mainly intermediate values of $k$ contribute, which
are largely unaffected by non-linear dynamics.\footnote{%
Note that the procedure used to infer the observational value of $\sigma_8 \Omega_m^{0.3}$  
assumes a $\Lambda$CDM cosmology, and properly accounting for the different 
cosmology considered here may lead to some deviations. To fully address this 
issue is beyond the scope of our analysis.
}
Specifically, $\sigma_8$ can be expressed as 
\begin{equation}
\sigma_8^2 = \frac{1}{2\pi^2}\int_0^\infty \text{d}k \,k^2 P(k)\,W^2\left(kR_8\right)\,,
\end{equation}
where $W(x)=3j_1(x)/x$ is the Fourier transform of the top-hat window function, 
$j_1$ is the first spherical Bessel function and $R_8 \equiv 8 \, h^{-1}\mathrm{Mpc}$.
Requiring the integration range to contribute 99\% to the value of $\sigma_8$, we find 
$0.025 h\, \text{Mpc}^{-1} \lesssim k \lesssim 0.5 h\, \text{Mpc}^{-1}$, 
which we indicate by the non-shaded region in Fig~\ref{fig:Pk}.

We first observe that on large scales, the spectrum is enhanced for our models. This is 
due to a larger value of $\Omega_\Lambda$, which enhances and shifts the spectrum 
towards larger scales~\cite{Lesgourgues:2006nd,Poulin:2016nat}. 
Secondly, for the range relevant for $\sigma_8$, we observe the spectrum to be suppressed. 
This is partially explained by a pure free streaming effect of the additional DR component
(see the dotted line indicating the case of a constant $\Delta N_{\rm eff}$), and partially
by the fact that perturbations evolve slightly different in our model than in $\Lambda$CDM,
see Section \ref{sec:pert}).

\begin{figure*}
\includegraphics[scale=0.8]{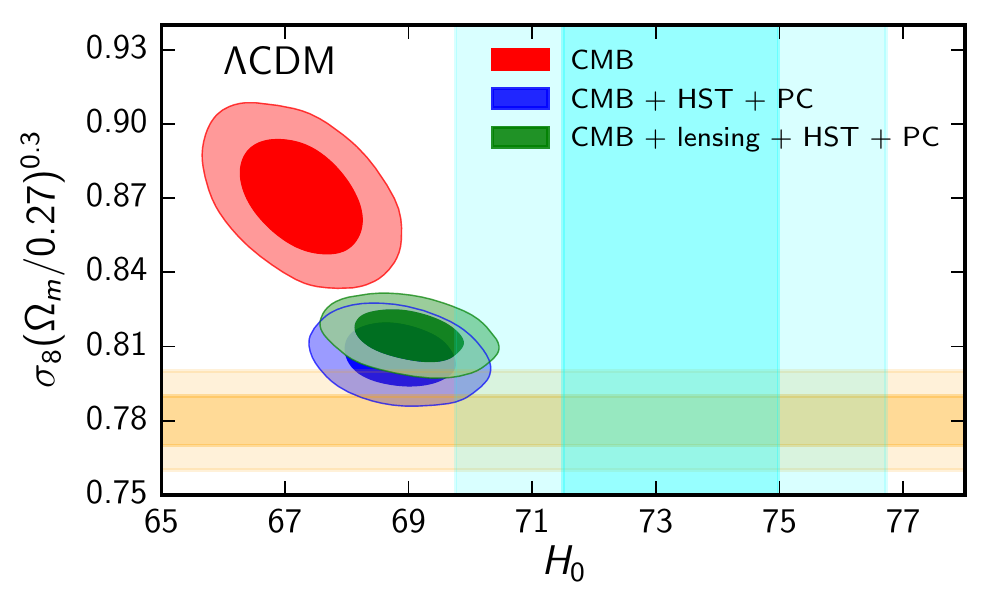}~~~~
\includegraphics[scale=0.8]{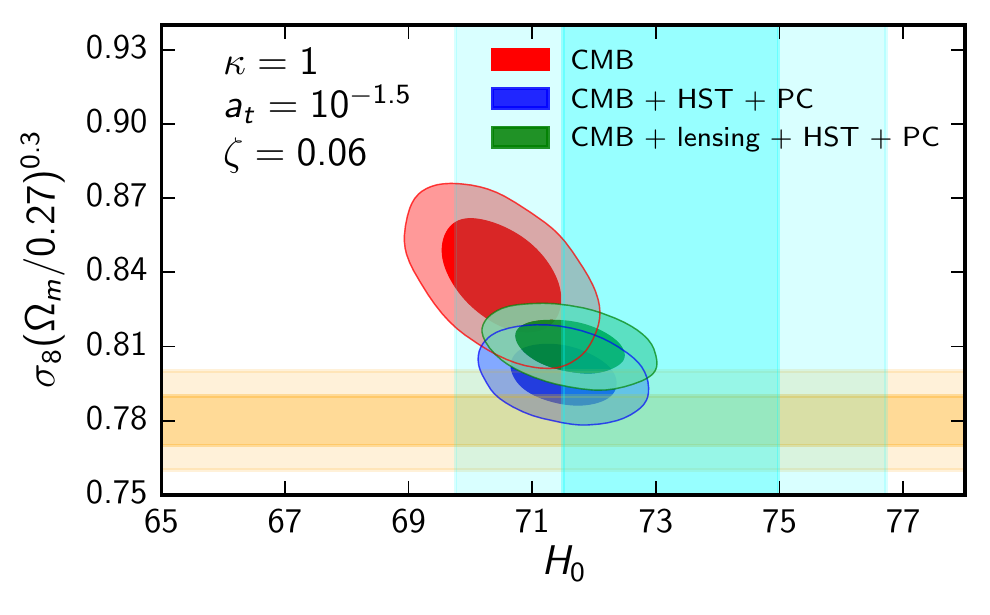}
\caption{
Best fit regions for $\Lambda$CDM ({\it left panel})
and our model with $\kappa=1$, $\zeta=0.06$ and $a_t=10^{-1.5}$ ({\it right panel}). The 
orange and cyan bands indicate the direct measurements of 
$\sigma_8 (\Omega_m/0.27)^{0.3} = 0.78 \pm 0.01$~\cite{Ade:2013lmv} and 
$H_0 =73.24\pm 1.74$~\cite{Riess:2016jrr} respectively.}
\label{fig:triangle_comp_LSS}
\end{figure*}

So far, we have included only  CMB data in our discussion. In this section we extend our analysis to 
post-CMB cosmology by including the following data sets:
\begin{itemize}
\item \textbf{CMB + Lensing:} Same as \textbf{CMB}, with  Planck lensing power spectrum 
reconstruction~\cite{Ade:2015zua}, using likelihoods as implemented in {\sf CosmoMC}
\item  \textbf{HST}: Direct measurements of the Hubble rate  $H_0=73.24 \pm 1.74  $ km/sec/Mpc by 
the Hubble space telescope~\cite{Riess:2016jrr}
\item \textbf{PC}:  Measurement of the power spectrum normalisation, 
$\sigma_8 (\Omega_m/0.27)^{0.30} = 0.782 \pm 0.010$,  from the Planck Clusters~\cite{Ade:2013lmv}.
\end{itemize}
In the right panel of Fig~\ref{fig:Pk}, we show how the matter power spectrum changes when using best-fit
values of $\Lambda$CDM parameters from a simultaneous fit to all these datasets rather than  \textbf{CMB}
alone. 
On scales relevant for $\sigma_8$, this mostly has the effect of slightly increasing the power with respect
to what is shown in the left panel of the same figure.
This is due to the fact that for fitting the CMB spectrum of the model to the data, a smaller DM density of 
our model needs to be compensated by a larger $A_s$.
Overall we thus typically expect a slightly larger value of $\sigma_8$ in our scenario, as compared to the
$\Lambda$CDM case. While this seemingly further increases the discrepancy between CMB and 
low-redshift observables, we will see that the simultaneous {\it decrease} in $\Omega_m$ overcompensates 
this effect, allowing for a slight alleviation of the observed tension. 

In Fig.~\ref{fig:triangle_comp_LSS}, we provide a first illustration of the tension in low- and high-redshift 
observables mentioned above. The left panel, in particular,  contrasts the $\Lambda$CDM best-fit region 
in the $H_0$ versus $\sigma_8(\Omega_m/0.27)^{0.30}$ plane obtained from \textbf{CMB} data only (red contours) 
with the direct measurements of these quantities by \textbf{HST} (cyan band) and \textbf{PC} (orange band). 
The blue contours show the preferred region in this plane when combining all these datasets.
(the green contours result when also adding the Planck lensing power spectrum reconstruction~\cite{Ade:2015zua}). 
The incompatibility between the different data sets is clearly visible and is in particular reflected in the fact that 
the red and blue ellipses do not overlap.

The right panel of Fig.~\ref{fig:triangle_comp_LSS} demonstrates how our conversion scenario may help to mitigate 
this discrepancy. For this purpose we show how the best-fit regions shift for specific values of our model parameters 
($\kappa=1$, $a_t=10^{-1.5}$, $\zeta=0.06$). We note that such an efficient DM conversion would appear
firmly excluded by the CMB limits shown in Fig.~\ref{fig:CMB_limits}, but we will discuss below how 
adding large-scale structure data strongly relaxes those constraints (and, depending on the choice of priors,
even {\it prefers} such large values of $\zeta$, see Appendix~\ref{app:Bayesian}.
For this model point, we find that the red ellipse, corresponding to the parameter region 
preferred by the CMB alone, moves downward and to the right, such that it overlaps with the blue ellipse obtained 
from combining all data sets at 95\%\,C.L.

\begin{figure*}
\includegraphics[width=\columnwidth]{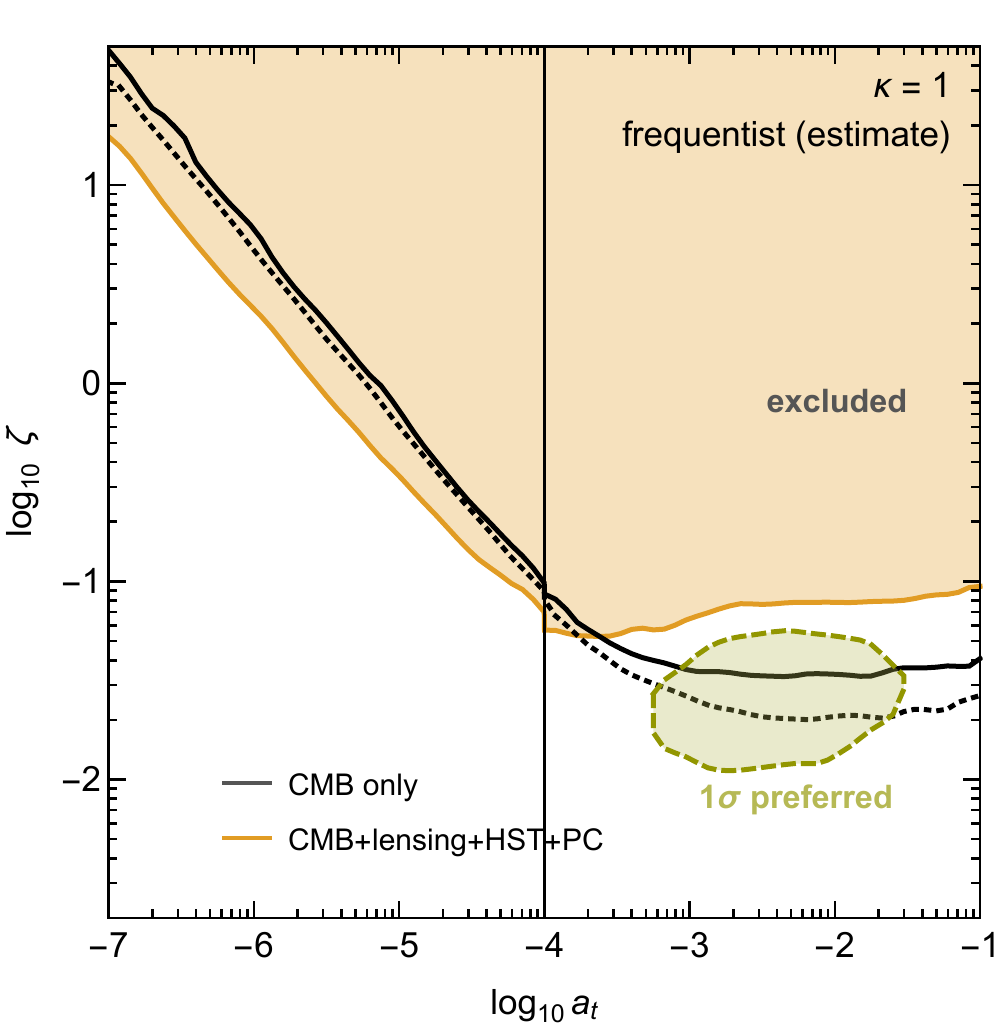}\quad
\includegraphics[width=\columnwidth]{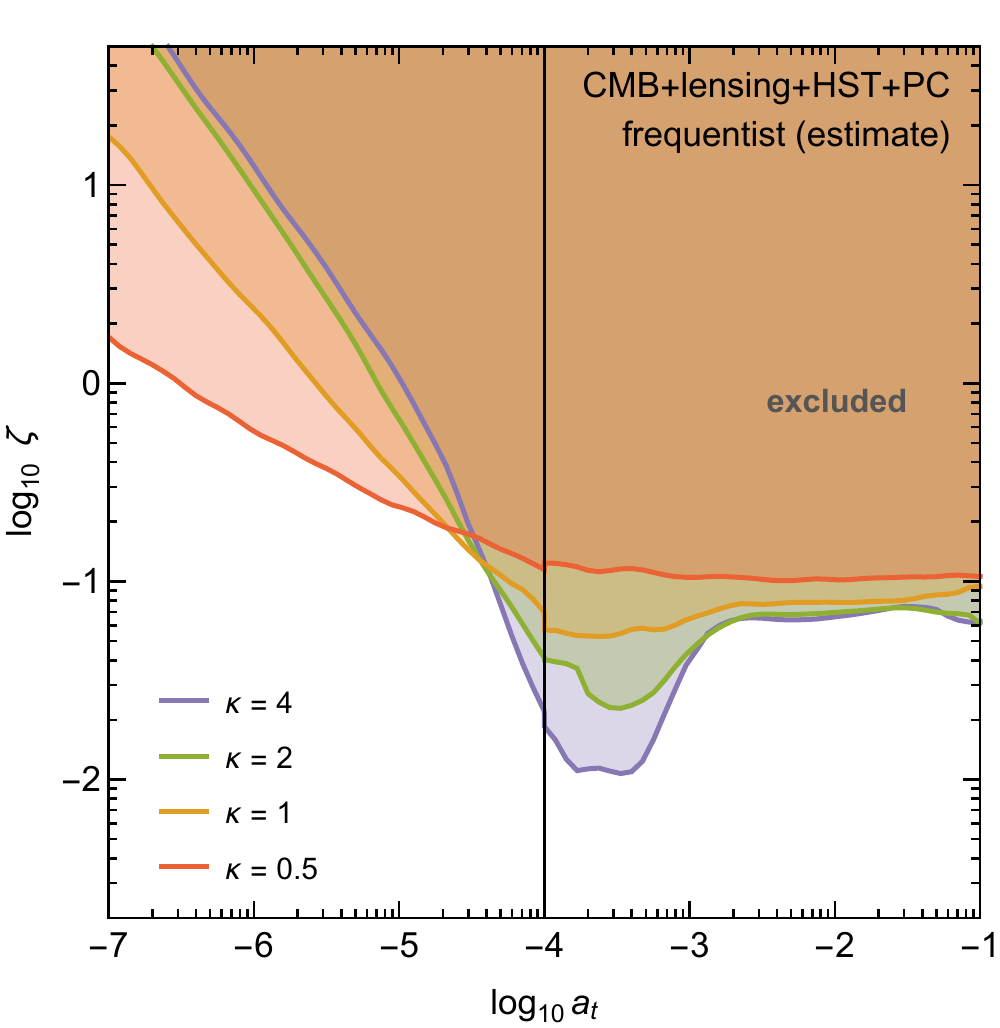}
\caption{{\it Left panel}: {Approximate frequentist constraints for our conversion scenario with $\kappa=1$, resulting from \textbf{CMB+Lensing+HST+PC} compared to the constraints obtained from \textbf{CMB} only (identical to the corresponding line in Fig.~\ref{fig:CMB_limits}). The regions above the solid lines are excluded at 99\%\,C.L. For \textbf{CMB} only we also show the 95\%\,C.L. exclusion limit (dotted), while for \textbf{CMB+Lensing+HST+PC} we find a \mbox{$\sim2\sigma$} `signal' preference and show the preferred parameter region at 68\% C.L. (dashed). {\it Right panel}: Frequentist exclusion limits at 99\%\,C.L. on the amount of converted DM from \textbf{CMB+Lensing+HST+PC} for different 
choices of $\kappa$.}\label{fig:constraints_LSS_vs_CMB}}
\end{figure*}

We can qualitatively understand this effect by recalling that 
$\Omega_\chi h^2$ is tightly constrained at recombination. The {\it decreasing}  DM component of our model at later 
times thus implies that we have to simultaneously {\it increase} the Hubble rate in order to remain compatible with 
CMB data. At the same time, the total matter density $\Omega_m=\Omega_\chi+\Omega_b$ also decreases, which 
shifts  $\sigma_8(\Omega_m/0.27)^{0.30}$ 
downwards, even though $\sigma_8$ increases slightly with respect to the $\Lambda$CDM case
(see Fig.~\ref{fig:Pk}). 
Including  \textbf{Lensing} (green contours) slightly enhances the tension with the $\sigma_8$ measurement again, 
but does not change the picture qualitatively.
We finally checked that adding Baryon Acoustic Oscillations measurements from the galaxy surveys  in 
Refs.~\cite{Beutler:2011hx,Ross:2014qpa,Gil-Marin:2015nqa} would not affect the left panel of 
Fig.~\ref{fig:triangle_comp_LSS}, but shift the blue contour in the right panel slightly to the left (to the point where 
the 1$\sigma$ contour does not quite overlap any more with the 1$\sigma$ band of the $H_0$ measurement). 

Since our model of DM conversion clearly has the potential to reduce the tension between CMB and LSS data, 
we can expect that the inclusion of the latter will also significantly modify the {\it constraints} discussed in 
Sec.~\ref{sec:cmb}.
In {the left panel of} Fig.~\ref{fig:constraints_LSS_vs_CMB} we demonstrate this for the case of $\kappa=1$.  The most prominent change compared to the 
bounds obtained from CMB data only is that constraints for large $a_t$ are substantially weaker. This is a direct 
consequence of the fact that in this region (and for $\zeta \sim 10^{-2}$) our model actually gives a better fit to data 
than $\Lambda$CDM (mostly by increasing the Hubble rate, as already 
indicated in Fig.~\ref{fig:triangle_comp_LSS}). 
At the same time, the limits for small values of $a_t$ strengthen because CMB and LSS independently
constrain a constant $\Delta N_{\rm eff}$. 
{In the right panel of Fig.~\ref{fig:constraints_LSS_vs_CMB}} we show the limits from \textbf{CMB + Lensing + HST + PC} for different 
choices of $\kappa$. In each case we observe a substantial weakening of the constraints for large $a_t$ compared 
to the limits obtained from CMB data only (see Figs.~\ref{fig:CMB_limits} and~\ref{fig:CMB_limits_frequentist}). 

At this stage the obvious question arises whether our model of DM conversion only reduces the tension
between CMB and LSS data, or whether one may even claim {\it positive evidence} for 
this model based on LSS data. From the frequentist perspective the preference is 
at the $\sim2\sigma$ level and hence not very
significant. We indicate in the right panel of Fig.~\ref{fig:constraints_LSS_vs_CMB} the parameter region preferred 
by the combination of CMB and LSS data at 68\% C.L.\footnote{%
To construct this parameter region, we again use the test statistic defined in Eq.~(\ref{eq:ts}). The preferred 
parameter region at 68\% C.L.\ is then given by the requirement $t < 2.28$. We refrain from attempting an 
exact reconstruction of the $2\sigma$ contour, which would require a higher sampling efficiency.
This parameter region is similar also in the other cases shown in {the right panel of} Fig.~\ref{fig:constraints_LSS_vs_CMB}, 
except for $\kappa = 1/2$, where the preference is slightly less than $2\sigma$ and hence the $1\sigma$ 
region is somewhat larger.
}
{From a Bayesian perspective, as discussed in more detail in Appendix~\ref{app:Bayesian}, 
the signal preference depends strongly on the adopted prior.}
%

\section{Sommerfeld-enhanced dark matter annihilation}
\label{sec:sommer}

In this section we discuss DM with Sommerfeld enhancement as an interesting scenario in which a fraction of DM is 
converted into DR over a well-defined period of time. The basic idea is that DM particles interact with each other via a 
mediator particle with mass small compared to the DM mass, $m_\text{med} \ll m_\chi$. The exchange of light 
mediators then generates a potential that modifies the wave function of DM particles, leading to an enhancement of 
the DM self-annihilation cross section $(\sigma v)_0$ at small velocities~\cite{Iengo:2009ni,Cassel:2009wt}:
\begin{equation}
 \sigma v = S(v) (\sigma v)_0 \; .
\end{equation}

As long as the Sommerfeld factor is small, $S(v) \approx 1$, the annihilation rate of a given DM particle decreases 
rapidly with decreasing redshift as the number density of DM particles decreases: 
$\Gamma_\text{ann} = \sigma v \, \rho_\chi / m_\chi \propto a^{-3}$. Since the Hubble rate decreases 
more slowly (proportional to $a^{-2}$ or $a^{-3/2}$ during radiation domination and matter domination, 
respectively), DM annihilations become less and less important in the late Universe. 

This situation can be reversed in the presence of a large Sommerfeld enhancement. As we will discuss in more detail 
below, in certain regions of parameter 
space one finds $S(v) \propto v^{-2}$ for small velocities. As long as DM particles are in local thermal equilibrium, 
their velocity is $v \propto T_\chi^{1/2} \propto a^{-1/2}$. After the DM particles have \emph{kinetically decoupled} 
from the heat bath, however, their momenta simply redshift as $v \propto a^{-1}$, such that 
$\Gamma_\text{ann} \propto a^{-1}$. In this case, the annihilation rate decreases more slowly than the Hubble rate 
and DM annihilations become increasingly important. This leads to a {\it second
period of DM annihilation} after the classical chemical freeze-out~\cite{Feng:2010zp,vandenAarssen:2012ag}. As a 
result, the comoving DM 
density may change appreciably at late times. For even smaller velocities, the Sommerfeld factor saturates and the 
DM annihilation rate reverts to its usual scaling proportional to $a^{-3}$, so that the comoving DM density again 
becomes constant.

\subsection{Model set-up}

To be more specific, we consider the case of a Dirac fermion DM particle $\chi$ coupled to a vector
 mediator $V^\mu$:
\begin{equation}
\label{eq:lagrangian}
 \mathcal{L} \supset g_\chi \, \bar{\chi}\gamma^\mu\chi V_\mu \; .
\end{equation}
The dominant DM annihilation channel in this set-up is the s-wave process $\chi \bar{\chi} \to V V$, for which one 
finds, in the limit of vanishing relative velocity and mediator mass, $(\sigma v)_0 = \pi \alpha^2 / m_\chi^2$  
with $\alpha = g_\chi^2 / (4\pi)$.\footnote{Similar results are found for the case of scalar DM. The case of a scalar 
mediator, on the other hand, is qualitatively different, as the annihilation into a pair of mediators is a $p$-wave 
process.} Although the mediators produced in DM annihilations could themselves act as DR, we assume that they 
subsequently decay into even lighter particles, such as sterile neutrinos. The advantage of such a set-up is that the 
resulting interactions between DM and DR can significantly delay the kinetic decoupling of 
DM~\cite{Aarssen:2012fx}
(while at the same time avoiding strong CMB constraints on visible decays~\cite{Bringmann:2016din}). Rather 
than specifying the coupling between the mediator and DR, however, we introduce here the kinetic decoupling 
temperature $T_\text{kd}$ as an additional free parameter to keep the discussion more model-independent.
In Sec.~\ref{sec:sommer_other} we will briefly get back to the range of decoupling temperatures that would 
be expected in the simplest extension to the model specified in Eq.~(\ref{eq:lagrangian}), and otherwise
refer to Ref.~\cite{Bringmann:2016ilk} for a detailed discussion of how late kinetic decoupling can be 
achieved in general.

\begin{figure}
\includegraphics[width=0.9\columnwidth]{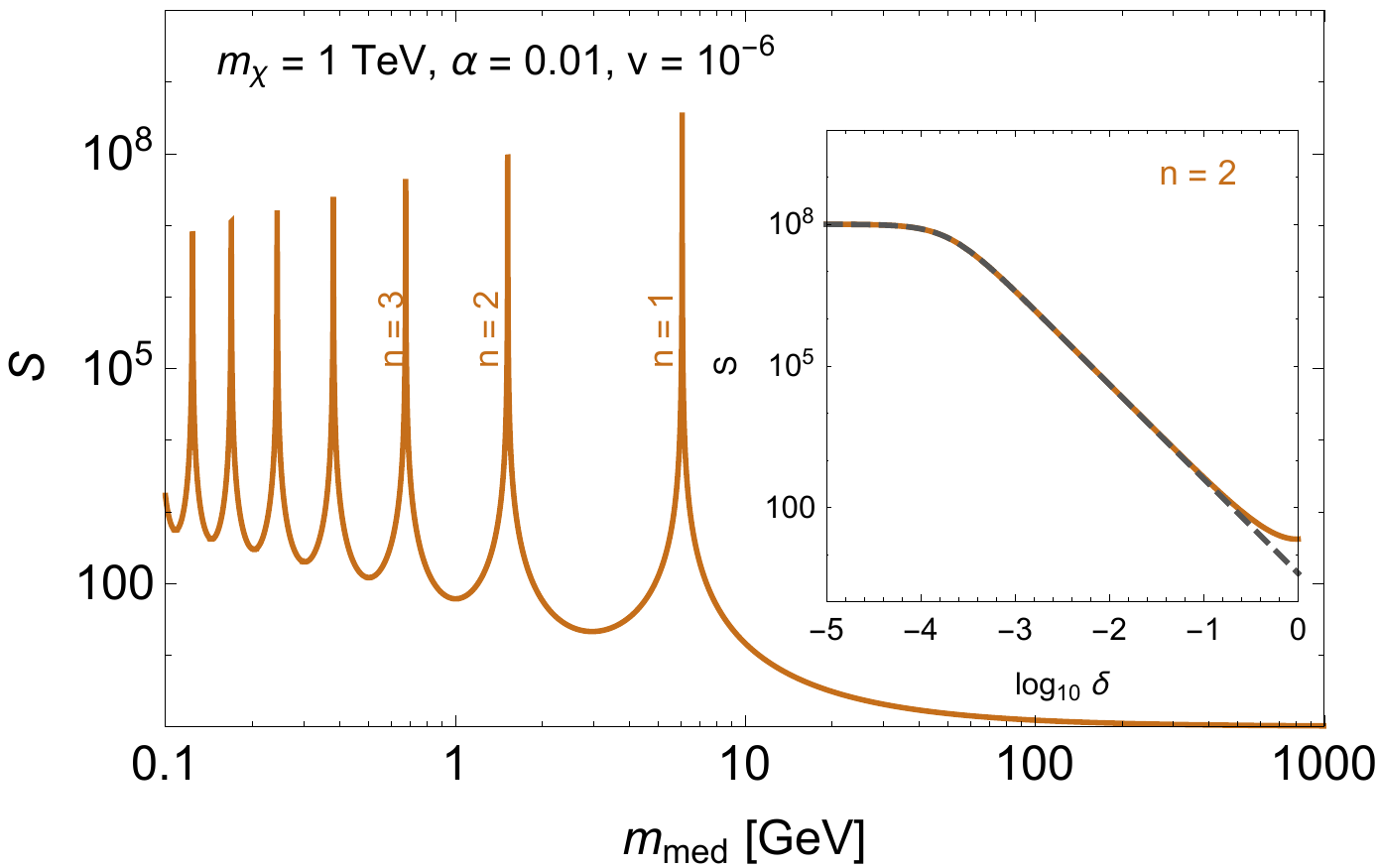}
\caption{Sommerfeld enhancement factor $S$ as a function of $m_\text{med}$ for fixed values of $m_\chi$, $\alpha$ 
and $v$. If the mediator mass satisfies Eq.~(\ref{eq:res}) for some integer $n$ the enhancement can be very large. 
In the inset we zoom into one specific resonance ($n = 2$) by replacing $m_\text{med}$ with $\delta$ as 
defined in Eq.~(\ref{eq:delta}).  For comparison we also show the approximation of the Sommerfeld enhancement 
factor given in Eq.~(\ref{eq:Sapprox}), which is valid for $\delta \ll 1/(n\pi)$ and $v \ll \alpha/(n^2 \pi)$.}
\label{fig:Sillustration}
\end{figure}

The exchange of vector mediators generates the Yukawa potential
\begin{equation}
 V(r) = \frac{\alpha \, e^{-r \, m_\text{med}}}{r} \; .
\end{equation}
The Sommerfeld factor can be calculated analytically by approximating the Yukawa potential with a Hulth\'{e}n 
potential, giving~\cite{Iengo:2009ni,Cassel:2009wt,Tulin:2013teo}
\begin{equation}
S = \frac{ 2\pi \, \alpha \sinh \left(\frac{6 m_\chi v}{\pi m_\text{med}}\right)}{v \left[-\cos\left(2 \pi \sqrt{\frac{6 m_\chi \alpha}{\pi^2 m_\text{med}} -\frac{9 m_\chi^2 v^2}{\pi^4 m_\text{med}^2}} \right) + 
\cosh \left(\frac{6 m_\chi v}{\pi m_\text{med}}\right)\right]} \; .
\end{equation}
We display this Sommerfeld enhancement factor in Fig.~\ref{fig:Sillustration} as a function of $m_\text{med}$ for 
fixed values of $m_\chi$, $\alpha$ and $v$.

In the limit of vanishing velocities, one finds that the denominator becomes very small if
\begin{equation}
m_\text{med} \approx \frac{6 m_\chi \alpha}{\pi^2 n^2}
\label{eq:res}
\end{equation}
for some integer $n$. To quantify how close a specific parameter point is to such a resonance, we define
\begin{equation}
 \delta \equiv \left|\frac{m_\text{med} - m_\text{med}^{(n)}}{m_\text{med}^{(n)}}\right| \equiv \left|1 - \frac{\pi^2 n^2 \, m_\text{med}}{6 m_\chi \alpha}\right| \; ,
 \label{eq:delta}
\end{equation}
where $m_\text{med}^{(n)}$ is the value of $m_\text{med}$ at the $n$th resonance and $n$ is chosen to 
minimise $\delta$. The inset in Fig.~\ref{fig:Sillustration} shows the Sommerfeld factor as a function of $\delta$ for $n = 2$.

If $\delta$ is sufficiently small, $\delta \ll 1/(n\pi)$, one finds that the Sommerfeld factor for small velocities, 
$v \ll \alpha/(n^2 \pi)$, can be written as
\begin{equation}
S(v) = \frac{4 \alpha^2}{n^2 v^2 + \alpha^2 \delta^2} \; .
\label{eq:Sapprox}
\end{equation}
The quality of this approximation can be inferred from the black dashed line in the inset of Fig.~\ref{fig:Sillustration}. We 
conclude that the Sommerfeld factor begins to grow as $1/v^2$ until $v \lesssim v_\text{sat} \equiv \alpha \delta / n$, at 
which point the Sommerfeld factor saturates at $S \approx 4 / \delta^2$.

\begin{figure*}
\includegraphics[width=0.9\columnwidth]{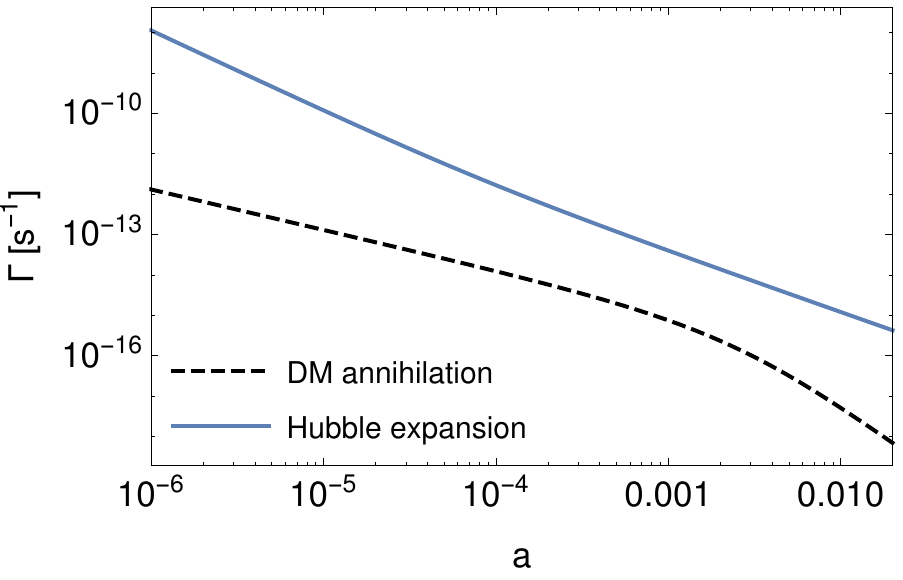} \hspace*{1cm} 
\includegraphics[width=0.9\columnwidth]{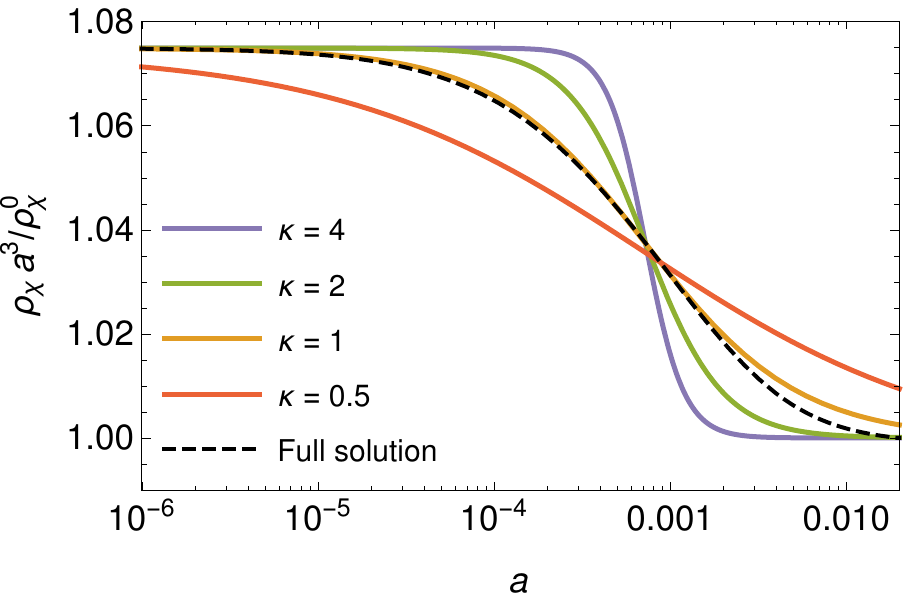}
\caption{{\it Left panel.}
Hubble expansion rate (solid line) compared to the annihilation rate of a given DM particle 
(dashed line) with 
$m_\chi = 250\,\text{GeV}$, $\alpha = 10^{-2}$, $n = 10$, $\delta = 10^{-5}$ 
(corresponding to a mediator mass $m_\text{med}\approx 13$\,MeV) and kinetic decoupling temperature
$T_\text{kd} = 1\,\text{keV}$.
{\it Right panel.} Resulting DM density evolution for the same parameter point as in the left panel (dashed line),
compared to the phenomenological transition scenarios introduced in Section \ref{sec:DMtoDR} for 
$a_t= 7.2 \cdot 10^{-4}$ and $\zeta=0.075$ (solid lines; see also Fig.~\ref{rhodmfig}).
}
\label{fig:rates}
\end{figure*}

An additional subtlety is that the Sommerfeld factor obtained from the naive solution of the Hulth\'{e}n potential can 
become so large that the annihilation cross section violates unitarity at very small velocities. To avoid this unitarity 
violation for very small $\delta$, we follow the prescription from Ref.~\cite{Blum:2016nrz} and consider the modified 
Sommerfeld factor
\begin{equation}
S(v) = \frac{4 \alpha^2}{n^2 \left(v + v_c\right)^2 + \alpha^2 \delta^2}
\label{eq:S}
\end{equation}
with $v_c = \alpha^4 / (4 n^2)$.

We emphasise that while Eq.~(\ref{eq:S}) provides a very good approximation to the Sommerfeld factor close to 
resonance at small velocities, it does not yield the correct description for large velocities or far away from a 
resonance. However, as argued above, DM annihilations will not be important in these regimes anyways, so that a 
more detailed modelling of the Sommerfeld factor is unnecessary for our purposes. We also note that the way in 
which we implement the restoration of unitarity for small $\delta$ is only approximate. While it ensures that the 
Sommerfeld factor does not exhibit unphysical behaviour for $v \to 0$, we expect a more detailed calculation to yield 
slightly different results for finite velocities.

\subsection{Evolution of dark matter density}

For the purpose of calculating DM annihilation rates, we are interested in the thermally averaged annihilation cross section
\begin{equation}
\langle \sigma v_\text{rel} \rangle = \langle S (\sigma v_\text{rel})_0 \rangle = \langle S \rangle (\sigma v_\text{rel})_0 \; ,
\end{equation}
where we have made use of the fact that $(\sigma v)_0$ is independent of velocity. To calculate the thermal average, 
we assume that the DM velocity distribution is given by a Maxwell-Boltzmann distribution with an effective 
temperature $T_\text{eff}$:
\begin{equation}
 f(v_\text{rel}) = \sqrt{\frac{x_\text{eff}^3}{4 \pi}} v_\text{rel}^2 \exp\left(- \frac{v_\text{rel}^2 \, x_\text{eff}}{4} \right) \; ,
\end{equation}
where we have introduced the dimensionless temperature $x_\text{eff} = m_\chi / T_\text{eff}$.
We note that the above ansatz is automatically satisfied for  parameter combinations close to a resonance because 
the same light mediator that causes the Sommerfeld enhancement also guarantees very efficient 
DM self-interactions~\cite{vandenAarssen:2012ag}.

In order to proceed, we need to express $x_\text{eff}$ as a function of the scale factor $a$. For this purpose, we assume 
that DM particles are no longer in kinetic equilibrium with the thermal bath. Denoting the temperature and scale factor
of kinetic decoupling by $T_\text{kd}$ and $a_\text{kd}$, respectively, we find
\begin{equation}
T_\text{eff} = T_\text{kd} \frac{a_\text{kd}^2}{a^2} = \frac{T_0^2}{T_\text{kd}} a^{-2} \; , 
\end{equation}
where $T_0$ is the present-day photon temperature.\footnote{%
Here we have made two additional assumptions. First 
we assume for simplicity that the temperature of the dark sector is the same as the temperature of the visible sector. 
Relaxing this assumption and introducing the temperature ratio of the two sectors as an additional free parameter 
does not change our results qualitatively. Second we assume that DM annihilations do not change the temperature of 
the dark sector. This is not necessarily a good approximation, since in the presence of Sommerfeld enhancement, 
DM particles with small velocities have higher probability to annihilate, leading effectively to an increase of the DM velocity 
dispersion. In principle, this effect can be included by solving a set of coupled differential 
equations~\cite{vandenAarssen:2012ag}. However, as long as the relative change of the DM density is small, 
we can neglect the resulting change in the dark sector temperature.
} 
We conclude that the thermally averaged Sommerfeld factor is proportional to $a^2$ for 
$a \lesssim a_\text{sat} \equiv T_0 / (v_\text{sat} \sqrt{T_\text{kd} m_\chi})$ and becomes constant for larger scale factors.

We show the corresponding DM annihilation rate $\Gamma_\text{ann}$ in comparison to the Hubble rate in the left 
panel of Fig.~\ref{fig:rates} for $m_\chi = 250\,\text{GeV}$, $\alpha = 10^{-2}$, $n = 10$, $\delta = 10^{-5}$ and 
$T_\text{kd} = 1\,\text{keV}$, corresponding to a mediator mass of $m_\text{med}\approx 13$\,MeV (the value 
of $\alpha$ was chosen such as to roughly result in the correct relic density from standard thermal freeze-out). 
For this choice of parameters the Sommerfeld factor saturates around $a \sim 10^{-3}$, staying significantly below the 
Hubble rate. 

To calculate the change of DM density resulting from this annihilation rate, we need to solve the 
Boltzmann equation
\begin{equation}
 \frac{\mathrm{d}\rho_\chi}{\mathrm{d}z} (1 + z) H(z) - 3 \rho_\chi H(z) - \frac12\langle \sigma v_\text{rel} \rangle \frac{\rho_\chi^2}{m_\chi} = 0
 \label{eq:boltz}
\end{equation}
with the boundary condition
\begin{equation}
 \rho_\chi(z_\text{CMB}) = \Omega_\chi \rho_c (1+z_\text{CMB})^3 \; ,
\end{equation}
where $z_\text{CMB} = 1100$ is the redshift at recombination and $\Omega_\chi = 0.1198 / h^2$ and $\rho_c 
=1.054\cdot 10^{-5}h^2\mathrm{GeV/cm^3}$
are the {present-day} DM abundance and critical density inferred from CMB 
observations {under the assumption of $\Lambda$CDM}.\footnote{%
There is some arbitrariness in the choice of $z_\text{CMB}$, but since we focus on the case 
where the DM density changes only slightly, the precise choice of $z_\text{CMB}$ does not affect our results.
} 
The factor of $1/2$ in front of the last term in Eq.~(\ref{eq:boltz}) accounts for the fact that DM consists of Dirac 
particles; in other words, $\rho_\chi$ refers here to the {\it total} density of both $\chi$ and $\bar\chi$ and thus has 
the same meaning as in the previous sections. The 
solution of this equation is shown in the right panel of Fig.~\ref{fig:rates} for the same choice of parameters as 
on the left (black dashed line).
We also show for comparison the DM density as a function of redshift for the phenomenological parametrisation 
introduced in Section~\ref{sec:DMtoDR}. 
We find that for $\kappa = 1$ (orange line) the model is very similar to the numerical solution of the 
Boltzmann equation, while for the other values of $\kappa$ the transition looks 
quite different. 

\subsection{CMB and LSS constraints}

We just concluded that late-time DM annihilations with resonant Sommerfeld enhancement provide a good 
example for the model discussed in the previous sections with $\kappa = 1$. We can therefore use the 
constraints obtained for that phenomenological model to place bounds on models with resonant Sommerfeld 
enhancement. In order to determine whether a specific parameter point is allowed or excluded by 
cosmological data, we thus need to determine the values of $a_t$ and $\zeta$ that provide the best fit to 
the numerical solution of the Boltzmann equation and then compare these values to the frequentist bounds shown in 
Fig.~\ref{fig:constraints_LSS_vs_CMB} (concretely, we determine $a_t$ as the scale factor where half of the 
conversion has happened).
Using frequentist bounds has the crucial advantage that we do not need to 
specify priors for the particle physics parameters of the model we consider. Moreover, even if priors for the particle 
physics parameters could be motivated, these would likely translate to non-trivial priors on $a_t$ and $\zeta$, 
meaning that the Bayesian limits derived in the previous section could not be directly applied.

From the discussion in the previous subsection, this translation to $a_t$ and $\zeta$ can be done for arbitrary 
combinations of $m_\chi$, $\alpha$, $n$, $\delta$ and $T_\text{kd}$ that satisfy the following conditions:
\begin{itemize}
 \item[{\it i)}] The parameter point lies in the resonant regime: $\delta \ll 1/(n\pi)$.
 \item[{\it ii)}] Kinetic decoupling happens before the Sommerfeld factor saturates: $T_\text{kd} \gg m_\chi \, v_\text{sat}^2$.
 \item[{\it iii)}] The DM annihilation rate stays significantly below the Hubble rate even for $a \approx a_\text{sat}$, 
 so that the total relative change of the DM density remains small: $\zeta \lesssim 1$.
\end{itemize}
While the last requirement is not strictly necessary, it becomes computationally very challenging to accurately calculate the 
evolution of the DM density for $\zeta > 1$ due to the need to account for changes in the temperature of the dark sector. As 
we will see below, parameter regions with $\zeta > 1$ are either robustly excluded or phenomenologically uninteresting, so 
that we do not consider these regions in more detail.

In the following, we will impose one additional requirement, namely that $\alpha$ is chosen in such a way that the DM 
abundance predicted from thermal freeze-out coincides with the solution of Eq.~(\ref{eq:boltz}) for very early times, 
i.e.~$a \ll a_\text{sat}$. This 
requires solving the Boltzmann equation iteratively until a self-consistent solution is found.\footnote{%
Following Ref.~\cite{Kahlhoefer:2017umn} we approximate the Sommerfeld enhancement factor during freeze-out by 
calculating the Sommerfeld factor for $v = \sqrt{\pi T_\text{fo} / m_\chi}$, where the freeze-out temperature 
$T_\text{fo}$ as a function of DM mass is taken from Ref.~\cite{Steigman:2012nb}.}
We note that such an iterative procedure is particularly important for the parameter region where unitarity
restoration plays a role, because in this case the saturated Sommerfeld factor is proportional to $\alpha^{-6}$.

\begin{figure*}
\includegraphics[width=0.45\textwidth]{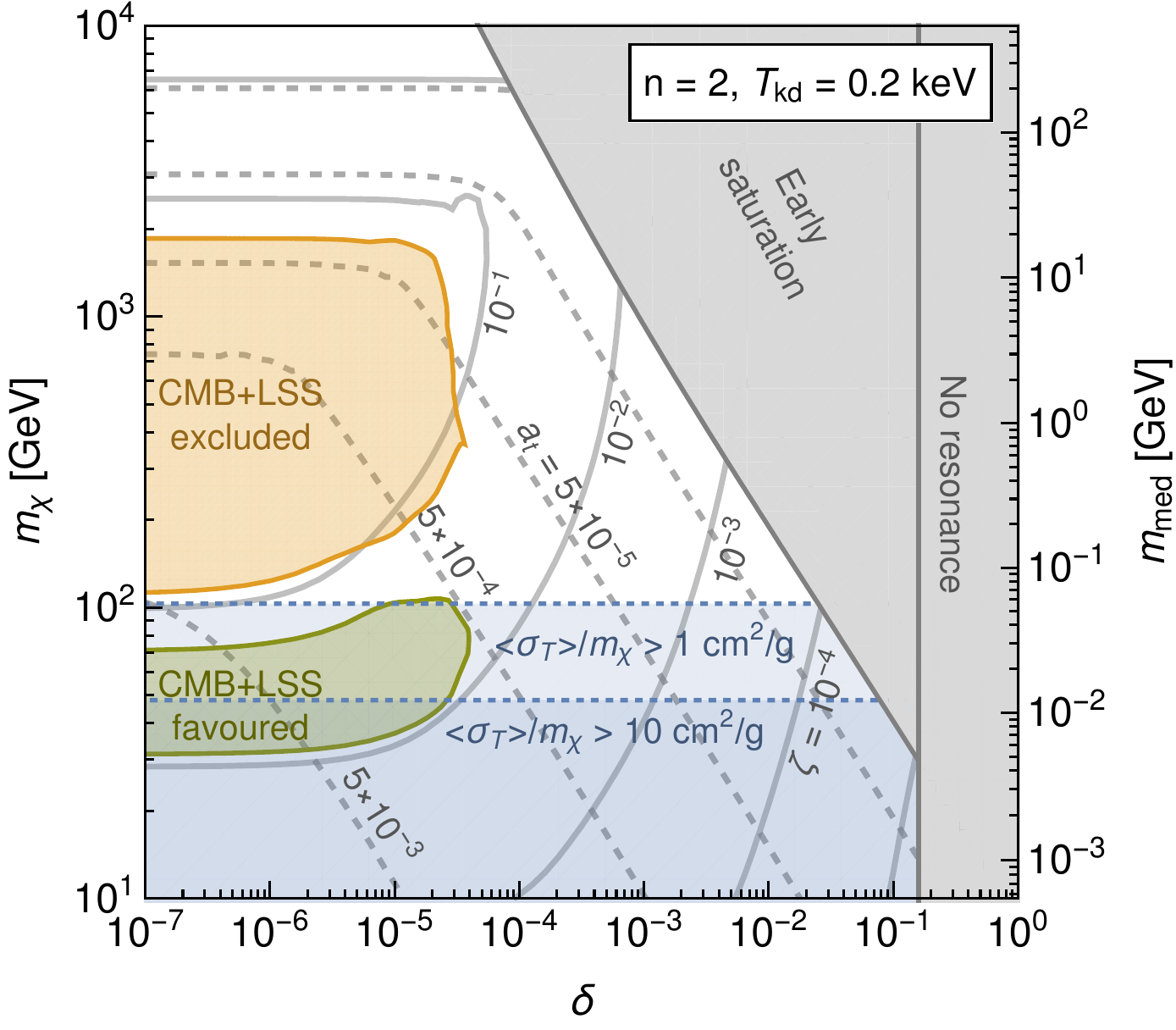} \quad \includegraphics[width=0.45\textwidth]{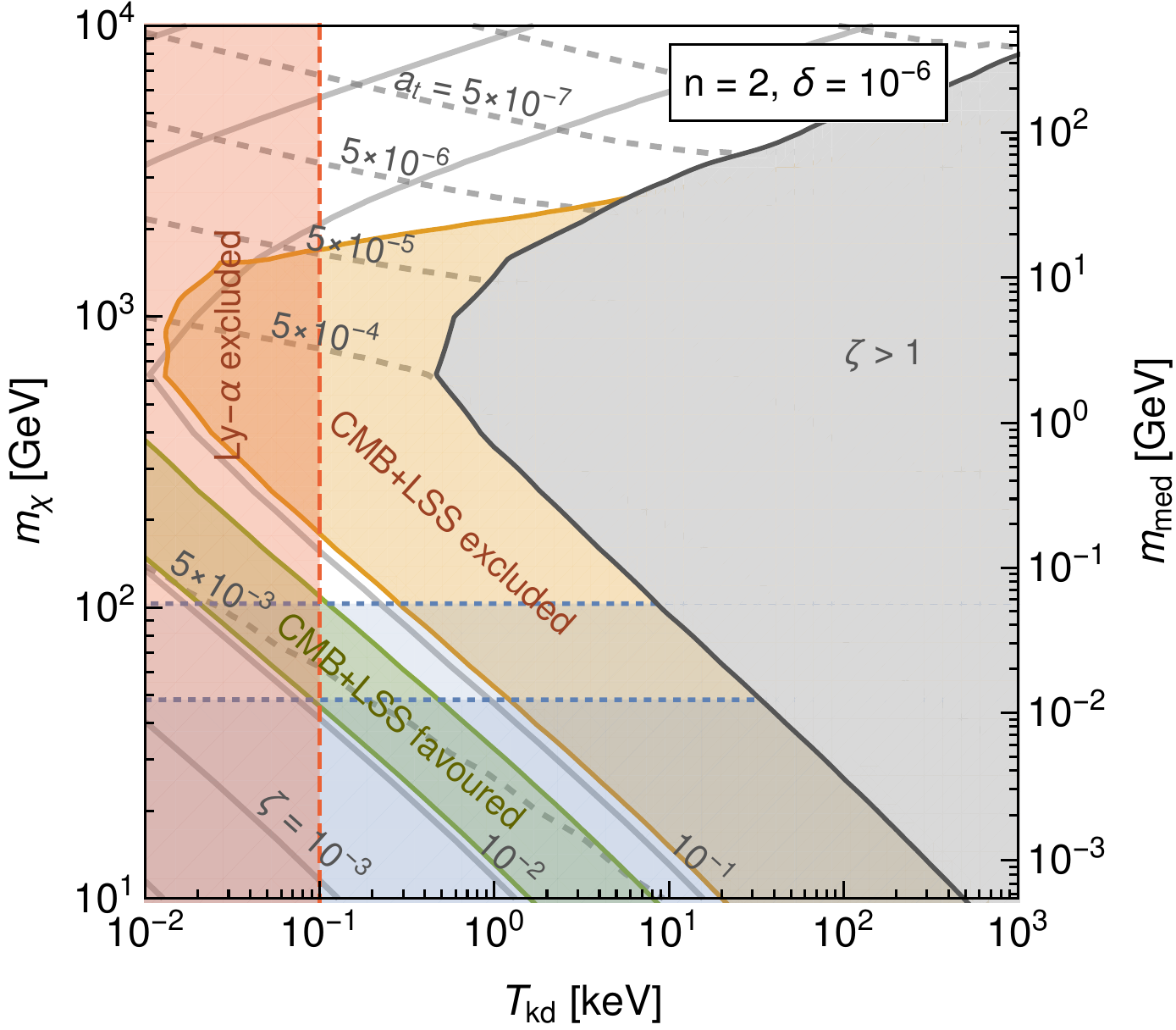}\\~\\
\includegraphics[width=0.45\textwidth]{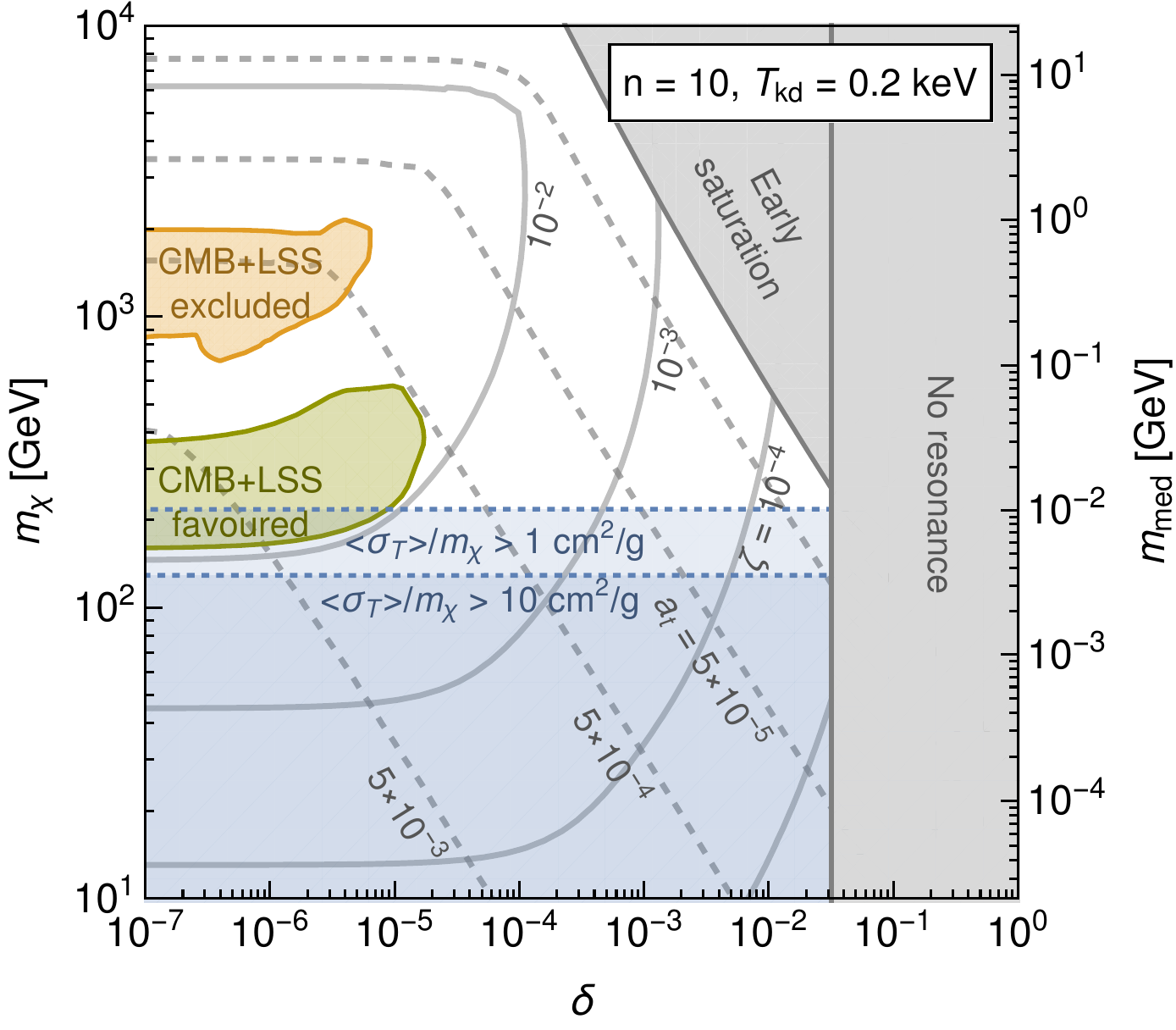}  \quad \includegraphics[width=0.45\textwidth]{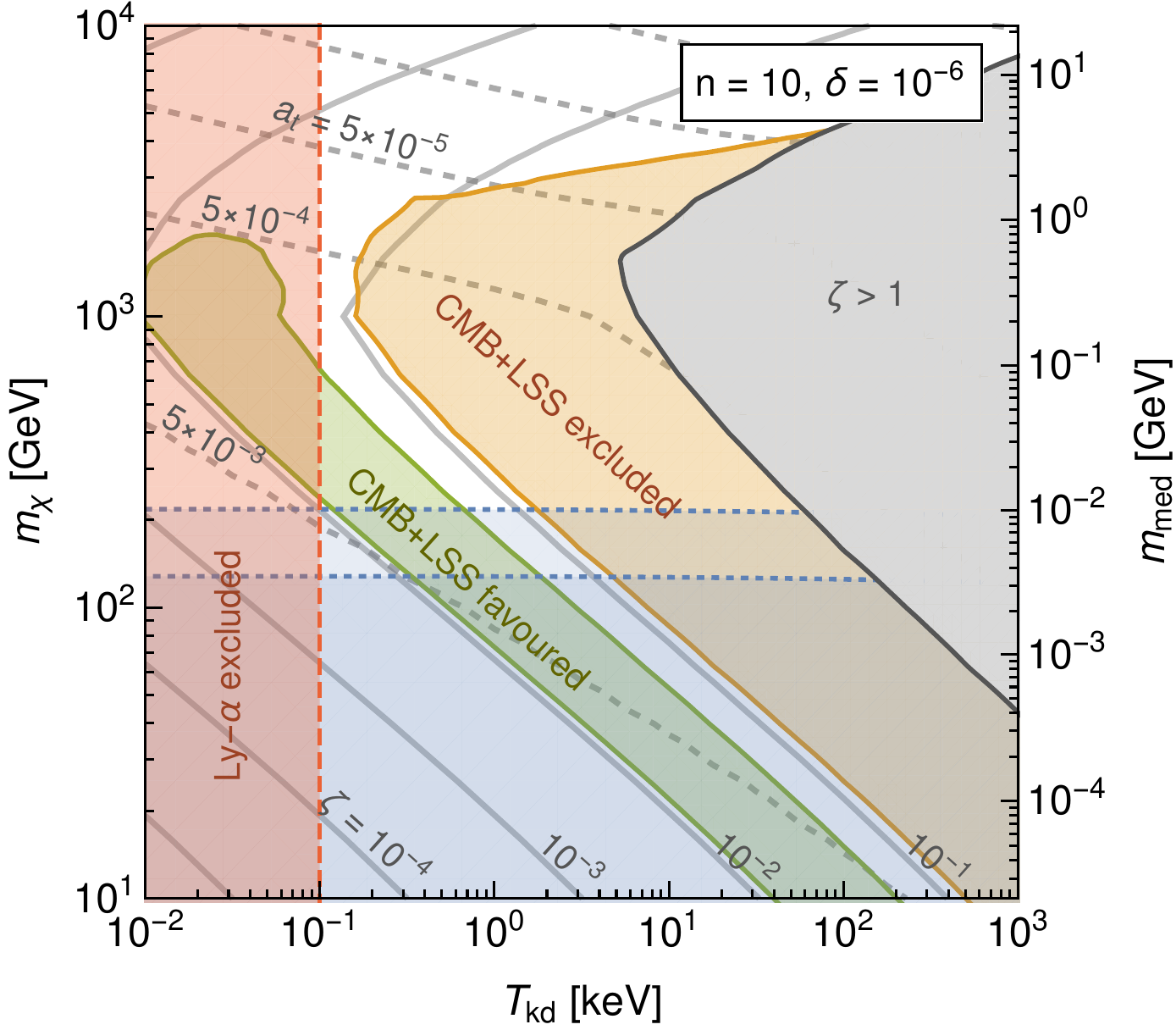}\\~\\
\includegraphics[width=0.45\textwidth]{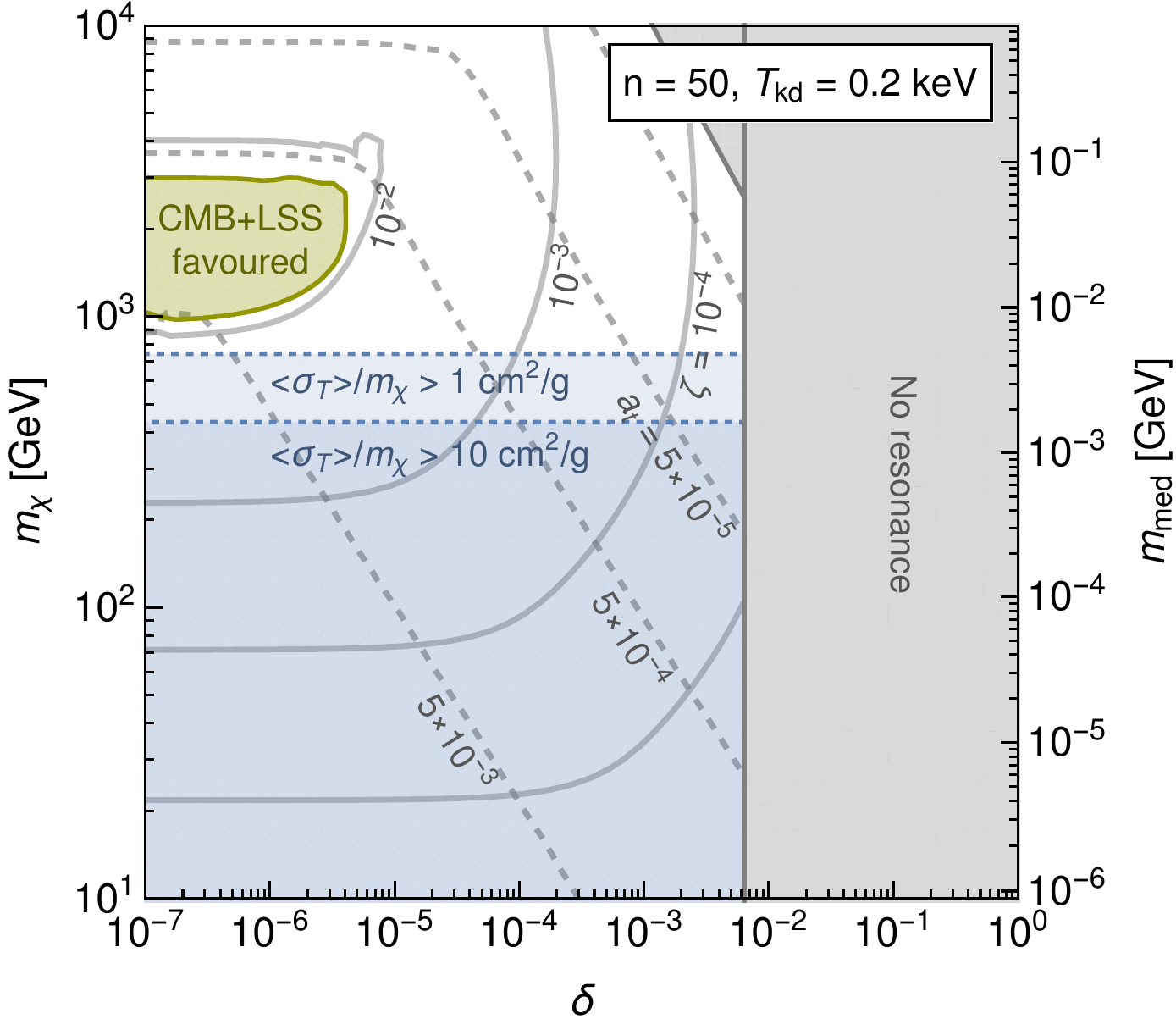}  \quad \includegraphics[width=0.45\textwidth]{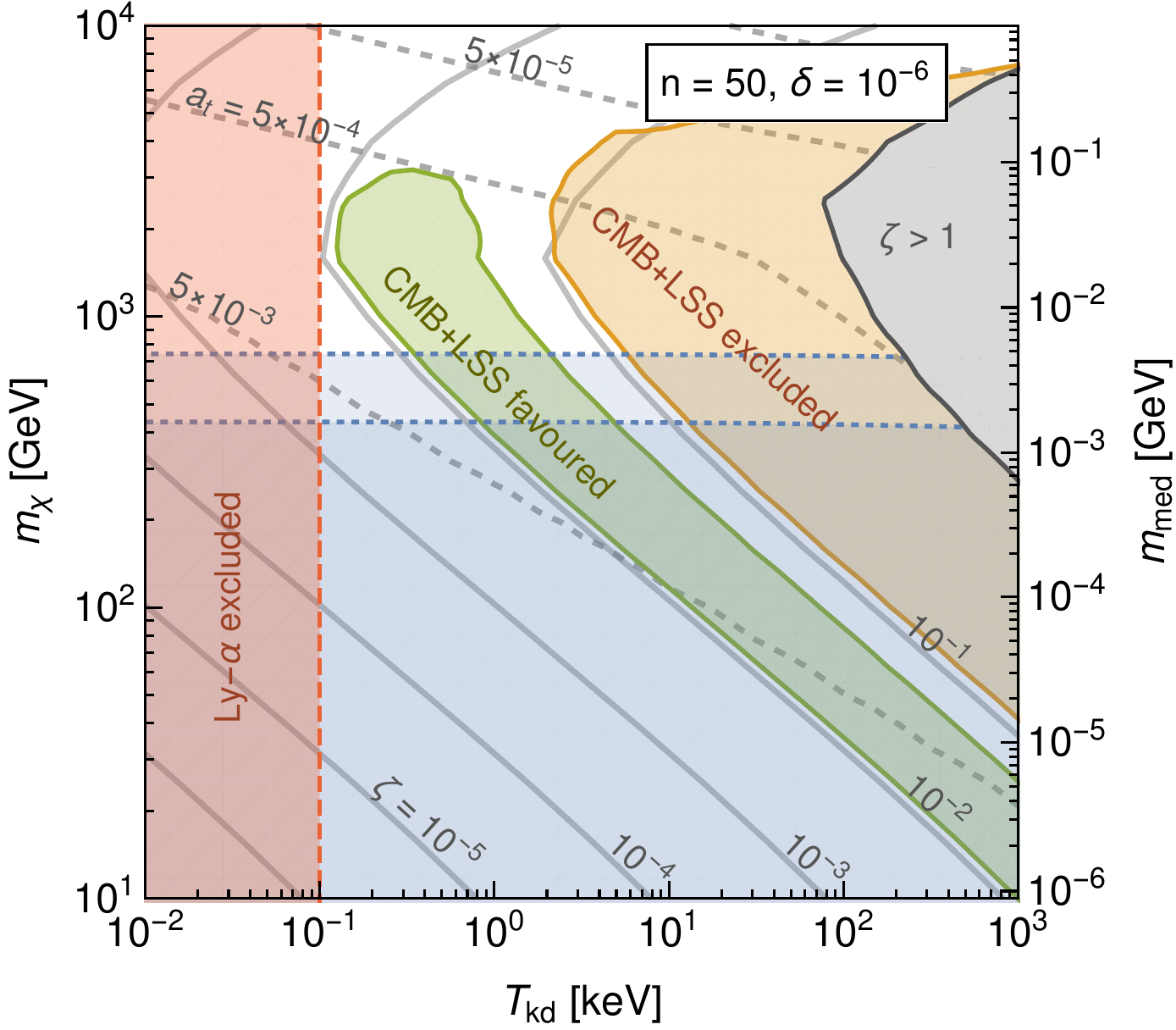}
\caption{Constraints on DM with Sommerfeld enhancement as a function of $m_\chi$ and $\delta$ (left) and as a 
function of $m_\chi$ and $T_\text{kd}$ (right). The different rows correspond to different resonances, i.e.\ different 
choices of $n$. The second $y$-axis in each panel indicates the mediator mass $m_\text{med}$ corresponding 
to $m_\chi$ for the specific resonance, cf.\ Eq.~(\ref{eq:res}).}
\label{fig:Sresult}
\end{figure*}

A final complication arises from the onset of non-linear structure formation around $z_\text{nl} \approx 50$. At this 
point the DM particles decouple from the Hubble flow, and their relative velocities start to increase. As a result the 
Sommerfeld enhancement factor drops and the comoving DM density very quickly becomes constant for 
$z \lesssim z_\text{nl}$, even if the 
Sommerfeld factor has not yet saturated. In this case, the functional form introduced in Eq.~(\ref{rhodmeq}) 
no longer provides a good 
description of the redshift dependence of the DM density for $a\gtrsim a_t$ (when choosing 
$a_t$ as the scale factor where half of the conversion has happened). 
However, as seen in Fig.~\ref{fig:constraints_LSS_vs_CMB}, if the conversion of 
DM to DR happens sufficiently after recombination, constraints are largely insensitive to the precise redshift 
dependence and only limit the total amount of DM converted. We can therefore continue to use the 
phenomenological parametrisation from above even in this regime. The cut-off of DM annihilations 
by non-linear structure formation can be shown to impose $a_t \gtrsim 7 \cdot 10^{-3}$ in our model.

Our results are summarised in Fig.~\ref{fig:Sresult} for $n = 2$ (top row), $n = 10$ (middle) and $n = 50$ 
(bottom row). In the left column we fix $T_\text{kd} = 0.2\,\mathrm{keV}$ and vary $\delta$, in the right 
column we fix $\delta = 10^{-6}$ and vary $T_\text{kd}$. 
The solid (dashed) lines in each panel indicate combinations of $m_\chi$ and $\delta$ corresponding to 
constant $\zeta$ (constant $a_t$). The yellow shaded region in each panel indicates the region of parameter space 
excluded by the constraints derived in this work, while the green shaded regions indicate the region favoured
by combining CMB and LSS data (as also shown in Fig.~\ref{fig:constraints_LSS_vs_CMB}).
The parameter regions that violate one or more of our basic conditions {\it i) -- iii)} are shaded in grey.

As expected, and as directly visible in the left panel of the figure, CMB and LSS data can only probe 
our model in the case of very small values of $\delta$, i.e.~for parameter regions rather 
close to a resonance. Interestingly, for each value of $n$, and a given kinetic decoupling temperature,
only a finite range of DM masses is excluded and preferred, respectively; these mass ranges move to 
larger values with increasing $n$. We note that 
the upper value of the excluded DM mass range is driven by the saturation of the Sommerfeld enhancement 
for very small velocities and $\delta$, so this is where the improvement of Eq.~(\ref{eq:Sapprox}) to 
Eq.~(\ref{eq:S}) is most relevant. Increasing $T_\text{kd}$, as in the right column, has the effect of lowering
the DM mass preferred by the data; the range of excluded DM mass is increased. We will soon see, however,
that too small DM masses inevitably lead to an unacceptably large DM self-scattering rate, so in practice
it is not very interesting to consider kinetic decoupling temperatures much larger than 1\,keV in this model.

\subsection{Discussion}

Let us briefly return to the treatment of perturbations in our conversion scenario. In 
Sec.~\ref{sec:pert} we argued that this is necessarily model-dependent, i.e.~not uniquely determined
by the choice of parameters $(\kappa,\zeta,a_t)$ that describe the evolution of the background densities.
Concretely we have so far always adopted the minimal option stated in Eq.~(\ref{eq:dQansatz}), i.e.
\be
  \delta \mathcal{Q}/Q=\delta_\chi\,.
\ee
For the case studied in this section the situation is different, because the conversion rate $\mathcal{Q}$
{\it is} associated to a concrete microphysics process, so that the Boltzmann equation directly determines the 
form of $\delta\mathcal{Q}$. The case of off-resonance Sommerfeld enhancement was discussed 
e.g.\ in Ref.~\cite{ArmendarizPicon:2012mu}, but the fully general case is rather involved. We will 
therefore estimate the impact of perturbations using a simplified treatment based on heuristic arguments. 

For annihilation processes with two DM particles in the initial state, we have 
$\mathcal{Q} \propto \langle \sigma v \rangle  \rho^2_\chi$, and thus
\begin{align}
\delta \mathcal{Q} &=\frac{\partial \mathcal{Q}}{\partial \langle \sigma v \rangle} \delta \langle \sigma v \rangle 
+ \frac{\partial \mathcal{Q}}{\partial \rho_\chi} \delta\rho_\chi 
=\frac{\delta \langle \sigma v \rangle}{\langle \sigma v \rangle} \mathcal{Q}  +  2 \mathcal{Q}\,\delta_\chi\,.
\end{align}
For $\sigma v \propto v^{-1}$, the perturbation $\delta \langle \sigma v \rangle$ at large scales is given by 
$\delta \langle \sigma v \rangle =\langle \sigma v \rangle \, \frac{h}{6}$~\cite{ArmendarizPicon:2012mu}.
Following the heuristic arguments given in that reference, this can be generalised to
$\delta \langle \sigma v \rangle =\beta \langle \sigma v \rangle \, \frac{h}{6}$ for a cross section scaling with
velocity as $\sigma v \propto v^{-\beta}$. Since we are mostly interested in parameter combinations very close 
to a resonance, where $\sigma v \propto v^{-2}$ this motivates us to change the prescription for perturbations to
\be
\delta \mathcal{Q}/Q=\delta_\chi \to \frac13 h + 2\delta_\chi\,.
\label{eq:sommer_pert}
\ee
This enters in both the evolution equation for DM perturbations, Eq.~(\ref{ddm}), and in those for the DR perturbations,
Eqs.~(\ref{ddr},\ref{vdr}).

\begin{figure}[t]
\includegraphics[scale=0.75]{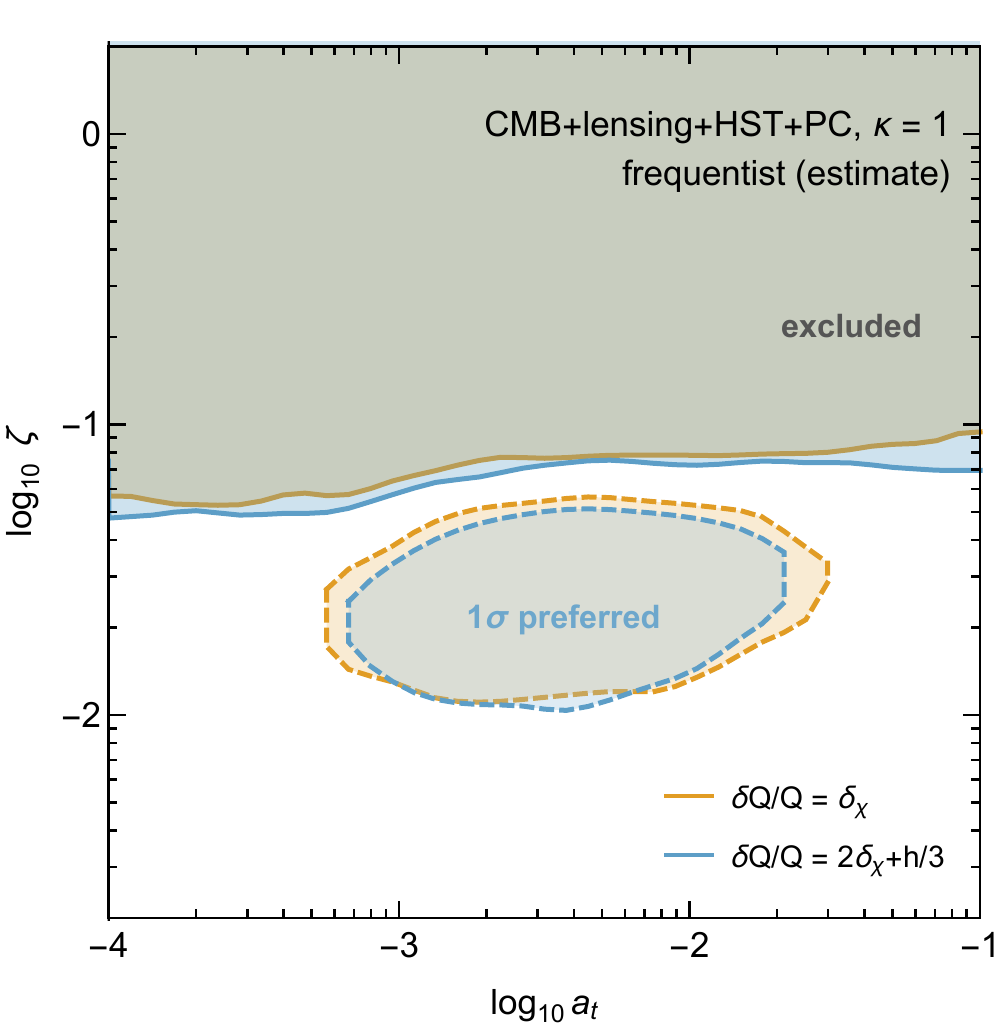}
\caption{Solid lines show the 99\%\,C.L.~approximate frequentist limits and dotted lines the 68\%\,C.L. preferred 
parameter region for our conversion scenario with $\kappa=1$, resulting from \textbf{CMB+Lensing+HST+PC}. The orange 
and green areas result from our standard treatment of perturbations and
are identical to those shown in the right panel of Fig.~\ref{fig:constraints_LSS_vs_CMB}.
The blue lines and area result when changing this prescription as stated in Eq.~(\ref{eq:sommer_pert}).}
\label{comp_pert_sommer}
\end{figure}

In Fig.~\ref{comp_pert_sommer} we demonstrate that this change hardly affects the frequentist limits and preferred region 
for the $\kappa=1$ model. This confirms our expectation from Sec.~\ref{sec:pert} that the impact of 
perturbations should typically be small, implying that one generally can directly adopt the results shown in  
the previous sections (in particular Figs.~\ref{fig:CMB_limits} and \ref{fig:constraints_LSS_vs_CMB}). 
We stress, however, that this remains a model-dependent statement,
which in principle has to be checked on a case-by-case basis (as we have done here).

Let us finally briefly discuss the case that the dark sector (i.e.\ DM and DR) is 
colder than the visible sector during thermal freeze-out, $\xi \equiv T_\text{dark} / T_\text{vis} < 1$. Such a situation 
occurs naturally if the two sectors only interact with each other at very high temperatures but then evolve 
independently. In fact, it is probably necessary to have $\xi < 1$ in order to avoid an unacceptably large contribution 
to $N_\text{eff}$ from DR (see, e.g., Ref.~\cite{Bringmann:2016ilk}). A non-trivial temperature ratio has three effects: 
it reduces the value of $\alpha$ required to reproduce the observed DM relic abundance, it reduces the 
velocity of DM particles for a given temperature of the visible sector, and it leads to earlier kinetic decoupling (see below). 
In combination, these three effects result in a larger Sommerfeld factor at early times but smaller saturation value, implying 
in particular that the saturation happens earlier~\cite{Binder:2017lkj}. A quantitative discussion of the resulting changes 
requires more specific assumptions and is thus beyond the scope of this work.

\subsection{Impact on small scales}
\label{sec:sommer_other}

The model that we have studied in this section has a number of further interesting properties, which allow
us to extend the discussion of Fig.~\ref{fig:Sresult} to additional cosmological and astrophysical observables.
First of all, an interaction as given in Eq.~(\ref{eq:lagrangian}) inevitably mediates a strong DM self-interaction
for the light mediators that we consider here. In the resonant regime, the 
self-interaction cross section can again be calculated by approximating the Yukawa potential by a Hulth\'{e}n 
potential. Close to a resonance (i.e.\ for $\delta \ll 1$), the phase shift from the scattering process is very close to 
$\pi/2$ and one therefore obtains the simple expression
\begin{equation}
\sigma_\mathrm{T} =  \frac{16 \pi}{m_\chi^2 \, v_\text{rel}^2} \; .
\label{eq:sigmat}
\end{equation}
For $m_\chi v / m_\text{med} \gtrsim 1$, corresponding to $n^2 v / \alpha \gtrsim 1$, the Hulth\'{e}n 
approximation becomes inaccurate and a better solution is obtained by fitting to numerical solutions of the 
Schroedinger equation. We adopt the parametrisation for this \emph{classical regime} from 
Ref.~\cite{Cyr-Racine:2015ihg}, noting that in this case the momentum transfer cross section scales 
approximately as $\sigma_\mathrm{T} \propto (n / m_\chi)^{3\text{--}4}$.

In all panels of Fig.~\ref{fig:Sresult}
we indicate the parameter regions $\langle \sigma_\mathrm{T} \rangle / m_\chi > 10\, \mathrm{cm^2 / g}$, where 
$\langle \sigma_\mathrm{T} \rangle$ denotes the velocity-averaged momentum transfer cross section for a typical 
relative velocity of $30\,\mathrm{km/s}$. This parameter region is robustly excluded by bounds from dwarf spheroidal 
galaxies and low-surface-brightness galaxies~\cite{Tulin:2017ara,Valli:2017ktb,Bondarenko:2017rfu} and
simply corresponds to an upper bound on the DM mass that only depends on $n$.
It is worth noting that DM self-interactions with
somewhat smaller cross sections, and correspondingly larger DM masses, have been independently 
invoked~\cite{Loeb:2010gj,Vogelsberger:2012ku,Peter:2012jh,Zavala:2012us,Aarssen:2012fx,Elbert:2014bma,Kamada:2016euw,Robertson:2017mgj} 
to mitigate a number of long-standing small-scale problems of structure formation, namely 
the cusp-versus-core~\cite{deBlok:1997zlw,Oh:2010ea,Walker:2011zu} and the too-big-to-fail 
problem~\cite{BoylanKolchin:2011de,Papastergis:2014aba} 
(as well as, more recently, the diversity problem~\cite{Oman:2015xda,Oman:2016zjn,Robertson:2017mgj}). 
As can be seen in the figure, such self-interaction 
rates can relatively easily be accommodated in our model for parameter values that also are favoured by 
the CMB+LSS data -- in particular for large values of $n$.

As already mentioned, these strong constraints from DM self-interactions imply rather small kinetic decoupling 
temperatures when compared to standard WIMP candidates. {Such a late kinetic decoupling introduces a small-scale cut-off in the matter power spectrum similar to warm DM~{\cite{Vogelsberger:2015gpr}}. Thus,} $T_\mathrm{kd}$ cannot
be too small without being in conflict with Lyman-$\alpha$ forest observations. In the right column 
of Fig.~\ref{fig:Sresult} we therefore also show a rough estimate of this bound, 
$T_\text{kd} \gtrsim 0.1\,\mathrm{keV}$ \cite{Bringmann:2016ilk,Huo:2017vef}. Kinetic decoupling 
temperatures close to this bound may {lead to the suppression of small-scale structure and thereby} alleviate yet another long-standing small-scale issue of 
$\Lambda$CDM cosmology, namely the missing satellites 
problem \cite{Moore:1999nt,Klypin:1999uc,Fattahi:2016nld} {(see however~\cite{Jethwa:2016gra,Huo:2017vef,Kim:2017iwr} for recent discussions of this issue)}. Fig.~\ref{fig:Sresult} thus
suggests that this is possible in the same parameter region that is favoured by large-scale data
and DM self-interactions at dwarf galaxy scales -- (almost) independent of which resonance, $n$, is 
considered.

At this point, however, we should recall that $T_\mathrm{kd}$ is not really a free parameter but is in principle
uniquely determined by the DM particle model. The simplest possibility would be to couple the mediator $V_\mu$
not only to DM but also to DR, with a coupling $g_\phi=\eta g_\chi$. This results in \cite{Bringmann:2016ilk}
\begin{align}
 T_\mathrm{kd}^\mathrm{simp} & \sim 0.3\,\mathrm{keV} \times \eta^{-1/2}\xi^{-3/2}\left(\frac{m_\chi}{\mathrm{TeV}} \right)^{-1/4}
 \left(\frac{m_\mathrm{med}}{\mathrm{MeV}} \right) \nonumber \\
 & \sim 0.7\,\mathrm{keV} \left(\frac{n}{100}\right)^2 \times \eta^{-1/2}\xi^{-3/2}\left(\frac{m_\chi}{\mathrm{TeV}} \right)^{7/4} \; ,
\end{align}
where $\xi$ denotes the temperature ratio of dark to visible sector. 
This clearly shows that it is in practice
difficult to achieve late kinetic decoupling for mediator masses above the MeV scale. Combining this insight
with the self-interaction bounds shown in the right column of Fig.~\ref{fig:Sresult}, we conclude that this
affects resonant annihilation for small $n$.

On the other hand, we make the interesting observation that for TeV-scale DM and 'high' resonances, with 
$n\gtrsim50$, it is in fact rather straightforward to {\it simultaneously} alleviate the missing satellites and
other small-scale problems, and at the same time reduce the $H_0$ and $\sigma_8$ tensions. Given that 
we adopted a rather minimal model set-up, this is an intriguing result. The fact that (various combinations of) 
these tensions between
observations and the cosmological concordance model can be simultaneously addressed  for similar 
models has been pointed out 
before~\cite{Loeb:2010gj,Vogelsberger:2012ku,Aarssen:2012fx,Dasgupta:2013zpn,Bringmann:2013vra,Huo:2017vef,Binder:2017lkj}; 
here we confirmed those claims, adding  the first full combined analysis of CMB and LSS data in this context.


\section{Conclusions}
\label{sec:discussion}

The cosmological concordance model rests on the somewhat bold assumption that the 
comoving DM density remains exactly constant while the Universe expands in volume by more 
than 20 orders of magnitude. In this article we have quantified how strongly deviations from this 
scenario are constrained observationally. In order to do so in as model-independent 
and conservative a way as possible, we have assumed a range of phenomenological 
transition scenarios (see Fig.~\ref{rhodmfig}) where DM is converted into a non-interacting 
form of radiation.

We find that all scenarios where the DM density is reduced by more than a few percent 
after matter-radiation equality are in strong tension with CMB observations 
(see Fig.~\ref{fig:CMB_limits}). For earlier transitions, on the other hand, 
a much larger fraction of DM can be converted; this is expected given that the relative contribution 
of DM to the total energy budget is correspondingly smaller. Adding low-redshift 
observables to the analysis relaxes the late-time constraints, 
cf.~Fig.~\ref{fig:constraints_LSS_vs_CMB}, allowing up to around 
10 percent of DM to be converted during matter domination. 

The reason for the weakening of the CMB constraints is that a late conversion from
DM to DR reduces the well-known tension between these essentially incompatible
datasets. We discussed in detail in what sense this implies positive evidence for
such a transition scenario, concluding that, from a frequentist perspective, the
preference is rather mild. We stressed, however, that a Bayesian analysis would
come to a very different conclusion for a prior choice that puts special emphasis
on late-time conversions (see Appendix~\ref{app:Bayesian}).

We argued that our parametrisation of possible transition scenarios from DM
to DR is very general and encompasses those previously discussed in the literature,
in particular the case of decaying DM (see again Fig.~\ref{rhodmfig}). Another 
interesting application would be primordial black hole DM, where merger events
inevitably transform part of the black hole mass to DR in the form of gravitational 
waves. In the last part of this work, Section \ref{sec:sommer}, we have discussed in detail
yet another scenario that can be mapped to our general parametrisation, namely 
DM coupled to DR via light mediator particles. For specific values of the mediator mass, 
cf.~Fig.~\ref{fig:Sresult}, this implies a strongly enhanced  DM
annihilation rate at late times that can mitigate the above-mentioned tension between CMB and 
large-scale structure data. Remarkably, as we have also discussed, such a simple 
scenario could simultaneously alleviate the most pressing
$\Lambda$CDM problems at {\it small} scales.

Turning this around, there is a surprising variety of astrophysical and cosmological 
observations that allow to test such a simple particle model even though it is almost 
fully confined to a dark sector, with negligible couplings to the Standard Model. The 
constraints derived in this work
are thus not only of general interest, in the sense that they quantify how well one
of the basic assumptions of the cosmological concordance model is tested observationally, 
but can very concretely help to test and discriminate a variety of (particle) DM models.

\vspace{0.5cm}

\begin{figure*}[t!]
\vspace{0.5cm}
\includegraphics[width=\columnwidth]{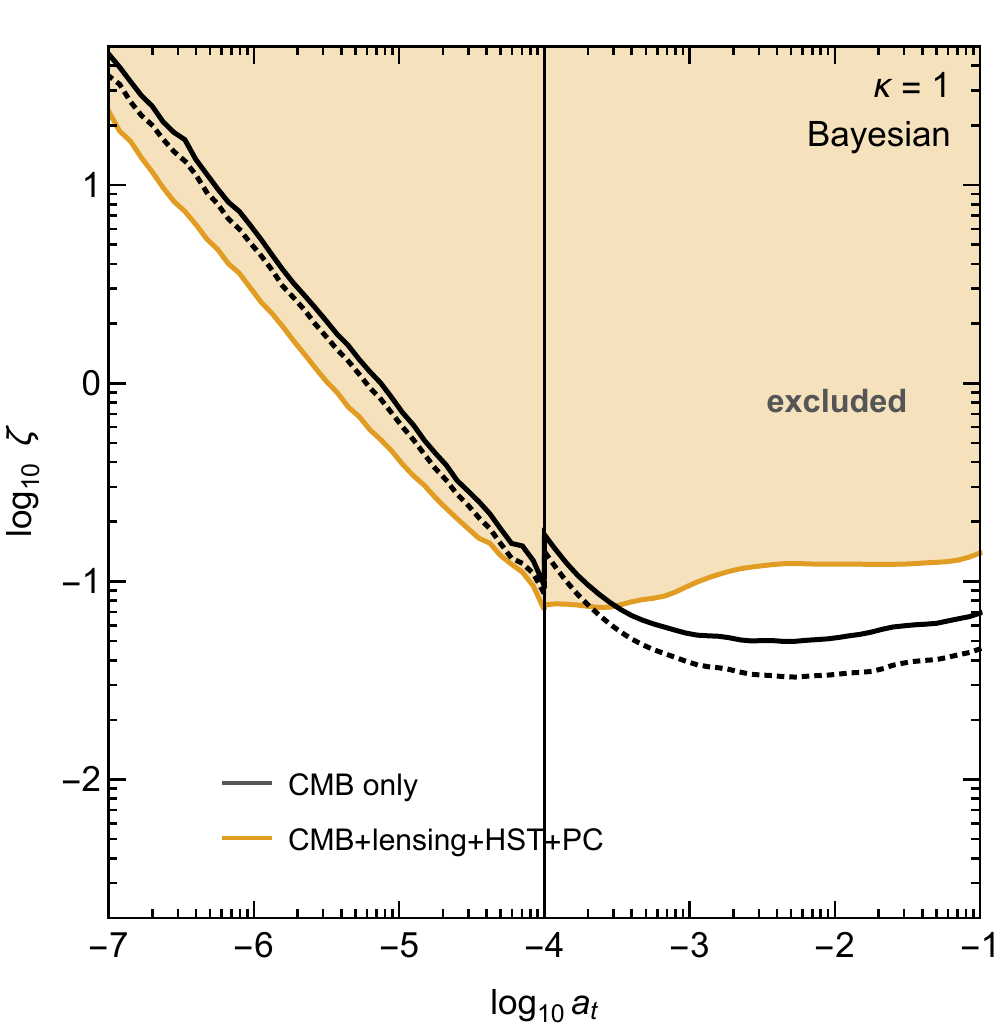}\quad
\includegraphics[width=\columnwidth]{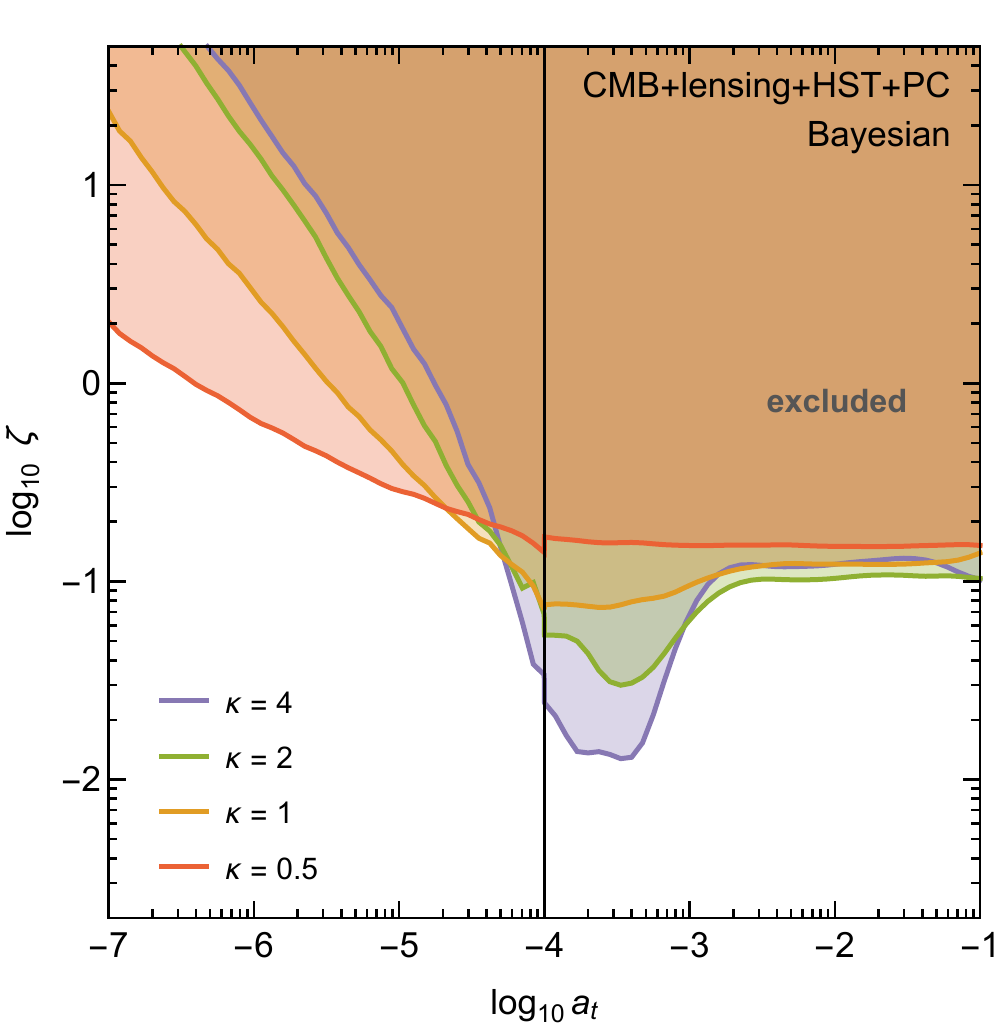}
\caption{Left panel: 95\%\,C.L.~(dotted lines) and 99\%\,C.L.~(solid lines) Bayesian limits for our conversion scenario with $\kappa=1$, resulting from \textbf{CMB+Lensing+HST+PC}. For comparison we also show the constraints obtained from \textbf{CMB} only (identical to the corresponding line in Fig.~\ref{fig:CMB_limits}). Right panel: Bayesian limits on the amount of converted DM from \textbf{CMB+Lensing+HST+PC} for different choices of $\kappa$; the coloured region above each line is excluded.
\label{fig:constraints_LSS_vs_CMB_Bayesian}}
\end{figure*}

\paragraph*{Acknowledgements.---}
We thank Tobias Binder, Thejs Brinckmann, Jens Chluba, Michael Gustafsson, Anders Kvellestad and Julien Lesgourgues 
for very useful discussions. This work is supported by the German Science Foundation (DFG) under the Collaborative 
Research Center (SFB) 676 ``Particles, Strings and the Early Universe'' and the Emmy Noether Grant No.\ KA 4662/1-1 as 
well as the ERC Starting Grant `NewAve' (638528). PW is partially supported by the University of Oslo through the 
Strategic Dark Matter Initiative (SDI).

\newpage
\appendix

\section{Bayesian exclusion limits}
\label{app:Bayesian}

{In this Appendix we complement the discussion in Section \ref{sec:lss} with a Bayesian perspective
on our general conversion scenario. We start by showing}
in Fig.~\ref{fig:constraints_LSS_vs_CMB_Bayesian} the Bayesian exclusion limits corresponding to the approximate frequentist exclusion limits shown in Fig.~\ref{fig:constraints_LSS_vs_CMB}. These limits are obtained using flat priors on $\log a_t$ and $\zeta$ (for $a_t>10^{-4}$) or $\Delta N_{\rm eff}^{\rm today}$ (for $a_t<10^{-4}$). In contrast to the $\sim2\sigma$ preference for our model found in the frequentist approach, a Bayesian model comparison actually favours $\Lambda$CDM, as the parameter region in which the extended model is preferred over $\Lambda$CDM is much smaller than the 
parameter region in which the model is strongly disfavoured. This conclusion nevertheless depends strongly on the priors assumed for our effective description and could be modified in a set-up where favourable values of $a_t$ and 
$\zeta$ occur naturally.

\begin{figure*}[t]
\includegraphics[width=0.91\textwidth,clip,trim=0 1 1 2]{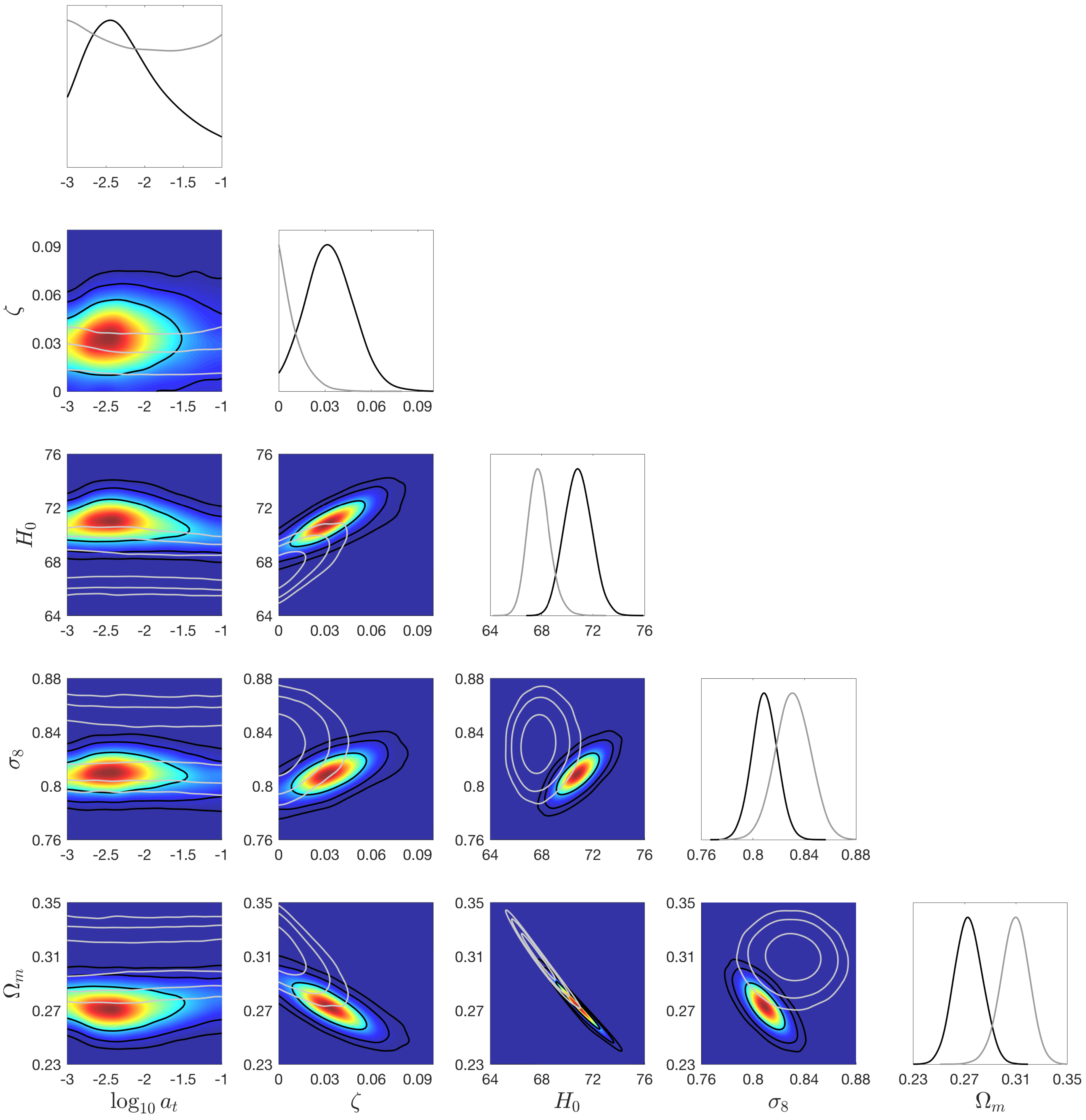}
\caption{Marginalised 2D and 1D posteriors resulting from {\bf CMB+Lensing}+{\bf HST}+{\bf PC} {for our conversion scenario with $\kappa=1$}, with closed 
contours indicating 68\%, 95\% and 99\%\,C.L., respectively. The white contours 
(in the 2D plots) and grey lines (in the 1D plots) are for {\bf CMB} only.}
\label{triangle_LSS_full}
\end{figure*}

In Fig.~\ref{triangle_LSS_full} we provide a supplementary perspective on our discussion so far,
{which also illustrates the point just made. We show}
the marginalised 1D and 2D posteriors for our model parameters as well as the $\Lambda$CDM
parameters most relevant in our context, namely $H_0$, $\sigma_8$ and $\Omega_m$. 
We do so both for {\bf CMB} only (grey lines and white contours) and {\bf CMB} + {\bf Lensing} + {\bf HST}
+ {\bf PC}  (black lines and coloured contours), respectively.
Here, we have
fixed  $\kappa=1$ and chosen a flat prior for $\zeta$; for $\log_{10}a_t$ we have chosen a flat prior
between $-3$ and $-1$, thus zooming in on the most relevant parameter region. For such a prior choice
we see a clear signal preference in the $\zeta$ vs.~$\log_{10} a_t$ plane, once we add LSS data,
which can directly be compared to Fig.~\ref{fig:constraints_LSS_vs_CMB}. 

It is also illuminating to see the correlation of our model
parameters with the cosmological observables considered here. 
For example, it becomes obvious that the degeneracy between matter density and Hubble rate 
is not broken when adding LSS data. This motivates previous statements that $\Omega_\chi h^2$ is 
very well constrained both in $\Lambda$CDM and in our scenario, and confirms our qualitative
discussion of Fig.~\ref{fig:triangle_comp_LSS}. There we argued that a larger Hubble rate at late 
times not only helps to reconcile direct measurements of $H_0$ but automatically, due to this degeneracy,   
the direct measurement of the parameter combination $\sigma_8\Omega_m^{0.3}$ as well. As a result of 
combining essentially incompatible data sets, the parameters that have been marginalised out 
are thus pushed towards values that reduce the tension between CMB and LSS data. 
In our case, as can explicitly be seen in the corresponding 2D posteriors in Fig.~\ref{triangle_LSS_full}, 
this independently results in large values for $a_t$ and $\zeta$.
Let us stress again that these conclusions are prior dependent; allowing $\log_{10} a_t$ to extend to much
smaller values, for example, fully erases the preference for a signal around $a_t\sim 10^{-2.5}$ and 
$\zeta\sim 0.03$.

\newpage


\begin{thebibliography}{95}%
\makeatletter
\providecommand \@ifxundefined [1]{%
 \@ifx{#1\undefined}
}%
\providecommand \@ifnum [1]{%
 \ifnum #1\expandafter \@firstoftwo
 \else \expandafter \@secondoftwo
 \fi
}%
\providecommand \@ifx [1]{%
 \ifx #1\expandafter \@firstoftwo
 \else \expandafter \@secondoftwo
 \fi
}%
\providecommand \natexlab [1]{#1}%
\providecommand \enquote  [1]{``#1''}%
\providecommand \bibnamefont  [1]{#1}%
\providecommand \bibfnamefont [1]{#1}%
\providecommand \citenamefont [1]{#1}%
\providecommand \href@noop [0]{\@secondoftwo}%
\providecommand \href [0]{\begingroup \@sanitize@url \@href}%
\providecommand \@href[1]{\@@startlink{#1}\@@href}%
\providecommand \@@href[1]{\endgroup#1\@@endlink}%
\providecommand \@sanitize@url [0]{\catcode `\\12\catcode `\$12\catcode
  `\&12\catcode `\#12\catcode `\^12\catcode `\_12\catcode `\%12\relax}%
\providecommand \@@startlink[1]{}%
\providecommand \@@endlink[0]{}%
\providecommand \url  [0]{\begingroup\@sanitize@url \@url }%
\providecommand \@url [1]{\endgroup\@href {#1}{\urlprefix }}%
\providecommand \urlprefix  [0]{URL }%
\providecommand \Eprint [0]{\href }%
\providecommand \doibase [0]{http://dx.doi.org/}%
\providecommand \selectlanguage [0]{\@gobble}%
\providecommand \bibinfo  [0]{\@secondoftwo}%
\providecommand \bibfield  [0]{\@secondoftwo}%
\providecommand \translation [1]{[#1]}%
\providecommand \BibitemOpen [0]{}%
\providecommand \bibitemStop [0]{}%
\providecommand \bibitemNoStop [0]{.\EOS\space}%
\providecommand \EOS [0]{\spacefactor3000\relax}%
\providecommand \BibitemShut  [1]{\csname bibitem#1\endcsname}%
\let\auto@bib@innerbib\@empty
\bibitem [{\citenamefont {Ade}\ \emph {et~al.}(2016{\natexlab{a}})\citenamefont
  {Ade} \emph {et~al.}}]{Ade:2015xua}%
  \BibitemOpen
  \bibfield  {author} {\bibinfo {author} {\bibfnamefont {P.~A.~R.}\
  \bibnamefont {Ade}} \emph {et~al.} (\bibinfo {collaboration} {Planck}),\
  }\href {\doibase 10.1051/0004-6361/201525830} {\bibfield  {journal} {\bibinfo
   {journal} {Astron. Astrophys.}\ }\textbf {\bibinfo {volume} {594}},\
  \bibinfo {pages} {A13} (\bibinfo {year} {2016}{\natexlab{a}})},\ \Eprint
  {http://arxiv.org/abs/1502.01589} {arXiv:1502.01589 [astro-ph.CO]}
  \BibitemShut {NoStop}%
\bibitem [{\citenamefont {Lee}\ and\ \citenamefont
  {Weinberg}(1977)}]{Lee:1977ua}%
  \BibitemOpen
  \bibfield  {author} {\bibinfo {author} {\bibfnamefont {B.~W.}\ \bibnamefont
  {Lee}}\ and\ \bibinfo {author} {\bibfnamefont {S.}~\bibnamefont {Weinberg}},\
  }\href {\doibase 10.1103/PhysRevLett.39.165} {\bibfield  {journal} {\bibinfo
  {journal} {Phys. Rev. Lett.}\ }\textbf {\bibinfo {volume} {39}},\ \bibinfo
  {pages} {165} (\bibinfo {year} {1977})}\BibitemShut {NoStop}%
\bibitem [{\citenamefont {Gondolo}\ and\ \citenamefont
  {Gelmini}(1991)}]{Gondolo:1990dk}%
  \BibitemOpen
  \bibfield  {author} {\bibinfo {author} {\bibfnamefont {P.}~\bibnamefont
  {Gondolo}}\ and\ \bibinfo {author} {\bibfnamefont {G.}~\bibnamefont
  {Gelmini}},\ }\href {\doibase 10.1016/0550-3213(91)90438-4} {\bibfield
  {journal} {\bibinfo  {journal} {Nucl. Phys.}\ }\textbf {\bibinfo {volume}
  {B360}},\ \bibinfo {pages} {145} (\bibinfo {year} {1991})}\BibitemShut
  {NoStop}%
\bibitem [{\citenamefont {Cirelli}\ \emph {et~al.}(2012)\citenamefont
  {Cirelli}, \citenamefont {Moulin}, \citenamefont {Panci}, \citenamefont
  {Serpico},\ and\ \citenamefont {Viana}}]{Cirelli:2012ut}%
  \BibitemOpen
  \bibfield  {author} {\bibinfo {author} {\bibfnamefont {M.}~\bibnamefont
  {Cirelli}}, \bibinfo {author} {\bibfnamefont {E.}~\bibnamefont {Moulin}},
  \bibinfo {author} {\bibfnamefont {P.}~\bibnamefont {Panci}}, \bibinfo
  {author} {\bibfnamefont {P.~D.}\ \bibnamefont {Serpico}}, \ and\ \bibinfo
  {author} {\bibfnamefont {A.}~\bibnamefont {Viana}},\ }\href {\doibase
  10.1103/PhysRevD.86.083506, 10.1103/PhysRevD.86.109901} {\bibfield  {journal}
  {\bibinfo  {journal} {Phys. Rev.}\ }\textbf {\bibinfo {volume} {D86}},\
  \bibinfo {pages} {083506} (\bibinfo {year} {2012})},\ \Eprint
  {http://arxiv.org/abs/1205.5283} {arXiv:1205.5283 [astro-ph.CO]} \BibitemShut
  {NoStop}%
\bibitem [{\citenamefont {Ibarra}\ \emph {et~al.}(2013)\citenamefont {Ibarra},
  \citenamefont {Tran},\ and\ \citenamefont {Weniger}}]{Ibarra:2013cra}%
  \BibitemOpen
  \bibfield  {author} {\bibinfo {author} {\bibfnamefont {A.}~\bibnamefont
  {Ibarra}}, \bibinfo {author} {\bibfnamefont {D.}~\bibnamefont {Tran}}, \ and\
  \bibinfo {author} {\bibfnamefont {C.}~\bibnamefont {Weniger}},\ }\href
  {\doibase 10.1142/S0217751X13300408} {\bibfield  {journal} {\bibinfo
  {journal} {Int. J. Mod. Phys.}\ }\textbf {\bibinfo {volume} {A28}},\ \bibinfo
  {pages} {1330040} (\bibinfo {year} {2013})},\ \Eprint
  {http://arxiv.org/abs/1307.6434} {arXiv:1307.6434 [hep-ph]} \BibitemShut
  {NoStop}%
\bibitem [{\citenamefont {Poulin}\ \emph {et~al.}(2017)\citenamefont {Poulin},
  \citenamefont {Lesgourgues},\ and\ \citenamefont {Serpico}}]{Poulin:2016anj}%
  \BibitemOpen
  \bibfield  {author} {\bibinfo {author} {\bibfnamefont {V.}~\bibnamefont
  {Poulin}}, \bibinfo {author} {\bibfnamefont {J.}~\bibnamefont {Lesgourgues}},
  \ and\ \bibinfo {author} {\bibfnamefont {P.~D.}\ \bibnamefont {Serpico}},\
  }\href {\doibase 10.1088/1475-7516/2017/03/043} {\bibfield  {journal}
  {\bibinfo  {journal} {JCAP}\ }\textbf {\bibinfo {volume} {1703}},\ \bibinfo
  {pages} {043} (\bibinfo {year} {2017})},\ \Eprint
  {http://arxiv.org/abs/1610.10051} {arXiv:1610.10051 [astro-ph.CO]}
  \BibitemShut {NoStop}%
\bibitem [{\citenamefont {Zentner}\ and\ \citenamefont
  {Walker}(2002)}]{Zentner:2001zr}%
  \BibitemOpen
  \bibfield  {author} {\bibinfo {author} {\bibfnamefont {A.~R.}\ \bibnamefont
  {Zentner}}\ and\ \bibinfo {author} {\bibfnamefont {T.~P.}\ \bibnamefont
  {Walker}},\ }\href {\doibase 10.1103/PhysRevD.65.063506} {\bibfield
  {journal} {\bibinfo  {journal} {Phys. Rev.}\ }\textbf {\bibinfo {volume}
  {D65}},\ \bibinfo {pages} {063506} (\bibinfo {year} {2002})},\ \Eprint
  {http://arxiv.org/abs/astro-ph/0110533} {arXiv:astro-ph/0110533 [astro-ph]}
  \BibitemShut {NoStop}%
\bibitem [{\citenamefont {Ichiki}\ \emph {et~al.}(2004)\citenamefont {Ichiki},
  \citenamefont {Oguri},\ and\ \citenamefont {Takahashi}}]{Ichiki:2004vi}%
  \BibitemOpen
  \bibfield  {author} {\bibinfo {author} {\bibfnamefont {K.}~\bibnamefont
  {Ichiki}}, \bibinfo {author} {\bibfnamefont {M.}~\bibnamefont {Oguri}}, \
  and\ \bibinfo {author} {\bibfnamefont {K.}~\bibnamefont {Takahashi}},\ }\href
  {\doibase 10.1103/PhysRevLett.93.071302} {\bibfield  {journal} {\bibinfo
  {journal} {Phys. Rev. Lett.}\ }\textbf {\bibinfo {volume} {93}},\ \bibinfo
  {pages} {071302} (\bibinfo {year} {2004})},\ \Eprint
  {http://arxiv.org/abs/astro-ph/0403164} {arXiv:astro-ph/0403164 [astro-ph]}
  \BibitemShut {NoStop}%
\bibitem [{\citenamefont {Lattanzi}\ and\ \citenamefont
  {Valle}(2007)}]{Lattanzi:2007ux}%
  \BibitemOpen
  \bibfield  {author} {\bibinfo {author} {\bibfnamefont {M.}~\bibnamefont
  {Lattanzi}}\ and\ \bibinfo {author} {\bibfnamefont {J.~W.~F.}\ \bibnamefont
  {Valle}},\ }\href {\doibase 10.1103/PhysRevLett.99.121301} {\bibfield
  {journal} {\bibinfo  {journal} {Phys. Rev. Lett.}\ }\textbf {\bibinfo
  {volume} {99}},\ \bibinfo {pages} {121301} (\bibinfo {year} {2007})},\
  \Eprint {http://arxiv.org/abs/0705.2406} {arXiv:0705.2406 [astro-ph]}
  \BibitemShut {NoStop}%
\bibitem [{\citenamefont {Peter}(2010)}]{Peter:2010au}%
  \BibitemOpen
  \bibfield  {author} {\bibinfo {author} {\bibfnamefont {A.~H.~G.}\
  \bibnamefont {Peter}},\ }\href {\doibase 10.1103/PhysRevD.81.083511}
  {\bibfield  {journal} {\bibinfo  {journal} {Phys. Rev.}\ }\textbf {\bibinfo
  {volume} {D81}},\ \bibinfo {pages} {083511} (\bibinfo {year} {2010})},\
  \Eprint {http://arxiv.org/abs/1001.3870} {arXiv:1001.3870 [astro-ph.CO]}
  \BibitemShut {NoStop}%
\bibitem [{\citenamefont {Wang}\ and\ \citenamefont
  {Zentner}(2010)}]{Wang:2010ma}%
  \BibitemOpen
  \bibfield  {author} {\bibinfo {author} {\bibfnamefont {M.-Y.}\ \bibnamefont
  {Wang}}\ and\ \bibinfo {author} {\bibfnamefont {A.~R.}\ \bibnamefont
  {Zentner}},\ }\href {\doibase 10.1103/PhysRevD.82.123507} {\bibfield
  {journal} {\bibinfo  {journal} {Phys. Rev.}\ }\textbf {\bibinfo {volume}
  {D82}},\ \bibinfo {pages} {123507} (\bibinfo {year} {2010})},\ \Eprint
  {http://arxiv.org/abs/1011.2774} {arXiv:1011.2774 [astro-ph.CO]} \BibitemShut
  {NoStop}%
\bibitem [{\citenamefont {Bjaelde}\ \emph {et~al.}(2012)\citenamefont
  {Bjaelde}, \citenamefont {Das},\ and\ \citenamefont {Moss}}]{Bjaelde:2012wi}%
  \BibitemOpen
  \bibfield  {author} {\bibinfo {author} {\bibfnamefont {O.~E.}\ \bibnamefont
  {Bjaelde}}, \bibinfo {author} {\bibfnamefont {S.}~\bibnamefont {Das}}, \ and\
  \bibinfo {author} {\bibfnamefont {A.}~\bibnamefont {Moss}},\ }\href {\doibase
  10.1088/1475-7516/2012/10/017} {\bibfield  {journal} {\bibinfo  {journal}
  {JCAP}\ }\textbf {\bibinfo {volume} {1210}},\ \bibinfo {pages} {017}
  (\bibinfo {year} {2012})},\ \Eprint {http://arxiv.org/abs/1205.0553}
  {arXiv:1205.0553 [astro-ph.CO]} \BibitemShut {NoStop}%
\bibitem [{\citenamefont {Allahverdi}\ \emph {et~al.}(2015)\citenamefont
  {Allahverdi}, \citenamefont {Dutta}, \citenamefont {Queiroz}, \citenamefont
  {Strigari},\ and\ \citenamefont {Wang}}]{Allahverdi:2014bva}%
  \BibitemOpen
  \bibfield  {author} {\bibinfo {author} {\bibfnamefont {R.}~\bibnamefont
  {Allahverdi}}, \bibinfo {author} {\bibfnamefont {B.}~\bibnamefont {Dutta}},
  \bibinfo {author} {\bibfnamefont {F.~S.}\ \bibnamefont {Queiroz}}, \bibinfo
  {author} {\bibfnamefont {L.~E.}\ \bibnamefont {Strigari}}, \ and\ \bibinfo
  {author} {\bibfnamefont {M.-Y.}\ \bibnamefont {Wang}},\ }\href {\doibase
  10.1103/PhysRevD.91.055033} {\bibfield  {journal} {\bibinfo  {journal} {Phys.
  Rev.}\ }\textbf {\bibinfo {volume} {D91}},\ \bibinfo {pages} {055033}
  (\bibinfo {year} {2015})},\ \Eprint {http://arxiv.org/abs/1412.4391}
  {arXiv:1412.4391 [hep-ph]} \BibitemShut {NoStop}%
\bibitem [{\citenamefont {Audren}\ \emph {et~al.}(2014)\citenamefont {Audren},
  \citenamefont {Lesgourgues}, \citenamefont {Mangano}, \citenamefont
  {Serpico},\ and\ \citenamefont {Tram}}]{Audren:2014bca}%
  \BibitemOpen
  \bibfield  {author} {\bibinfo {author} {\bibfnamefont {B.}~\bibnamefont
  {Audren}}, \bibinfo {author} {\bibfnamefont {J.}~\bibnamefont {Lesgourgues}},
  \bibinfo {author} {\bibfnamefont {G.}~\bibnamefont {Mangano}}, \bibinfo
  {author} {\bibfnamefont {P.~D.}\ \bibnamefont {Serpico}}, \ and\ \bibinfo
  {author} {\bibfnamefont {T.}~\bibnamefont {Tram}},\ }\href {\doibase
  10.1088/1475-7516/2014/12/028} {\bibfield  {journal} {\bibinfo  {journal}
  {JCAP}\ }\textbf {\bibinfo {volume} {1412}},\ \bibinfo {pages} {028}
  (\bibinfo {year} {2014})},\ \Eprint {http://arxiv.org/abs/1407.2418}
  {arXiv:1407.2418 [astro-ph.CO]} \BibitemShut {NoStop}%
\bibitem [{\citenamefont {Poulin}\ \emph {et~al.}(2016)\citenamefont {Poulin},
  \citenamefont {Serpico},\ and\ \citenamefont {Lesgourgues}}]{Poulin:2016nat}%
  \BibitemOpen
  \bibfield  {author} {\bibinfo {author} {\bibfnamefont {V.}~\bibnamefont
  {Poulin}}, \bibinfo {author} {\bibfnamefont {P.~D.}\ \bibnamefont {Serpico}},
  \ and\ \bibinfo {author} {\bibfnamefont {J.}~\bibnamefont {Lesgourgues}},\
  }\href {\doibase 10.1088/1475-7516/2016/08/036} {\bibfield  {journal}
  {\bibinfo  {journal} {JCAP}\ }\textbf {\bibinfo {volume} {1608}},\ \bibinfo
  {pages} {036} (\bibinfo {year} {2016})},\ \Eprint
  {http://arxiv.org/abs/1606.02073} {arXiv:1606.02073 [astro-ph.CO]}
  \BibitemShut {NoStop}%
\bibitem [{\citenamefont {Enqvist}\ \emph {et~al.}(2015)\citenamefont
  {Enqvist}, \citenamefont {Nadathur}, \citenamefont {Sekiguchi},\ and\
  \citenamefont {Takahashi}}]{Enqvist:2015ara}%
  \BibitemOpen
  \bibfield  {author} {\bibinfo {author} {\bibfnamefont {K.}~\bibnamefont
  {Enqvist}}, \bibinfo {author} {\bibfnamefont {S.}~\bibnamefont {Nadathur}},
  \bibinfo {author} {\bibfnamefont {T.}~\bibnamefont {Sekiguchi}}, \ and\
  \bibinfo {author} {\bibfnamefont {T.}~\bibnamefont {Takahashi}},\ }\href
  {\doibase 10.1088/1475-7516/2015/09/067} {\bibfield  {journal} {\bibinfo
  {journal} {JCAP}\ }\textbf {\bibinfo {volume} {1509}},\ \bibinfo {pages}
  {067} (\bibinfo {year} {2015})},\ \Eprint {http://arxiv.org/abs/1505.05511}
  {arXiv:1505.05511 [astro-ph.CO]} \BibitemShut {NoStop}%
\bibitem [{\citenamefont {Berezhiani}\ \emph {et~al.}(2015)\citenamefont
  {Berezhiani}, \citenamefont {Dolgov},\ and\ \citenamefont
  {Tkachev}}]{Berezhiani:2015yta}%
  \BibitemOpen
  \bibfield  {author} {\bibinfo {author} {\bibfnamefont {Z.}~\bibnamefont
  {Berezhiani}}, \bibinfo {author} {\bibfnamefont {A.~D.}\ \bibnamefont
  {Dolgov}}, \ and\ \bibinfo {author} {\bibfnamefont {I.~I.}\ \bibnamefont
  {Tkachev}},\ }\href {\doibase 10.1103/PhysRevD.92.061303} {\bibfield
  {journal} {\bibinfo  {journal} {Phys. Rev.}\ }\textbf {\bibinfo {volume}
  {D92}},\ \bibinfo {pages} {061303} (\bibinfo {year} {2015})},\ \Eprint
  {http://arxiv.org/abs/1505.03644} {arXiv:1505.03644 [astro-ph.CO]}
  \BibitemShut {NoStop}%
\bibitem [{\citenamefont {Blackadder}\ and\ \citenamefont
  {Koushiappas}(2016)}]{Blackadder:2015uta}%
  \BibitemOpen
  \bibfield  {author} {\bibinfo {author} {\bibfnamefont {G.}~\bibnamefont
  {Blackadder}}\ and\ \bibinfo {author} {\bibfnamefont {S.~M.}\ \bibnamefont
  {Koushiappas}},\ }\href {\doibase 10.1103/PhysRevD.93.023510} {\bibfield
  {journal} {\bibinfo  {journal} {Phys. Rev.}\ }\textbf {\bibinfo {volume}
  {D93}},\ \bibinfo {pages} {023510} (\bibinfo {year} {2016})},\ \Eprint
  {http://arxiv.org/abs/1510.06026} {arXiv:1510.06026 [astro-ph.CO]}
  \BibitemShut {NoStop}%
\bibitem [{\citenamefont {Pourtsidou}\ and\ \citenamefont
  {Tram}(2016)}]{Pourtsidou:2016ico}%
  \BibitemOpen
  \bibfield  {author} {\bibinfo {author} {\bibfnamefont {A.}~\bibnamefont
  {Pourtsidou}}\ and\ \bibinfo {author} {\bibfnamefont {T.}~\bibnamefont
  {Tram}},\ }\href {\doibase 10.1103/PhysRevD.94.043518} {\bibfield  {journal}
  {\bibinfo  {journal} {Phys. Rev.}\ }\textbf {\bibinfo {volume} {D94}},\
  \bibinfo {pages} {043518} (\bibinfo {year} {2016})},\ \Eprint
  {http://arxiv.org/abs/1604.04222} {arXiv:1604.04222 [astro-ph.CO]}
  \BibitemShut {NoStop}%
\bibitem [{\citenamefont {Chudaykin}\ \emph {et~al.}(2016)\citenamefont
  {Chudaykin}, \citenamefont {Gorbunov},\ and\ \citenamefont
  {Tkachev}}]{Chudaykin:2016yfk}%
  \BibitemOpen
  \bibfield  {author} {\bibinfo {author} {\bibfnamefont {A.}~\bibnamefont
  {Chudaykin}}, \bibinfo {author} {\bibfnamefont {D.}~\bibnamefont {Gorbunov}},
  \ and\ \bibinfo {author} {\bibfnamefont {I.}~\bibnamefont {Tkachev}},\ }\href
  {\doibase 10.1103/PhysRevD.94.023528} {\bibfield  {journal} {\bibinfo
  {journal} {Phys. Rev.}\ }\textbf {\bibinfo {volume} {D94}},\ \bibinfo {pages}
  {023528} (\bibinfo {year} {2016})},\ \Eprint
  {http://arxiv.org/abs/1602.08121} {arXiv:1602.08121 [astro-ph.CO]}
  \BibitemShut {NoStop}%
\bibitem [{\citenamefont {Hamaguchi}\ \emph {et~al.}(2017)\citenamefont
  {Hamaguchi}, \citenamefont {Nakayama},\ and\ \citenamefont
  {Tang}}]{Hamaguchi:2017ihw}%
  \BibitemOpen
  \bibfield  {author} {\bibinfo {author} {\bibfnamefont {K.}~\bibnamefont
  {Hamaguchi}}, \bibinfo {author} {\bibfnamefont {K.}~\bibnamefont {Nakayama}},
  \ and\ \bibinfo {author} {\bibfnamefont {Y.}~\bibnamefont {Tang}},\ }\href
  {\doibase 10.1016/j.physletb.2017.06.071} {\bibfield  {journal} {\bibinfo
  {journal} {Phys. Lett.}\ }\textbf {\bibinfo {volume} {B772}},\ \bibinfo
  {pages} {415} (\bibinfo {year} {2017})},\ \Eprint
  {http://arxiv.org/abs/1705.04521} {arXiv:1705.04521 [hep-ph]} \BibitemShut
  {NoStop}%
\bibitem [{\citenamefont {Dent}\ \emph {et~al.}(2010)\citenamefont {Dent},
  \citenamefont {Dutta},\ and\ \citenamefont {Scherrer}}]{Dent:2009bv}%
  \BibitemOpen
  \bibfield  {author} {\bibinfo {author} {\bibfnamefont {J.~B.}\ \bibnamefont
  {Dent}}, \bibinfo {author} {\bibfnamefont {S.}~\bibnamefont {Dutta}}, \ and\
  \bibinfo {author} {\bibfnamefont {R.~J.}\ \bibnamefont {Scherrer}},\ }\href
  {\doibase 10.1016/j.physletb.2010.03.018} {\bibfield  {journal} {\bibinfo
  {journal} {Phys. Lett.}\ }\textbf {\bibinfo {volume} {B687}},\ \bibinfo
  {pages} {275} (\bibinfo {year} {2010})},\ \Eprint
  {http://arxiv.org/abs/0909.4128} {arXiv:0909.4128 [astro-ph.CO]} \BibitemShut
  {NoStop}%
\bibitem [{\citenamefont {Zavala}\ \emph {et~al.}(2010)\citenamefont {Zavala},
  \citenamefont {Vogelsberger},\ and\ \citenamefont {White}}]{Zavala:2009mi}%
  \BibitemOpen
  \bibfield  {author} {\bibinfo {author} {\bibfnamefont {J.}~\bibnamefont
  {Zavala}}, \bibinfo {author} {\bibfnamefont {M.}~\bibnamefont
  {Vogelsberger}}, \ and\ \bibinfo {author} {\bibfnamefont {S.~D.~M.}\
  \bibnamefont {White}},\ }\href {\doibase 10.1103/PhysRevD.81.083502}
  {\bibfield  {journal} {\bibinfo  {journal} {Phys. Rev.}\ }\textbf {\bibinfo
  {volume} {D81}},\ \bibinfo {pages} {083502} (\bibinfo {year} {2010})},\
  \Eprint {http://arxiv.org/abs/0910.5221} {arXiv:0910.5221 [astro-ph.CO]}
  \BibitemShut {NoStop}%
\bibitem [{\citenamefont {Feng}\ \emph {et~al.}(2010)\citenamefont {Feng},
  \citenamefont {Kaplinghat},\ and\ \citenamefont {Yu}}]{Feng:2010zp}%
  \BibitemOpen
  \bibfield  {author} {\bibinfo {author} {\bibfnamefont {J.~L.}\ \bibnamefont
  {Feng}}, \bibinfo {author} {\bibfnamefont {M.}~\bibnamefont {Kaplinghat}}, \
  and\ \bibinfo {author} {\bibfnamefont {H.-B.}\ \bibnamefont {Yu}},\ }\href
  {\doibase 10.1103/PhysRevD.82.083525} {\bibfield  {journal} {\bibinfo
  {journal} {Phys. Rev.}\ }\textbf {\bibinfo {volume} {D82}},\ \bibinfo {pages}
  {083525} (\bibinfo {year} {2010})},\ \Eprint {http://arxiv.org/abs/1005.4678}
  {arXiv:1005.4678 [hep-ph]} \BibitemShut {NoStop}%
\bibitem [{\citenamefont {van~den Aarssen}\ \emph
  {et~al.}(2012{\natexlab{a}})\citenamefont {van~den Aarssen}, \citenamefont
  {Bringmann},\ and\ \citenamefont {Goedecke}}]{vandenAarssen:2012ag}%
  \BibitemOpen
  \bibfield  {author} {\bibinfo {author} {\bibfnamefont {L.~G.}\ \bibnamefont
  {van~den Aarssen}}, \bibinfo {author} {\bibfnamefont {T.}~\bibnamefont
  {Bringmann}}, \ and\ \bibinfo {author} {\bibfnamefont {Y.~C.}\ \bibnamefont
  {Goedecke}},\ }\href {\doibase 10.1103/PhysRevD.85.123512} {\bibfield
  {journal} {\bibinfo  {journal} {Phys. Rev.}\ }\textbf {\bibinfo {volume}
  {D85}},\ \bibinfo {pages} {123512} (\bibinfo {year} {2012}{\natexlab{a}})},\
  \Eprint {http://arxiv.org/abs/1202.5456} {arXiv:1202.5456 [hep-ph]}
  \BibitemShut {NoStop}%
\bibitem [{\citenamefont {Armendariz-Picon}\ and\ \citenamefont
  {Neelakanta}(2012)}]{ArmendarizPicon:2012mu}%
  \BibitemOpen
  \bibfield  {author} {\bibinfo {author} {\bibfnamefont {C.}~\bibnamefont
  {Armendariz-Picon}}\ and\ \bibinfo {author} {\bibfnamefont {J.~T.}\
  \bibnamefont {Neelakanta}},\ }\href {\doibase 10.1088/1475-7516/2012/12/009}
  {\bibfield  {journal} {\bibinfo  {journal} {JCAP}\ }\textbf {\bibinfo
  {volume} {1212}},\ \bibinfo {pages} {009} (\bibinfo {year} {2012})},\ \Eprint
  {http://arxiv.org/abs/1210.3017} {arXiv:1210.3017 [astro-ph.CO]} \BibitemShut
  {NoStop}%
\bibitem [{\citenamefont {Binder}\ \emph {et~al.}(2018)\citenamefont {Binder},
  \citenamefont {Gustafsson}, \citenamefont {Kamada}, \citenamefont {Sandner},\
  and\ \citenamefont {Wiesner}}]{Binder:2017lkj}%
  \BibitemOpen
  \bibfield  {author} {\bibinfo {author} {\bibfnamefont {T.}~\bibnamefont
  {Binder}}, \bibinfo {author} {\bibfnamefont {M.}~\bibnamefont {Gustafsson}},
  \bibinfo {author} {\bibfnamefont {A.}~\bibnamefont {Kamada}}, \bibinfo
  {author} {\bibfnamefont {S.~M.~R.}\ \bibnamefont {Sandner}}, \ and\ \bibinfo
  {author} {\bibfnamefont {M.}~\bibnamefont {Wiesner}},\ }\href {\doibase
  10.1103/PhysRevD.97.123004} {\bibfield  {journal} {\bibinfo  {journal} {Phys.
  Rev.}\ }\textbf {\bibinfo {volume} {D97}},\ \bibinfo {pages} {123004}
  (\bibinfo {year} {2018})},\ \Eprint {http://arxiv.org/abs/1712.01246}
  {arXiv:1712.01246 [astro-ph.CO]} \BibitemShut {NoStop}%
\bibitem [{\citenamefont {Nakamura}\ \emph {et~al.}(1997)\citenamefont
  {Nakamura}, \citenamefont {Sasaki}, \citenamefont {Tanaka},\ and\
  \citenamefont {Thorne}}]{Nakamura:1997sm}%
  \BibitemOpen
  \bibfield  {author} {\bibinfo {author} {\bibfnamefont {T.}~\bibnamefont
  {Nakamura}}, \bibinfo {author} {\bibfnamefont {M.}~\bibnamefont {Sasaki}},
  \bibinfo {author} {\bibfnamefont {T.}~\bibnamefont {Tanaka}}, \ and\ \bibinfo
  {author} {\bibfnamefont {K.~S.}\ \bibnamefont {Thorne}},\ }\href {\doibase
  10.1086/310886} {\bibfield  {journal} {\bibinfo  {journal} {Astrophys. J.}\
  }\textbf {\bibinfo {volume} {487}},\ \bibinfo {pages} {L139} (\bibinfo {year}
  {1997})},\ \Eprint {http://arxiv.org/abs/astro-ph/9708060}
  {arXiv:astro-ph/9708060 [astro-ph]} \BibitemShut {NoStop}%
\bibitem [{\citenamefont {Raidal}\ \emph {et~al.}(2017)\citenamefont {Raidal},
  \citenamefont {Vaskonen},\ and\ \citenamefont {Veermae}}]{Raidal:2017mfl}%
  \BibitemOpen
  \bibfield  {author} {\bibinfo {author} {\bibfnamefont {M.}~\bibnamefont
  {Raidal}}, \bibinfo {author} {\bibfnamefont {V.}~\bibnamefont {Vaskonen}}, \
  and\ \bibinfo {author} {\bibfnamefont {H.}~\bibnamefont {Veermae}},\ }\href
  {\doibase 10.1088/1475-7516/2017/09/037} {\bibfield  {journal} {\bibinfo
  {journal} {JCAP}\ }\textbf {\bibinfo {volume} {1709}},\ \bibinfo {pages}
  {037} (\bibinfo {year} {2017})},\ \Eprint {http://arxiv.org/abs/1707.01480}
  {arXiv:1707.01480 [astro-ph.CO]} \BibitemShut {NoStop}%
\bibitem [{\citenamefont {Abbott}\ \emph {et~al.}(2016)\citenamefont {Abbott}
  \emph {et~al.}}]{Abbott:2016blz}%
  \BibitemOpen
  \bibfield  {author} {\bibinfo {author} {\bibfnamefont {B.~P.}\ \bibnamefont
  {Abbott}} \emph {et~al.} (\bibinfo {collaboration} {Virgo, LIGO
  Scientific}),\ }\href {\doibase 10.1103/PhysRevLett.116.061102} {\bibfield
  {journal} {\bibinfo  {journal} {Phys. Rev. Lett.}\ }\textbf {\bibinfo
  {volume} {116}},\ \bibinfo {pages} {061102} (\bibinfo {year} {2016})},\
  \Eprint {http://arxiv.org/abs/1602.03837} {arXiv:1602.03837 [gr-qc]}
  \BibitemShut {NoStop}%
\bibitem [{\citenamefont {Camarena}\ and\ \citenamefont
  {Marra}(2016)}]{Torres:2016fmj}%
  \BibitemOpen
  \bibfield  {author} {\bibinfo {author} {\bibfnamefont {D.}~\bibnamefont
  {Camarena}}\ and\ \bibinfo {author} {\bibfnamefont {V.}~\bibnamefont
  {Marra}},\ }\href {\doibase 10.1140/epjc/s10052-016-4517-7} {\bibfield
  {journal} {\bibinfo  {journal} {Eur. Phys. J.}\ }\textbf {\bibinfo {volume}
  {C76}},\ \bibinfo {pages} {644} (\bibinfo {year} {2016})},\ \Eprint
  {http://arxiv.org/abs/1608.08824} {arXiv:1608.08824 [astro-ph.CO]}
  \BibitemShut {NoStop}%
\bibitem [{\citenamefont {Hufnagel}\ \emph {et~al.}(2018)\citenamefont
  {Hufnagel}, \citenamefont {Schmidt-Hoberg},\ and\ \citenamefont
  {Wild}}]{Hufnagel:2017dgo}%
  \BibitemOpen
  \bibfield  {author} {\bibinfo {author} {\bibfnamefont {M.}~\bibnamefont
  {Hufnagel}}, \bibinfo {author} {\bibfnamefont {K.}~\bibnamefont
  {Schmidt-Hoberg}}, \ and\ \bibinfo {author} {\bibfnamefont {S.}~\bibnamefont
  {Wild}},\ }\href {\doibase 10.1088/1475-7516/2018/02/044} {\bibfield
  {journal} {\bibinfo  {journal} {JCAP}\ }\textbf {\bibinfo {volume} {1802}},\
  \bibinfo {pages} {044} (\bibinfo {year} {2018})},\ \Eprint
  {http://arxiv.org/abs/1712.03972} {arXiv:1712.03972 [hep-ph]} \BibitemShut
  {NoStop}%
\bibitem [{\citenamefont {Ma}\ and\ \citenamefont
  {Bertschinger}(1995)}]{Ma:1995ey}%
  \BibitemOpen
  \bibfield  {author} {\bibinfo {author} {\bibfnamefont {C.-P.}\ \bibnamefont
  {Ma}}\ and\ \bibinfo {author} {\bibfnamefont {E.}~\bibnamefont
  {Bertschinger}},\ }\href {\doibase 10.1086/176550} {\bibfield  {journal}
  {\bibinfo  {journal} {Astrophys. J.}\ }\textbf {\bibinfo {volume} {455}},\
  \bibinfo {pages} {7} (\bibinfo {year} {1995})},\ \Eprint
  {http://arxiv.org/abs/astro-ph/9506072} {arXiv:astro-ph/9506072 [astro-ph]}
  \BibitemShut {NoStop}%
\bibitem [{\citenamefont {Lewis}\ \emph {et~al.}(2000)\citenamefont {Lewis},
  \citenamefont {Challinor},\ and\ \citenamefont {Lasenby}}]{Lewis:1999bs}%
  \BibitemOpen
  \bibfield  {author} {\bibinfo {author} {\bibfnamefont {A.}~\bibnamefont
  {Lewis}}, \bibinfo {author} {\bibfnamefont {A.}~\bibnamefont {Challinor}}, \
  and\ \bibinfo {author} {\bibfnamefont {A.}~\bibnamefont {Lasenby}},\ }\href
  {\doibase 10.1086/309179} {\bibfield  {journal} {\bibinfo  {journal}
  {Astrophys. J.}\ }\textbf {\bibinfo {volume} {538}},\ \bibinfo {pages} {473}
  (\bibinfo {year} {2000})},\ \Eprint {http://arxiv.org/abs/astro-ph/9911177}
  {arXiv:astro-ph/9911177 [astro-ph]} \BibitemShut {NoStop}%
\bibitem [{\citenamefont {Howlett}\ \emph {et~al.}(2012)\citenamefont
  {Howlett}, \citenamefont {Lewis}, \citenamefont {Hall},\ and\ \citenamefont
  {Challinor}}]{Howlett:2012mh}%
  \BibitemOpen
  \bibfield  {author} {\bibinfo {author} {\bibfnamefont {C.}~\bibnamefont
  {Howlett}}, \bibinfo {author} {\bibfnamefont {A.}~\bibnamefont {Lewis}},
  \bibinfo {author} {\bibfnamefont {A.}~\bibnamefont {Hall}}, \ and\ \bibinfo
  {author} {\bibfnamefont {A.}~\bibnamefont {Challinor}},\ }\href {\doibase
  10.1088/1475-7516/2012/04/027} {\bibfield  {journal} {\bibinfo  {journal}
  {JCAP}\ }\textbf {\bibinfo {volume} {1204}},\ \bibinfo {pages} {027}
  (\bibinfo {year} {2012})},\ \Eprint {http://arxiv.org/abs/1201.3654}
  {arXiv:1201.3654 [astro-ph.CO]} \BibitemShut {NoStop}%
\bibitem [{\citenamefont {Weinberg}(2004)}]{Weinberg:2003ur}%
  \BibitemOpen
  \bibfield  {author} {\bibinfo {author} {\bibfnamefont {S.}~\bibnamefont
  {Weinberg}},\ }\href {\doibase 10.1103/PhysRevD.69.023503} {\bibfield
  {journal} {\bibinfo  {journal} {Phys. Rev.}\ }\textbf {\bibinfo {volume}
  {D69}},\ \bibinfo {pages} {023503} (\bibinfo {year} {2004})},\ \Eprint
  {http://arxiv.org/abs/astro-ph/0306304} {arXiv:astro-ph/0306304 [astro-ph]}
  \BibitemShut {NoStop}%
\bibitem [{\citenamefont {Lewis}\ and\ \citenamefont
  {Bridle}(2002)}]{Lewis:2002ah}%
  \BibitemOpen
  \bibfield  {author} {\bibinfo {author} {\bibfnamefont {A.}~\bibnamefont
  {Lewis}}\ and\ \bibinfo {author} {\bibfnamefont {S.}~\bibnamefont {Bridle}},\
  }\href {\doibase 10.1103/PhysRevD.66.103511} {\bibfield  {journal} {\bibinfo
  {journal} {Phys. Rev.}\ }\textbf {\bibinfo {volume} {D66}},\ \bibinfo {pages}
  {103511} (\bibinfo {year} {2002})},\ \Eprint
  {http://arxiv.org/abs/astro-ph/0205436} {arXiv:astro-ph/0205436 [astro-ph]}
  \BibitemShut {NoStop}%
\bibitem [{\citenamefont {Lewis}(2013)}]{Lewis:2013hha}%
  \BibitemOpen
  \bibfield  {author} {\bibinfo {author} {\bibfnamefont {A.}~\bibnamefont
  {Lewis}},\ }\href {\doibase 10.1103/PhysRevD.87.103529} {\bibfield  {journal}
  {\bibinfo  {journal} {Phys. Rev.}\ }\textbf {\bibinfo {volume} {D87}},\
  \bibinfo {pages} {103529} (\bibinfo {year} {2013})},\ \Eprint
  {http://arxiv.org/abs/1304.4473} {arXiv:1304.4473 [astro-ph.CO]} \BibitemShut
  {NoStop}%
\bibitem [{\citenamefont {Ade}\ \emph {et~al.}(2014{\natexlab{a}})\citenamefont
  {Ade} \emph {et~al.}}]{Ade:2013zuv}%
  \BibitemOpen
  \bibfield  {author} {\bibinfo {author} {\bibfnamefont {P.~A.~R.}\
  \bibnamefont {Ade}} \emph {et~al.} (\bibinfo {collaboration} {Planck}),\
  }\href {\doibase 10.1051/0004-6361/201321591} {\bibfield  {journal} {\bibinfo
   {journal} {Astron. Astrophys.}\ }\textbf {\bibinfo {volume} {571}},\
  \bibinfo {pages} {A16} (\bibinfo {year} {2014}{\natexlab{a}})},\ \Eprint
  {http://arxiv.org/abs/1303.5076} {arXiv:1303.5076 [astro-ph.CO]} \BibitemShut
  {NoStop}%
\bibitem [{\citenamefont {Aghanim}\ \emph {et~al.}(2016)\citenamefont {Aghanim}
  \emph {et~al.}}]{Aghanim:2015xee}%
  \BibitemOpen
  \bibfield  {author} {\bibinfo {author} {\bibfnamefont {N.}~\bibnamefont
  {Aghanim}} \emph {et~al.} (\bibinfo {collaboration} {Planck}),\ }\href
  {\doibase 10.1051/0004-6361/201526926} {\bibfield  {journal} {\bibinfo
  {journal} {Astron. Astrophys.}\ }\textbf {\bibinfo {volume} {594}},\ \bibinfo
  {pages} {A11} (\bibinfo {year} {2016})},\ \Eprint
  {http://arxiv.org/abs/1507.02704} {arXiv:1507.02704 [astro-ph.CO]}
  \BibitemShut {NoStop}%
\bibitem [{\citenamefont {Ade}\ \emph {et~al.}(2016{\natexlab{b}})\citenamefont
  {Ade} \emph {et~al.}}]{Ade:2015zua}%
  \BibitemOpen
  \bibfield  {author} {\bibinfo {author} {\bibfnamefont {P.~A.~R.}\
  \bibnamefont {Ade}} \emph {et~al.} (\bibinfo {collaboration} {Planck}),\
  }\href {\doibase 10.1051/0004-6361/201525941} {\bibfield  {journal} {\bibinfo
   {journal} {Astron. Astrophys.}\ }\textbf {\bibinfo {volume} {594}},\
  \bibinfo {pages} {A15} (\bibinfo {year} {2016}{\natexlab{b}})},\ \Eprint
  {http://arxiv.org/abs/1502.01591} {arXiv:1502.01591 [astro-ph.CO]}
  \BibitemShut {NoStop}%
\bibitem [{\citenamefont {Neal}(2005)}]{Neal}%
  \BibitemOpen
  \bibfield  {author} {\bibinfo {author} {\bibfnamefont {R.~M.}\ \bibnamefont
  {Neal}},\ }\href@noop {} {\  (\bibinfo {year} {2005})},\ \Eprint
  {http://arxiv.org/abs/math/0502099} {math/0502099 [math.ST]} \BibitemShut
  {NoStop}%
\bibitem [{\citenamefont {Gelman}\ and\ \citenamefont
  {Rubin}(1992)}]{Gelman:1992zz}%
  \BibitemOpen
  \bibfield  {author} {\bibinfo {author} {\bibfnamefont {A.}~\bibnamefont
  {Gelman}}\ and\ \bibinfo {author} {\bibfnamefont {D.~B.}\ \bibnamefont
  {Rubin}},\ }\href {\doibase 10.1214/ss/1177011136} {\bibfield  {journal}
  {\bibinfo  {journal} {Statist. Sci.}\ }\textbf {\bibinfo {volume} {7}},\
  \bibinfo {pages} {457} (\bibinfo {year} {1992})}\BibitemShut {NoStop}%
\bibitem [{\citenamefont {Peebles}(1966)}]{Peebles:1966zz}%
  \BibitemOpen
  \bibfield  {author} {\bibinfo {author} {\bibfnamefont {P.~J.~E.}\
  \bibnamefont {Peebles}},\ }\href {\doibase 10.1086/148918} {\bibfield
  {journal} {\bibinfo  {journal} {Astrophys. J.}\ }\textbf {\bibinfo {volume}
  {146}},\ \bibinfo {pages} {542} (\bibinfo {year} {1966})}\BibitemShut
  {NoStop}%
\bibitem [{\citenamefont {Hou}\ \emph {et~al.}(2013)\citenamefont {Hou},
  \citenamefont {Keisler}, \citenamefont {Knox}, \citenamefont {Millea},\ and\
  \citenamefont {Reichardt}}]{Hou:2011ec}%
  \BibitemOpen
  \bibfield  {author} {\bibinfo {author} {\bibfnamefont {Z.}~\bibnamefont
  {Hou}}, \bibinfo {author} {\bibfnamefont {R.}~\bibnamefont {Keisler}},
  \bibinfo {author} {\bibfnamefont {L.}~\bibnamefont {Knox}}, \bibinfo {author}
  {\bibfnamefont {M.}~\bibnamefont {Millea}}, \ and\ \bibinfo {author}
  {\bibfnamefont {C.}~\bibnamefont {Reichardt}},\ }\href {\doibase
  10.1103/PhysRevD.87.083008} {\bibfield  {journal} {\bibinfo  {journal} {Phys.
  Rev.}\ }\textbf {\bibinfo {volume} {D87}},\ \bibinfo {pages} {083008}
  (\bibinfo {year} {2013})},\ \Eprint {http://arxiv.org/abs/1104.2333}
  {arXiv:1104.2333 [astro-ph.CO]} \BibitemShut {NoStop}%
\bibitem [{\citenamefont {Hamann}(2012)}]{Hamann:2011hu}%
  \BibitemOpen
  \bibfield  {author} {\bibinfo {author} {\bibfnamefont {J.}~\bibnamefont
  {Hamann}},\ }\href {\doibase 10.1088/1475-7516/2012/03/021} {\bibfield
  {journal} {\bibinfo  {journal} {JCAP}\ }\textbf {\bibinfo {volume} {1203}},\
  \bibinfo {pages} {021} (\bibinfo {year} {2012})},\ \Eprint
  {http://arxiv.org/abs/1110.4271} {arXiv:1110.4271 [astro-ph.CO]} \BibitemShut
  {NoStop}%
\bibitem [{\citenamefont {Raveri}(2016)}]{Raveri:2015maa}%
  \BibitemOpen
  \bibfield  {author} {\bibinfo {author} {\bibfnamefont {M.}~\bibnamefont
  {Raveri}},\ }\href {\doibase 10.1103/PhysRevD.93.043522} {\bibfield
  {journal} {\bibinfo  {journal} {Phys. Rev.}\ }\textbf {\bibinfo {volume}
  {D93}},\ \bibinfo {pages} {043522} (\bibinfo {year} {2016})},\ \Eprint
  {http://arxiv.org/abs/1510.00688} {arXiv:1510.00688 [astro-ph.CO]}
  \BibitemShut {NoStop}%
\bibitem [{\citenamefont {Bernal}\ \emph {et~al.}(2016)\citenamefont {Bernal},
  \citenamefont {Verde},\ and\ \citenamefont {Riess}}]{Bernal:2016gxb}%
  \BibitemOpen
  \bibfield  {author} {\bibinfo {author} {\bibfnamefont {J.~L.}\ \bibnamefont
  {Bernal}}, \bibinfo {author} {\bibfnamefont {L.}~\bibnamefont {Verde}}, \
  and\ \bibinfo {author} {\bibfnamefont {A.~G.}\ \bibnamefont {Riess}},\ }\href
  {\doibase 10.1088/1475-7516/2016/10/019} {\bibfield  {journal} {\bibinfo
  {journal} {JCAP}\ }\textbf {\bibinfo {volume} {1610}},\ \bibinfo {pages}
  {019} (\bibinfo {year} {2016})},\ \Eprint {http://arxiv.org/abs/1607.05617}
  {arXiv:1607.05617 [astro-ph.CO]} \BibitemShut {NoStop}%
\bibitem [{\citenamefont {Riess}\ \emph {et~al.}(2016)\citenamefont {Riess}
  \emph {et~al.}}]{Riess:2016jrr}%
  \BibitemOpen
  \bibfield  {author} {\bibinfo {author} {\bibfnamefont {A.~G.}\ \bibnamefont
  {Riess}} \emph {et~al.},\ }\href {\doibase 10.3847/0004-637X/826/1/56}
  {\bibfield  {journal} {\bibinfo  {journal} {Astrophys. J.}\ }\textbf
  {\bibinfo {volume} {826}},\ \bibinfo {pages} {56} (\bibinfo {year} {2016})},\
  \Eprint {http://arxiv.org/abs/1604.01424} {arXiv:1604.01424 [astro-ph.CO]}
  \BibitemShut {NoStop}%
\bibitem [{\citenamefont {Mehta}\ \emph {et~al.}(2012)\citenamefont {Mehta},
  \citenamefont {Cuesta}, \citenamefont {Xu}, \citenamefont {Eisenstein},\ and\
  \citenamefont {Padmanabhan}}]{Mehta:2012hh}%
  \BibitemOpen
  \bibfield  {author} {\bibinfo {author} {\bibfnamefont {K.~T.}\ \bibnamefont
  {Mehta}}, \bibinfo {author} {\bibfnamefont {A.~J.}\ \bibnamefont {Cuesta}},
  \bibinfo {author} {\bibfnamefont {X.}~\bibnamefont {Xu}}, \bibinfo {author}
  {\bibfnamefont {D.~J.}\ \bibnamefont {Eisenstein}}, \ and\ \bibinfo {author}
  {\bibfnamefont {N.}~\bibnamefont {Padmanabhan}},\ }\href {\doibase
  10.1111/j.1365-2966.2012.21112.x} {\bibfield  {journal} {\bibinfo  {journal}
  {Mon. Not. Roy. Astron. Soc.}\ }\textbf {\bibinfo {volume} {427}},\ \bibinfo
  {pages} {2168} (\bibinfo {year} {2012})},\ \Eprint
  {http://arxiv.org/abs/1202.0092} {arXiv:1202.0092 [astro-ph.CO]} \BibitemShut
  {NoStop}%
\bibitem [{\citenamefont {Wyman}\ \emph {et~al.}(2014)\citenamefont {Wyman},
  \citenamefont {Rudd}, \citenamefont {Vanderveld},\ and\ \citenamefont
  {Hu}}]{Wyman:2013lza}%
  \BibitemOpen
  \bibfield  {author} {\bibinfo {author} {\bibfnamefont {M.}~\bibnamefont
  {Wyman}}, \bibinfo {author} {\bibfnamefont {D.~H.}\ \bibnamefont {Rudd}},
  \bibinfo {author} {\bibfnamefont {R.~A.}\ \bibnamefont {Vanderveld}}, \ and\
  \bibinfo {author} {\bibfnamefont {W.}~\bibnamefont {Hu}},\ }\href {\doibase
  10.1103/PhysRevLett.112.051302} {\bibfield  {journal} {\bibinfo  {journal}
  {Phys. Rev. Lett.}\ }\textbf {\bibinfo {volume} {112}},\ \bibinfo {pages}
  {051302} (\bibinfo {year} {2014})},\ \Eprint {http://arxiv.org/abs/1307.7715}
  {arXiv:1307.7715 [astro-ph.CO]} \BibitemShut {NoStop}%
\bibitem [{\citenamefont {Hamann}\ and\ \citenamefont
  {Hasenkamp}(2013)}]{Hamann:2013iba}%
  \BibitemOpen
  \bibfield  {author} {\bibinfo {author} {\bibfnamefont {J.}~\bibnamefont
  {Hamann}}\ and\ \bibinfo {author} {\bibfnamefont {J.}~\bibnamefont
  {Hasenkamp}},\ }\href {\doibase 10.1088/1475-7516/2013/10/044} {\bibfield
  {journal} {\bibinfo  {journal} {JCAP}\ }\textbf {\bibinfo {volume} {1310}},\
  \bibinfo {pages} {044} (\bibinfo {year} {2013})},\ \Eprint
  {http://arxiv.org/abs/1308.3255} {arXiv:1308.3255 [astro-ph.CO]} \BibitemShut
  {NoStop}%
\bibitem [{\citenamefont {Battye}\ and\ \citenamefont
  {Moss}(2014)}]{Battye:2013xqa}%
  \BibitemOpen
  \bibfield  {author} {\bibinfo {author} {\bibfnamefont {R.~A.}\ \bibnamefont
  {Battye}}\ and\ \bibinfo {author} {\bibfnamefont {A.}~\bibnamefont {Moss}},\
  }\href {\doibase 10.1103/PhysRevLett.112.051303} {\bibfield  {journal}
  {\bibinfo  {journal} {Phys. Rev. Lett.}\ }\textbf {\bibinfo {volume} {112}},\
  \bibinfo {pages} {051303} (\bibinfo {year} {2014})},\ \Eprint
  {http://arxiv.org/abs/1308.5870} {arXiv:1308.5870 [astro-ph.CO]} \BibitemShut
  {NoStop}%
\bibitem [{\citenamefont {Costanzi}\ \emph {et~al.}(2014)\citenamefont
  {Costanzi}, \citenamefont {Sartoris}, \citenamefont {Viel},\ and\
  \citenamefont {Borgani}}]{Costanzi:2014tna}%
  \BibitemOpen
  \bibfield  {author} {\bibinfo {author} {\bibfnamefont {M.}~\bibnamefont
  {Costanzi}}, \bibinfo {author} {\bibfnamefont {B.}~\bibnamefont {Sartoris}},
  \bibinfo {author} {\bibfnamefont {M.}~\bibnamefont {Viel}}, \ and\ \bibinfo
  {author} {\bibfnamefont {S.}~\bibnamefont {Borgani}},\ }\href {\doibase
  10.1088/1475-7516/2014/10/081} {\bibfield  {journal} {\bibinfo  {journal}
  {JCAP}\ }\textbf {\bibinfo {volume} {1410}},\ \bibinfo {pages} {081}
  (\bibinfo {year} {2014})},\ \Eprint {http://arxiv.org/abs/1407.8338}
  {arXiv:1407.8338 [astro-ph.CO]} \BibitemShut {NoStop}%
\bibitem [{\citenamefont {Ade}\ \emph {et~al.}(2014{\natexlab{b}})\citenamefont
  {Ade} \emph {et~al.}}]{Ade:2013lmv}%
  \BibitemOpen
  \bibfield  {author} {\bibinfo {author} {\bibfnamefont {P.~A.~R.}\
  \bibnamefont {Ade}} \emph {et~al.} (\bibinfo {collaboration} {Planck}),\
  }\href {\doibase 10.1051/0004-6361/201321521} {\bibfield  {journal} {\bibinfo
   {journal} {Astron. Astrophys.}\ }\textbf {\bibinfo {volume} {571}},\
  \bibinfo {pages} {A20} (\bibinfo {year} {2014}{\natexlab{b}})},\ \Eprint
  {http://arxiv.org/abs/1303.5080} {arXiv:1303.5080 [astro-ph.CO]} \BibitemShut
  {NoStop}%
\bibitem [{\citenamefont {Lesgourgues}\ and\ \citenamefont
  {Pastor}(2006)}]{Lesgourgues:2006nd}%
  \BibitemOpen
  \bibfield  {author} {\bibinfo {author} {\bibfnamefont {J.}~\bibnamefont
  {Lesgourgues}}\ and\ \bibinfo {author} {\bibfnamefont {S.}~\bibnamefont
  {Pastor}},\ }\href {\doibase 10.1016/j.physrep.2006.04.001} {\bibfield
  {journal} {\bibinfo  {journal} {Phys. Rept.}\ }\textbf {\bibinfo {volume}
  {429}},\ \bibinfo {pages} {307} (\bibinfo {year} {2006})},\ \Eprint
  {http://arxiv.org/abs/astro-ph/0603494} {arXiv:astro-ph/0603494 [astro-ph]}
  \BibitemShut {NoStop}%
\bibitem [{\citenamefont {Beutler}\ \emph {et~al.}(2011)\citenamefont
  {Beutler}, \citenamefont {Blake}, \citenamefont {Colless}, \citenamefont
  {Jones}, \citenamefont {Staveley-Smith}, \citenamefont {Campbell},
  \citenamefont {Parker}, \citenamefont {Saunders},\ and\ \citenamefont
  {Watson}}]{Beutler:2011hx}%
  \BibitemOpen
  \bibfield  {author} {\bibinfo {author} {\bibfnamefont {F.}~\bibnamefont
  {Beutler}}, \bibinfo {author} {\bibfnamefont {C.}~\bibnamefont {Blake}},
  \bibinfo {author} {\bibfnamefont {M.}~\bibnamefont {Colless}}, \bibinfo
  {author} {\bibfnamefont {D.~H.}\ \bibnamefont {Jones}}, \bibinfo {author}
  {\bibfnamefont {L.}~\bibnamefont {Staveley-Smith}}, \bibinfo {author}
  {\bibfnamefont {L.}~\bibnamefont {Campbell}}, \bibinfo {author}
  {\bibfnamefont {Q.}~\bibnamefont {Parker}}, \bibinfo {author} {\bibfnamefont
  {W.}~\bibnamefont {Saunders}}, \ and\ \bibinfo {author} {\bibfnamefont
  {F.}~\bibnamefont {Watson}},\ }\href {\doibase
  10.1111/j.1365-2966.2011.19250.x} {\bibfield  {journal} {\bibinfo  {journal}
  {Mon. Not. Roy. Astron. Soc.}\ }\textbf {\bibinfo {volume} {416}},\ \bibinfo
  {pages} {3017} (\bibinfo {year} {2011})},\ \Eprint
  {http://arxiv.org/abs/1106.3366} {arXiv:1106.3366 [astro-ph.CO]} \BibitemShut
  {NoStop}%
\bibitem [{\citenamefont {Ross}\ \emph {et~al.}(2015)\citenamefont {Ross},
  \citenamefont {Samushia}, \citenamefont {Howlett}, \citenamefont {Percival},
  \citenamefont {Burden},\ and\ \citenamefont {Manera}}]{Ross:2014qpa}%
  \BibitemOpen
  \bibfield  {author} {\bibinfo {author} {\bibfnamefont {A.~J.}\ \bibnamefont
  {Ross}}, \bibinfo {author} {\bibfnamefont {L.}~\bibnamefont {Samushia}},
  \bibinfo {author} {\bibfnamefont {C.}~\bibnamefont {Howlett}}, \bibinfo
  {author} {\bibfnamefont {W.~J.}\ \bibnamefont {Percival}}, \bibinfo {author}
  {\bibfnamefont {A.}~\bibnamefont {Burden}}, \ and\ \bibinfo {author}
  {\bibfnamefont {M.}~\bibnamefont {Manera}},\ }\href {\doibase
  10.1093/mnras/stv154} {\bibfield  {journal} {\bibinfo  {journal} {Mon. Not.
  Roy. Astron. Soc.}\ }\textbf {\bibinfo {volume} {449}},\ \bibinfo {pages}
  {835} (\bibinfo {year} {2015})},\ \Eprint {http://arxiv.org/abs/1409.3242}
  {arXiv:1409.3242 [astro-ph.CO]} \BibitemShut {NoStop}%
\bibitem [{\citenamefont {Gil-Marín}\ \emph {et~al.}(2016)\citenamefont
  {Gil-Marín} \emph {et~al.}}]{Gil-Marin:2015nqa}%
  \BibitemOpen
  \bibfield  {author} {\bibinfo {author} {\bibfnamefont {H.}~\bibnamefont
  {Gil-Marín}} \emph {et~al.},\ }\href {\doibase 10.1093/mnras/stw1264}
  {\bibfield  {journal} {\bibinfo  {journal} {Mon. Not. Roy. Astron. Soc.}\
  }\textbf {\bibinfo {volume} {460}},\ \bibinfo {pages} {4210} (\bibinfo {year}
  {2016})},\ \Eprint {http://arxiv.org/abs/1509.06373} {arXiv:1509.06373
  [astro-ph.CO]} \BibitemShut {NoStop}%
\bibitem [{\citenamefont {Iengo}(2009)}]{Iengo:2009ni}%
  \BibitemOpen
  \bibfield  {author} {\bibinfo {author} {\bibfnamefont {R.}~\bibnamefont
  {Iengo}},\ }\href {\doibase 10.1088/1126-6708/2009/05/024} {\bibfield
  {journal} {\bibinfo  {journal} {JHEP}\ }\textbf {\bibinfo {volume} {05}},\
  \bibinfo {pages} {024} (\bibinfo {year} {2009})},\ \Eprint
  {http://arxiv.org/abs/0902.0688} {arXiv:0902.0688 [hep-ph]} \BibitemShut
  {NoStop}%
\bibitem [{\citenamefont {Cassel}(2010)}]{Cassel:2009wt}%
  \BibitemOpen
  \bibfield  {author} {\bibinfo {author} {\bibfnamefont {S.}~\bibnamefont
  {Cassel}},\ }\href {\doibase 10.1088/0954-3899/37/10/105009} {\bibfield
  {journal} {\bibinfo  {journal} {J. Phys.}\ }\textbf {\bibinfo {volume}
  {G37}},\ \bibinfo {pages} {105009} (\bibinfo {year} {2010})},\ \Eprint
  {http://arxiv.org/abs/0903.5307} {arXiv:0903.5307 [hep-ph]} \BibitemShut
  {NoStop}%
\bibitem [{\citenamefont {van~den Aarssen}\ \emph
  {et~al.}(2012{\natexlab{b}})\citenamefont {van~den Aarssen}, \citenamefont
  {Bringmann},\ and\ \citenamefont {Pfrommer}}]{Aarssen:2012fx}%
  \BibitemOpen
  \bibfield  {author} {\bibinfo {author} {\bibfnamefont {L.~G.}\ \bibnamefont
  {van~den Aarssen}}, \bibinfo {author} {\bibfnamefont {T.}~\bibnamefont
  {Bringmann}}, \ and\ \bibinfo {author} {\bibfnamefont {C.}~\bibnamefont
  {Pfrommer}},\ }\href {\doibase 10.1103/PhysRevLett.109.231301} {\bibfield
  {journal} {\bibinfo  {journal} {Phys. Rev. Lett.}\ }\textbf {\bibinfo
  {volume} {109}},\ \bibinfo {pages} {231301} (\bibinfo {year}
  {2012}{\natexlab{b}})},\ \Eprint {http://arxiv.org/abs/1205.5809}
  {arXiv:1205.5809 [astro-ph.CO]} \BibitemShut {NoStop}%
\bibitem [{\citenamefont {Bringmann}\ \emph {et~al.}(2017)\citenamefont
  {Bringmann}, \citenamefont {Kahlhoefer}, \citenamefont {Schmidt-Hoberg},\
  and\ \citenamefont {Walia}}]{Bringmann:2016din}%
  \BibitemOpen
  \bibfield  {author} {\bibinfo {author} {\bibfnamefont {T.}~\bibnamefont
  {Bringmann}}, \bibinfo {author} {\bibfnamefont {F.}~\bibnamefont
  {Kahlhoefer}}, \bibinfo {author} {\bibfnamefont {K.}~\bibnamefont
  {Schmidt-Hoberg}}, \ and\ \bibinfo {author} {\bibfnamefont {P.}~\bibnamefont
  {Walia}},\ }\href {\doibase 10.1103/PhysRevLett.118.141802} {\bibfield
  {journal} {\bibinfo  {journal} {Phys. Rev. Lett.}\ }\textbf {\bibinfo
  {volume} {118}},\ \bibinfo {pages} {141802} (\bibinfo {year} {2017})},\
  \Eprint {http://arxiv.org/abs/1612.00845} {arXiv:1612.00845 [hep-ph]}
  \BibitemShut {NoStop}%
\bibitem [{\citenamefont {Bringmann}\ \emph {et~al.}(2016)\citenamefont
  {Bringmann}, \citenamefont {Ihle}, \citenamefont {Kersten},\ and\
  \citenamefont {Walia}}]{Bringmann:2016ilk}%
  \BibitemOpen
  \bibfield  {author} {\bibinfo {author} {\bibfnamefont {T.}~\bibnamefont
  {Bringmann}}, \bibinfo {author} {\bibfnamefont {H.~T.}\ \bibnamefont {Ihle}},
  \bibinfo {author} {\bibfnamefont {J.}~\bibnamefont {Kersten}}, \ and\
  \bibinfo {author} {\bibfnamefont {P.}~\bibnamefont {Walia}},\ }\href
  {\doibase 10.1103/PhysRevD.94.103529} {\bibfield  {journal} {\bibinfo
  {journal} {Phys. Rev.}\ }\textbf {\bibinfo {volume} {D94}},\ \bibinfo {pages}
  {103529} (\bibinfo {year} {2016})},\ \Eprint
  {http://arxiv.org/abs/1603.04884} {arXiv:1603.04884 [hep-ph]} \BibitemShut
  {NoStop}%
\bibitem [{\citenamefont {Tulin}\ \emph {et~al.}(2013)\citenamefont {Tulin},
  \citenamefont {Yu},\ and\ \citenamefont {Zurek}}]{Tulin:2013teo}%
  \BibitemOpen
  \bibfield  {author} {\bibinfo {author} {\bibfnamefont {S.}~\bibnamefont
  {Tulin}}, \bibinfo {author} {\bibfnamefont {H.-B.}\ \bibnamefont {Yu}}, \
  and\ \bibinfo {author} {\bibfnamefont {K.~M.}\ \bibnamefont {Zurek}},\ }\href
  {\doibase 10.1103/PhysRevD.87.115007} {\bibfield  {journal} {\bibinfo
  {journal} {Phys. Rev.}\ }\textbf {\bibinfo {volume} {D87}},\ \bibinfo {pages}
  {115007} (\bibinfo {year} {2013})},\ \Eprint {http://arxiv.org/abs/1302.3898}
  {arXiv:1302.3898 [hep-ph]} \BibitemShut {NoStop}%
\bibitem [{\citenamefont {Blum}\ \emph {et~al.}(2016)\citenamefont {Blum},
  \citenamefont {Sato},\ and\ \citenamefont {Slatyer}}]{Blum:2016nrz}%
  \BibitemOpen
  \bibfield  {author} {\bibinfo {author} {\bibfnamefont {K.}~\bibnamefont
  {Blum}}, \bibinfo {author} {\bibfnamefont {R.}~\bibnamefont {Sato}}, \ and\
  \bibinfo {author} {\bibfnamefont {T.~R.}\ \bibnamefont {Slatyer}},\ }\href
  {\doibase 10.1088/1475-7516/2016/06/021} {\bibfield  {journal} {\bibinfo
  {journal} {JCAP}\ }\textbf {\bibinfo {volume} {1606}},\ \bibinfo {pages}
  {021} (\bibinfo {year} {2016})},\ \Eprint {http://arxiv.org/abs/1603.01383}
  {arXiv:1603.01383 [hep-ph]} \BibitemShut {NoStop}%
\bibitem [{\citenamefont {Kahlhoefer}\ \emph {et~al.}(2017)\citenamefont
  {Kahlhoefer}, \citenamefont {Schmidt-Hoberg},\ and\ \citenamefont
  {Wild}}]{Kahlhoefer:2017umn}%
  \BibitemOpen
  \bibfield  {author} {\bibinfo {author} {\bibfnamefont {F.}~\bibnamefont
  {Kahlhoefer}}, \bibinfo {author} {\bibfnamefont {K.}~\bibnamefont
  {Schmidt-Hoberg}}, \ and\ \bibinfo {author} {\bibfnamefont {S.}~\bibnamefont
  {Wild}},\ }\href {\doibase 10.1088/1475-7516/2017/08/003} {\bibfield
  {journal} {\bibinfo  {journal} {JCAP}\ }\textbf {\bibinfo {volume} {1708}},\
  \bibinfo {pages} {003} (\bibinfo {year} {2017})},\ \Eprint
  {http://arxiv.org/abs/1704.02149} {arXiv:1704.02149 [hep-ph]} \BibitemShut
  {NoStop}%
\bibitem [{\citenamefont {Steigman}\ \emph {et~al.}(2012)\citenamefont
  {Steigman}, \citenamefont {Dasgupta},\ and\ \citenamefont
  {Beacom}}]{Steigman:2012nb}%
  \BibitemOpen
  \bibfield  {author} {\bibinfo {author} {\bibfnamefont {G.}~\bibnamefont
  {Steigman}}, \bibinfo {author} {\bibfnamefont {B.}~\bibnamefont {Dasgupta}},
  \ and\ \bibinfo {author} {\bibfnamefont {J.~F.}\ \bibnamefont {Beacom}},\
  }\href {\doibase 10.1103/PhysRevD.86.023506} {\bibfield  {journal} {\bibinfo
  {journal} {Phys. Rev.}\ }\textbf {\bibinfo {volume} {D86}},\ \bibinfo {pages}
  {023506} (\bibinfo {year} {2012})},\ \Eprint {http://arxiv.org/abs/1204.3622}
  {arXiv:1204.3622 [hep-ph]} \BibitemShut {NoStop}%
\bibitem [{\citenamefont {Cyr-Racine}\ \emph {et~al.}(2016)\citenamefont
  {Cyr-Racine}, \citenamefont {Sigurdson}, \citenamefont {Zavala},
  \citenamefont {Bringmann}, \citenamefont {Vogelsberger},\ and\ \citenamefont
  {Pfrommer}}]{Cyr-Racine:2015ihg}%
  \BibitemOpen
  \bibfield  {author} {\bibinfo {author} {\bibfnamefont {F.-Y.}\ \bibnamefont
  {Cyr-Racine}}, \bibinfo {author} {\bibfnamefont {K.}~\bibnamefont
  {Sigurdson}}, \bibinfo {author} {\bibfnamefont {J.}~\bibnamefont {Zavala}},
  \bibinfo {author} {\bibfnamefont {T.}~\bibnamefont {Bringmann}}, \bibinfo
  {author} {\bibfnamefont {M.}~\bibnamefont {Vogelsberger}}, \ and\ \bibinfo
  {author} {\bibfnamefont {C.}~\bibnamefont {Pfrommer}},\ }\href {\doibase
  10.1103/PhysRevD.93.123527} {\bibfield  {journal} {\bibinfo  {journal} {Phys.
  Rev.}\ }\textbf {\bibinfo {volume} {D93}},\ \bibinfo {pages} {123527}
  (\bibinfo {year} {2016})},\ \Eprint {http://arxiv.org/abs/1512.05344}
  {arXiv:1512.05344 [astro-ph.CO]} \BibitemShut {NoStop}%
\bibitem [{\citenamefont {Tulin}\ and\ \citenamefont
  {Yu}(2018)}]{Tulin:2017ara}%
  \BibitemOpen
  \bibfield  {author} {\bibinfo {author} {\bibfnamefont {S.}~\bibnamefont
  {Tulin}}\ and\ \bibinfo {author} {\bibfnamefont {H.-B.}\ \bibnamefont {Yu}},\
  }\href {\doibase 10.1016/j.physrep.2017.11.004} {\bibfield  {journal}
  {\bibinfo  {journal} {Phys. Rept.}\ }\textbf {\bibinfo {volume} {730}},\
  \bibinfo {pages} {1} (\bibinfo {year} {2018})},\ \Eprint
  {http://arxiv.org/abs/1705.02358} {arXiv:1705.02358 [hep-ph]} \BibitemShut
  {NoStop}%
\bibitem [{\citenamefont {Valli}\ and\ \citenamefont
  {Yu}(2017)}]{Valli:2017ktb}%
  \BibitemOpen
  \bibfield  {author} {\bibinfo {author} {\bibfnamefont {M.}~\bibnamefont
  {Valli}}\ and\ \bibinfo {author} {\bibfnamefont {H.-B.}\ \bibnamefont {Yu}},\
  }\href@noop {} {\  (\bibinfo {year} {2017})},\ \Eprint
  {http://arxiv.org/abs/1711.03502} {arXiv:1711.03502 [astro-ph.GA]}
  \BibitemShut {NoStop}%
\bibitem [{\citenamefont {Bondarenko}\ \emph {et~al.}(2018)\citenamefont
  {Bondarenko}, \citenamefont {Boyarsky}, \citenamefont {Bringmann},\ and\
  \citenamefont {Sokolenko}}]{Bondarenko:2017rfu}%
  \BibitemOpen
  \bibfield  {author} {\bibinfo {author} {\bibfnamefont {K.}~\bibnamefont
  {Bondarenko}}, \bibinfo {author} {\bibfnamefont {A.}~\bibnamefont
  {Boyarsky}}, \bibinfo {author} {\bibfnamefont {T.}~\bibnamefont {Bringmann}},
  \ and\ \bibinfo {author} {\bibfnamefont {A.}~\bibnamefont {Sokolenko}},\
  }\href {\doibase 10.1088/1475-7516/2018/04/049} {\bibfield  {journal}
  {\bibinfo  {journal} {JCAP}\ }\textbf {\bibinfo {volume} {1804}},\ \bibinfo
  {pages} {049} (\bibinfo {year} {2018})},\ \Eprint
  {http://arxiv.org/abs/1712.06602} {arXiv:1712.06602 [astro-ph.CO]}
  \BibitemShut {NoStop}%
\bibitem [{\citenamefont {Loeb}\ and\ \citenamefont
  {Weiner}(2011)}]{Loeb:2010gj}%
  \BibitemOpen
  \bibfield  {author} {\bibinfo {author} {\bibfnamefont {A.}~\bibnamefont
  {Loeb}}\ and\ \bibinfo {author} {\bibfnamefont {N.}~\bibnamefont {Weiner}},\
  }\href {\doibase 10.1103/PhysRevLett.106.171302} {\bibfield  {journal}
  {\bibinfo  {journal} {Phys. Rev. Lett.}\ }\textbf {\bibinfo {volume} {106}},\
  \bibinfo {pages} {171302} (\bibinfo {year} {2011})},\ \Eprint
  {http://arxiv.org/abs/1011.6374} {arXiv:1011.6374 [astro-ph.CO]} \BibitemShut
  {NoStop}%
\bibitem [{\citenamefont {Vogelsberger}\ \emph {et~al.}(2012)\citenamefont
  {Vogelsberger}, \citenamefont {Zavala},\ and\ \citenamefont
  {Loeb}}]{Vogelsberger:2012ku}%
  \BibitemOpen
  \bibfield  {author} {\bibinfo {author} {\bibfnamefont {M.}~\bibnamefont
  {Vogelsberger}}, \bibinfo {author} {\bibfnamefont {J.}~\bibnamefont
  {Zavala}}, \ and\ \bibinfo {author} {\bibfnamefont {A.}~\bibnamefont
  {Loeb}},\ }\href {\doibase 10.1111/j.1365-2966.2012.21182.x} {\bibfield
  {journal} {\bibinfo  {journal} {Mon. Not. Roy. Astron. Soc.}\ }\textbf
  {\bibinfo {volume} {423}},\ \bibinfo {pages} {3740} (\bibinfo {year}
  {2012})},\ \Eprint {http://arxiv.org/abs/1201.5892} {arXiv:1201.5892
  [astro-ph.CO]} \BibitemShut {NoStop}%
\bibitem [{\citenamefont {Peter}\ \emph {et~al.}(2013)\citenamefont {Peter},
  \citenamefont {Rocha}, \citenamefont {Bullock},\ and\ \citenamefont
  {Kaplinghat}}]{Peter:2012jh}%
  \BibitemOpen
  \bibfield  {author} {\bibinfo {author} {\bibfnamefont {A.~H.~G.}\
  \bibnamefont {Peter}}, \bibinfo {author} {\bibfnamefont {M.}~\bibnamefont
  {Rocha}}, \bibinfo {author} {\bibfnamefont {J.~S.}\ \bibnamefont {Bullock}},
  \ and\ \bibinfo {author} {\bibfnamefont {M.}~\bibnamefont {Kaplinghat}},\
  }\href {\doibase 10.1093/mnras/sts535} {\bibfield  {journal} {\bibinfo
  {journal} {Mon. Not. Roy. Astron. Soc.}\ }\textbf {\bibinfo {volume} {430}},\
  \bibinfo {pages} {105} (\bibinfo {year} {2013})},\ \Eprint
  {http://arxiv.org/abs/1208.3026} {arXiv:1208.3026 [astro-ph.CO]} \BibitemShut
  {NoStop}%
\bibitem [{\citenamefont {Zavala}\ \emph {et~al.}(2013)\citenamefont {Zavala},
  \citenamefont {Vogelsberger},\ and\ \citenamefont {Walker}}]{Zavala:2012us}%
  \BibitemOpen
  \bibfield  {author} {\bibinfo {author} {\bibfnamefont {J.}~\bibnamefont
  {Zavala}}, \bibinfo {author} {\bibfnamefont {M.}~\bibnamefont
  {Vogelsberger}}, \ and\ \bibinfo {author} {\bibfnamefont {M.~G.}\
  \bibnamefont {Walker}},\ }\href {\doibase 10.1093/mnrasl/sls053} {\bibfield
  {journal} {\bibinfo  {journal} {Monthly Notices of the Royal Astronomical
  Society: Letters}\ }\textbf {\bibinfo {volume} {431}},\ \bibinfo {pages}
  {L20} (\bibinfo {year} {2013})},\ \Eprint {http://arxiv.org/abs/1211.6426}
  {arXiv:1211.6426 [astro-ph.CO]} \BibitemShut {NoStop}%
\bibitem [{\citenamefont {Elbert}\ \emph {et~al.}(2015)\citenamefont {Elbert},
  \citenamefont {Bullock}, \citenamefont {Garrison-Kimmel}, \citenamefont
  {Rocha}, \citenamefont {Oñorbe},\ and\ \citenamefont
  {Peter}}]{Elbert:2014bma}%
  \BibitemOpen
  \bibfield  {author} {\bibinfo {author} {\bibfnamefont {O.~D.}\ \bibnamefont
  {Elbert}}, \bibinfo {author} {\bibfnamefont {J.~S.}\ \bibnamefont {Bullock}},
  \bibinfo {author} {\bibfnamefont {S.}~\bibnamefont {Garrison-Kimmel}},
  \bibinfo {author} {\bibfnamefont {M.}~\bibnamefont {Rocha}}, \bibinfo
  {author} {\bibfnamefont {J.}~\bibnamefont {Oñorbe}}, \ and\ \bibinfo
  {author} {\bibfnamefont {A.~H.~G.}\ \bibnamefont {Peter}},\ }\href {\doibase
  10.1093/mnras/stv1470} {\bibfield  {journal} {\bibinfo  {journal} {Mon. Not.
  Roy. Astron. Soc.}\ }\textbf {\bibinfo {volume} {453}},\ \bibinfo {pages}
  {29} (\bibinfo {year} {2015})},\ \Eprint {http://arxiv.org/abs/1412.1477}
  {arXiv:1412.1477 [astro-ph.GA]} \BibitemShut {NoStop}%
\bibitem [{\citenamefont {Kamada}\ \emph {et~al.}(2017)\citenamefont {Kamada},
  \citenamefont {Kaplinghat}, \citenamefont {Pace},\ and\ \citenamefont
  {Yu}}]{Kamada:2016euw}%
  \BibitemOpen
  \bibfield  {author} {\bibinfo {author} {\bibfnamefont {A.}~\bibnamefont
  {Kamada}}, \bibinfo {author} {\bibfnamefont {M.}~\bibnamefont {Kaplinghat}},
  \bibinfo {author} {\bibfnamefont {A.~B.}\ \bibnamefont {Pace}}, \ and\
  \bibinfo {author} {\bibfnamefont {H.-B.}\ \bibnamefont {Yu}},\ }\href
  {\doibase 10.1103/PhysRevLett.119.111102} {\bibfield  {journal} {\bibinfo
  {journal} {Phys. Rev. Lett.}\ }\textbf {\bibinfo {volume} {119}},\ \bibinfo
  {pages} {111102} (\bibinfo {year} {2017})},\ \Eprint
  {http://arxiv.org/abs/1611.02716} {arXiv:1611.02716 [astro-ph.GA]}
  \BibitemShut {NoStop}%
\bibitem [{\citenamefont {Robertson}\ \emph {et~al.}(2017)\citenamefont
  {Robertson}, \citenamefont {Massey}, \citenamefont {Eke}, \citenamefont
  {Tulin}, \citenamefont {Yu}, \citenamefont {Bahé}, \citenamefont {Barnes},
  \citenamefont {Vecchia},\ and\ \citenamefont {Kay}}]{Robertson:2017mgj}%
  \BibitemOpen
  \bibfield  {author} {\bibinfo {author} {\bibfnamefont {A.}~\bibnamefont
  {Robertson}}, \bibinfo {author} {\bibfnamefont {R.}~\bibnamefont {Massey}},
  \bibinfo {author} {\bibfnamefont {V.}~\bibnamefont {Eke}}, \bibinfo {author}
  {\bibfnamefont {S.}~\bibnamefont {Tulin}}, \bibinfo {author} {\bibfnamefont
  {H.-B.}\ \bibnamefont {Yu}}, \bibinfo {author} {\bibfnamefont
  {Y.}~\bibnamefont {Bahé}}, \bibinfo {author} {\bibfnamefont {D.~J.}\
  \bibnamefont {Barnes}}, \bibinfo {author} {\bibfnamefont {C.~D.}\
  \bibnamefont {Vecchia}}, \ and\ \bibinfo {author} {\bibfnamefont {S.~T.}\
  \bibnamefont {Kay}},\ }\href@noop {} {\  (\bibinfo {year} {2017})},\ \Eprint
  {http://arxiv.org/abs/1711.09096} {arXiv:1711.09096 [astro-ph.CO]}
  \BibitemShut {NoStop}%
\bibitem [{\citenamefont {de~Blok}\ and\ \citenamefont
  {McGaugh}(1997)}]{deBlok:1997zlw}%
  \BibitemOpen
  \bibfield  {author} {\bibinfo {author} {\bibfnamefont {W.~J.~G.}\
  \bibnamefont {de~Blok}}\ and\ \bibinfo {author} {\bibfnamefont {S.~S.}\
  \bibnamefont {McGaugh}},\ }\href {\doibase 10.1093/mnras/290.3.533}
  {\bibfield  {journal} {\bibinfo  {journal} {Mon. Not. Roy. Astron. Soc.}\
  }\textbf {\bibinfo {volume} {290}},\ \bibinfo {pages} {533} (\bibinfo {year}
  {1997})},\ \Eprint {http://arxiv.org/abs/astro-ph/9704274}
  {arXiv:astro-ph/9704274 [astro-ph]} \BibitemShut {NoStop}%
\bibitem [{\citenamefont {Oh}\ \emph {et~al.}(2011)\citenamefont {Oh},
  \citenamefont {de~Blok}, \citenamefont {Brinks}, \citenamefont {Walter},\
  and\ \citenamefont {Kennicutt}}]{Oh:2010ea}%
  \BibitemOpen
  \bibfield  {author} {\bibinfo {author} {\bibfnamefont {S.-H.}\ \bibnamefont
  {Oh}}, \bibinfo {author} {\bibfnamefont {W.~J.~G.}\ \bibnamefont {de~Blok}},
  \bibinfo {author} {\bibfnamefont {E.}~\bibnamefont {Brinks}}, \bibinfo
  {author} {\bibfnamefont {F.}~\bibnamefont {Walter}}, \ and\ \bibinfo {author}
  {\bibfnamefont {R.~C.}\ \bibnamefont {Kennicutt}, \bibfnamefont {Jr}},\
  }\href {\doibase 10.1088/0004-6256/141/6/193} {\bibfield  {journal} {\bibinfo
   {journal} {Astron. J.}\ }\textbf {\bibinfo {volume} {141}},\ \bibinfo
  {pages} {193} (\bibinfo {year} {2011})},\ \Eprint
  {http://arxiv.org/abs/1011.0899} {arXiv:1011.0899 [astro-ph.CO]} \BibitemShut
  {NoStop}%
\bibitem [{\citenamefont {Walker}\ and\ \citenamefont
  {Penarrubia}(2011)}]{Walker:2011zu}%
  \BibitemOpen
  \bibfield  {author} {\bibinfo {author} {\bibfnamefont {M.~G.}\ \bibnamefont
  {Walker}}\ and\ \bibinfo {author} {\bibfnamefont {J.}~\bibnamefont
  {Penarrubia}},\ }\href {\doibase 10.1088/0004-637X/742/1/20} {\bibfield
  {journal} {\bibinfo  {journal} {Astrophys. J.}\ }\textbf {\bibinfo {volume}
  {742}},\ \bibinfo {pages} {20} (\bibinfo {year} {2011})},\ \Eprint
  {http://arxiv.org/abs/1108.2404} {arXiv:1108.2404 [astro-ph.CO]} \BibitemShut
  {NoStop}%
\bibitem [{\citenamefont {Boylan-Kolchin}\ \emph {et~al.}(2011)\citenamefont
  {Boylan-Kolchin}, \citenamefont {Bullock},\ and\ \citenamefont
  {Kaplinghat}}]{BoylanKolchin:2011de}%
  \BibitemOpen
  \bibfield  {author} {\bibinfo {author} {\bibfnamefont {M.}~\bibnamefont
  {Boylan-Kolchin}}, \bibinfo {author} {\bibfnamefont {J.~S.}\ \bibnamefont
  {Bullock}}, \ and\ \bibinfo {author} {\bibfnamefont {M.}~\bibnamefont
  {Kaplinghat}},\ }\href {\doibase 10.1111/j.1745-3933.2011.01074.x} {\bibfield
   {journal} {\bibinfo  {journal} {Mon. Not. Roy. Astron. Soc.}\ }\textbf
  {\bibinfo {volume} {415}},\ \bibinfo {pages} {L40} (\bibinfo {year}
  {2011})},\ \Eprint {http://arxiv.org/abs/1103.0007} {arXiv:1103.0007
  [astro-ph.CO]} \BibitemShut {NoStop}%
\bibitem [{\citenamefont {Papastergis}\ \emph {et~al.}(2015)\citenamefont
  {Papastergis}, \citenamefont {Giovanelli}, \citenamefont {Haynes},\ and\
  \citenamefont {Shankar}}]{Papastergis:2014aba}%
  \BibitemOpen
  \bibfield  {author} {\bibinfo {author} {\bibfnamefont {E.}~\bibnamefont
  {Papastergis}}, \bibinfo {author} {\bibfnamefont {R.}~\bibnamefont
  {Giovanelli}}, \bibinfo {author} {\bibfnamefont {M.~P.}\ \bibnamefont
  {Haynes}}, \ and\ \bibinfo {author} {\bibfnamefont {F.}~\bibnamefont
  {Shankar}},\ }\href {\doibase 10.1051/0004-6361/201424909} {\bibfield
  {journal} {\bibinfo  {journal} {Astron. Astrophys.}\ }\textbf {\bibinfo
  {volume} {574}},\ \bibinfo {pages} {A113} (\bibinfo {year} {2015})},\ \Eprint
  {http://arxiv.org/abs/1407.4665} {arXiv:1407.4665 [astro-ph.GA]} \BibitemShut
  {NoStop}%
\bibitem [{\citenamefont {Oman}\ \emph {et~al.}(2015)\citenamefont {Oman} \emph
  {et~al.}}]{Oman:2015xda}%
  \BibitemOpen
  \bibfield  {author} {\bibinfo {author} {\bibfnamefont {K.~A.}\ \bibnamefont
  {Oman}} \emph {et~al.},\ }\href {\doibase 10.1093/mnras/stv1504} {\bibfield
  {journal} {\bibinfo  {journal} {Mon. Not. Roy. Astron. Soc.}\ }\textbf
  {\bibinfo {volume} {452}},\ \bibinfo {pages} {3650} (\bibinfo {year}
  {2015})},\ \Eprint {http://arxiv.org/abs/1504.01437} {arXiv:1504.01437
  [astro-ph.GA]} \BibitemShut {NoStop}%
\bibitem [{\citenamefont {Oman}\ \emph {et~al.}(2016)\citenamefont {Oman},
  \citenamefont {Navarro}, \citenamefont {Sales}, \citenamefont {Fattahi},
  \citenamefont {Frenk}, \citenamefont {Sawala}, \citenamefont {Schaller},\
  and\ \citenamefont {White}}]{Oman:2016zjn}%
  \BibitemOpen
  \bibfield  {author} {\bibinfo {author} {\bibfnamefont {K.~A.}\ \bibnamefont
  {Oman}}, \bibinfo {author} {\bibfnamefont {J.~F.}\ \bibnamefont {Navarro}},
  \bibinfo {author} {\bibfnamefont {L.~V.}\ \bibnamefont {Sales}}, \bibinfo
  {author} {\bibfnamefont {A.}~\bibnamefont {Fattahi}}, \bibinfo {author}
  {\bibfnamefont {C.~S.}\ \bibnamefont {Frenk}}, \bibinfo {author}
  {\bibfnamefont {T.}~\bibnamefont {Sawala}}, \bibinfo {author} {\bibfnamefont
  {M.}~\bibnamefont {Schaller}}, \ and\ \bibinfo {author} {\bibfnamefont
  {S.~D.~M.}\ \bibnamefont {White}},\ }\href {\doibase 10.1093/mnras/stw1251}
  {\bibfield  {journal} {\bibinfo  {journal} {Mon. Not. Roy. Astron. Soc.}\
  }\textbf {\bibinfo {volume} {460}},\ \bibinfo {pages} {3610} (\bibinfo {year}
  {2016})},\ \Eprint {http://arxiv.org/abs/1601.01026} {arXiv:1601.01026
  [astro-ph.GA]} \BibitemShut {NoStop}%
\bibitem [{\citenamefont {Vogelsberger}\ \emph {et~al.}(2016)\citenamefont
  {Vogelsberger}, \citenamefont {Zavala}, \citenamefont {Cyr-Racine},
  \citenamefont {Pfrommer}, \citenamefont {Bringmann},\ and\ \citenamefont
  {Sigurdson}}]{Vogelsberger:2015gpr}%
  \BibitemOpen
  \bibfield  {author} {\bibinfo {author} {\bibfnamefont {M.}~\bibnamefont
  {Vogelsberger}}, \bibinfo {author} {\bibfnamefont {J.}~\bibnamefont
  {Zavala}}, \bibinfo {author} {\bibfnamefont {F.-Y.}\ \bibnamefont
  {Cyr-Racine}}, \bibinfo {author} {\bibfnamefont {C.}~\bibnamefont
  {Pfrommer}}, \bibinfo {author} {\bibfnamefont {T.}~\bibnamefont {Bringmann}},
  \ and\ \bibinfo {author} {\bibfnamefont {K.}~\bibnamefont {Sigurdson}},\
  }\href {\doibase 10.1093/mnras/stw1076} {\bibfield  {journal} {\bibinfo
  {journal} {Mon. Not. Roy. Astron. Soc.}\ }\textbf {\bibinfo {volume} {460}},\
  \bibinfo {pages} {1399} (\bibinfo {year} {2016})},\ \Eprint
  {http://arxiv.org/abs/1512.05349} {arXiv:1512.05349 [astro-ph.CO]}
  \BibitemShut {NoStop}%
\bibitem [{\citenamefont {Huo}\ \emph {et~al.}(2018)\citenamefont {Huo},
  \citenamefont {Kaplinghat}, \citenamefont {Pan},\ and\ \citenamefont
  {Yu}}]{Huo:2017vef}%
  \BibitemOpen
  \bibfield  {author} {\bibinfo {author} {\bibfnamefont {R.}~\bibnamefont
  {Huo}}, \bibinfo {author} {\bibfnamefont {M.}~\bibnamefont {Kaplinghat}},
  \bibinfo {author} {\bibfnamefont {Z.}~\bibnamefont {Pan}}, \ and\ \bibinfo
  {author} {\bibfnamefont {H.-B.}\ \bibnamefont {Yu}},\ }\href {\doibase
  10.1016/j.physletb.2018.06.024} {\bibfield  {journal} {\bibinfo  {journal}
  {Phys. Lett.}\ }\textbf {\bibinfo {volume} {B783}},\ \bibinfo {pages} {76}
  (\bibinfo {year} {2018})},\ \Eprint {http://arxiv.org/abs/1709.09717}
  {arXiv:1709.09717 [hep-ph]} \BibitemShut {NoStop}%
\bibitem [{\citenamefont {Moore}\ \emph {et~al.}(1999)\citenamefont {Moore},
  \citenamefont {Ghigna}, \citenamefont {Governato}, \citenamefont {Lake},
  \citenamefont {Quinn}, \citenamefont {Stadel},\ and\ \citenamefont
  {Tozzi}}]{Moore:1999nt}%
  \BibitemOpen
  \bibfield  {author} {\bibinfo {author} {\bibfnamefont {B.}~\bibnamefont
  {Moore}}, \bibinfo {author} {\bibfnamefont {S.}~\bibnamefont {Ghigna}},
  \bibinfo {author} {\bibfnamefont {F.}~\bibnamefont {Governato}}, \bibinfo
  {author} {\bibfnamefont {G.}~\bibnamefont {Lake}}, \bibinfo {author}
  {\bibfnamefont {T.~R.}\ \bibnamefont {Quinn}}, \bibinfo {author}
  {\bibfnamefont {J.}~\bibnamefont {Stadel}}, \ and\ \bibinfo {author}
  {\bibfnamefont {P.}~\bibnamefont {Tozzi}},\ }\href {\doibase 10.1086/312287}
  {\bibfield  {journal} {\bibinfo  {journal} {Astrophys. J.}\ }\textbf
  {\bibinfo {volume} {524}},\ \bibinfo {pages} {L19} (\bibinfo {year}
  {1999})},\ \Eprint {http://arxiv.org/abs/astro-ph/9907411}
  {arXiv:astro-ph/9907411 [astro-ph]} \BibitemShut {NoStop}%
\bibitem [{\citenamefont {Klypin}\ \emph {et~al.}(1999)\citenamefont {Klypin},
  \citenamefont {Kravtsov}, \citenamefont {Valenzuela},\ and\ \citenamefont
  {Prada}}]{Klypin:1999uc}%
  \BibitemOpen
  \bibfield  {author} {\bibinfo {author} {\bibfnamefont {A.~A.}\ \bibnamefont
  {Klypin}}, \bibinfo {author} {\bibfnamefont {A.~V.}\ \bibnamefont
  {Kravtsov}}, \bibinfo {author} {\bibfnamefont {O.}~\bibnamefont
  {Valenzuela}}, \ and\ \bibinfo {author} {\bibfnamefont {F.}~\bibnamefont
  {Prada}},\ }\href {\doibase 10.1086/307643} {\bibfield  {journal} {\bibinfo
  {journal} {Astrophys. J.}\ }\textbf {\bibinfo {volume} {522}},\ \bibinfo
  {pages} {82} (\bibinfo {year} {1999})},\ \Eprint
  {http://arxiv.org/abs/astro-ph/9901240} {arXiv:astro-ph/9901240 [astro-ph]}
  \BibitemShut {NoStop}%
\bibitem [{\citenamefont {Fattahi}\ \emph {et~al.}(2016)\citenamefont
  {Fattahi}, \citenamefont {Navarro}, \citenamefont {Sawala}, \citenamefont
  {Frenk}, \citenamefont {Sales}, \citenamefont {Oman}, \citenamefont
  {Schaller},\ and\ \citenamefont {Wang}}]{Fattahi:2016nld}%
  \BibitemOpen
  \bibfield  {author} {\bibinfo {author} {\bibfnamefont {A.}~\bibnamefont
  {Fattahi}}, \bibinfo {author} {\bibfnamefont {J.~F.}\ \bibnamefont
  {Navarro}}, \bibinfo {author} {\bibfnamefont {T.}~\bibnamefont {Sawala}},
  \bibinfo {author} {\bibfnamefont {C.~S.}\ \bibnamefont {Frenk}}, \bibinfo
  {author} {\bibfnamefont {L.~V.}\ \bibnamefont {Sales}}, \bibinfo {author}
  {\bibfnamefont {K.}~\bibnamefont {Oman}}, \bibinfo {author} {\bibfnamefont
  {M.}~\bibnamefont {Schaller}}, \ and\ \bibinfo {author} {\bibfnamefont
  {J.}~\bibnamefont {Wang}},\ }\href@noop {} {\  (\bibinfo {year} {2016})},\
  \Eprint {http://arxiv.org/abs/1607.06479} {arXiv:1607.06479 [astro-ph.GA]}
  \BibitemShut {NoStop}%
\bibitem [{\citenamefont {Jethwa}\ \emph {et~al.}(2018)\citenamefont {Jethwa},
  \citenamefont {Erkal},\ and\ \citenamefont {Belokurov}}]{Jethwa:2016gra}%
  \BibitemOpen
  \bibfield  {author} {\bibinfo {author} {\bibfnamefont {P.}~\bibnamefont
  {Jethwa}}, \bibinfo {author} {\bibfnamefont {D.}~\bibnamefont {Erkal}}, \
  and\ \bibinfo {author} {\bibfnamefont {V.}~\bibnamefont {Belokurov}},\ }\href
  {\doibase 10.1093/mnras/stx2330} {\bibfield  {journal} {\bibinfo  {journal}
  {Mon. Not. Roy. Astron. Soc.}\ }\textbf {\bibinfo {volume} {473}},\ \bibinfo
  {pages} {2060} (\bibinfo {year} {2018})},\ \Eprint
  {http://arxiv.org/abs/1612.07834} {arXiv:1612.07834 [astro-ph.GA]}
  \BibitemShut {NoStop}%
\bibitem [{\citenamefont {Kim}\ \emph {et~al.}(2017)\citenamefont {Kim},
  \citenamefont {Peter},\ and\ \citenamefont {Hargis}}]{Kim:2017iwr}%
  \BibitemOpen
  \bibfield  {author} {\bibinfo {author} {\bibfnamefont {S.~Y.}\ \bibnamefont
  {Kim}}, \bibinfo {author} {\bibfnamefont {A.~H.~G.}\ \bibnamefont {Peter}}, \
  and\ \bibinfo {author} {\bibfnamefont {J.~R.}\ \bibnamefont {Hargis}},\
  }\href@noop {} {\  (\bibinfo {year} {2017})},\ \Eprint
  {http://arxiv.org/abs/1711.06267} {arXiv:1711.06267 [astro-ph.CO]}
  \BibitemShut {NoStop}%
\bibitem [{\citenamefont {Dasgupta}\ and\ \citenamefont
  {Kopp}(2014)}]{Dasgupta:2013zpn}%
  \BibitemOpen
  \bibfield  {author} {\bibinfo {author} {\bibfnamefont {B.}~\bibnamefont
  {Dasgupta}}\ and\ \bibinfo {author} {\bibfnamefont {J.}~\bibnamefont
  {Kopp}},\ }\href {\doibase 10.1103/PhysRevLett.112.031803} {\bibfield
  {journal} {\bibinfo  {journal} {Phys. Rev. Lett.}\ }\textbf {\bibinfo
  {volume} {112}},\ \bibinfo {pages} {031803} (\bibinfo {year} {2014})},\
  \Eprint {http://arxiv.org/abs/1310.6337} {arXiv:1310.6337 [hep-ph]}
  \BibitemShut {NoStop}%
\bibitem [{\citenamefont {Bringmann}\ \emph {et~al.}(2014)\citenamefont
  {Bringmann}, \citenamefont {Hasenkamp},\ and\ \citenamefont
  {Kersten}}]{Bringmann:2013vra}%
  \BibitemOpen
  \bibfield  {author} {\bibinfo {author} {\bibfnamefont {T.}~\bibnamefont
  {Bringmann}}, \bibinfo {author} {\bibfnamefont {J.}~\bibnamefont
  {Hasenkamp}}, \ and\ \bibinfo {author} {\bibfnamefont {J.}~\bibnamefont
  {Kersten}},\ }\href {\doibase 10.1088/1475-7516/2014/07/042} {\bibfield
  {journal} {\bibinfo  {journal} {JCAP}\ }\textbf {\bibinfo {volume} {1407}},\
  \bibinfo {pages} {042} (\bibinfo {year} {2014})},\ \Eprint
  {http://arxiv.org/abs/1312.4947} {arXiv:1312.4947 [hep-ph]} \BibitemShut
  {NoStop}%
\end{thebibliography}
\end{document}